\pdfoutput=1
\documentclass{jfm}
\usepackage{graphicx}
\usepackage{braket}
\usepackage{tabularx}
\usepackage{epstopdf, epsfig}
\usepackage{pifont}
\usepackage[dvipsnames]{xcolor}
\usepackage{gensymb}
\usepackage{url}
\usepackage{multirow}
\usepackage{hhline}

\definecolor{lblue1}{RGB}{179, 179, 255}
\definecolor{dblue1}{RGB}{0, 0, 150}
\definecolor{lgreen1}{RGB}{106, 255, 107}
\definecolor{dgreen1}{RGB}{0, 150, 0}

\DeclareSymbolFont{extraup}{U}{zavm}{m}{n}
\DeclareMathSymbol{\varheart}{\mathalpha}{extraup}{86}
\DeclareMathSymbol{\vardiamond}{\mathalpha}{extraup}{87}

\shorttitle{Coherent structures and secondary flow in turbulent square duct}
\shortauthor{M. Atzori, R. Vinuesa, A. Lozano-Dur\'an and P. Schlatter}

\title{Coherent structures and secondary flow in turbulent square duct}

\author{M. Atzori\aff{1}
  \corresp{\email{atzori@mech.kth.se}},
  R. Vinuesa\aff{1}, A. Lozano-Dur\'an\aff{2}
 \and P. Schlatter\aff{1}}

\affiliation{\aff{1}Linn\'e FLOW Centre, KTH Mechanics, SE-100 44 Stockholm
\aff{2}Center for Turbulence Research, Stanford University, CA, USA}

\begin{document}

\maketitle

\begin{abstract}
The aim of the present work is to investigate the role of coherent structures in the generation of the secondary flow in a turbulent square duct. 
The coherent structures are defined as connected regions of flow where the product of the instantaneous fluctuations of two velocity components is higher than a threshold based on the long-time turbulence statistics, in the spirit of the three-dimensional quadrant analysis proposed by Lozano-Dur\'an \textit{et al} (\textit{J.~Fluid Mech.}, vol. 694, 2012, pp. 100--130). 
We consider both the direct contribution of the structures to the mean in-plane velocity components and their geometrical properties. 
The instantaneous phenomena taking place in the turbulent duct are compared with turbulent channel flow at Reynolds numbers of $Re_{\tau}=180$ and $360$, based on friction velocity at the center-plane and channel half height.
In the core region of the duct, the fractional contribution of intense events to the wall-normal component of the mean velocity is in very good agreement with that in the channel, despite the presence of the secondary flow in the former. 
Additionally, the shapes of the three-dimensional objects do not differ significantly in both flows. 
On the other hand, in the corner region of the duct, the proximity of the walls affects both the geometrical properties of the coherent structures and the contribution to the mean component of the vertical velocity, which is less relevant than that of the complementary portion of the flow not included in such objects. 
Our results show however that strong Reynolds shear-stress events, despite the differences observed between channel and duct, do not contribute directly to the secondary motion, and thus other phenomena need to be considered instead.
\end{abstract}

\begin{keywords}
turbulence simulation, turbulent flows, turbulent boundary layers 
\end{keywords}

\section{Introduction}
Turbulent flows with inhomogeneity in both cross-stream directions exhibit non-vanishing values of the mean streamwise vorticity and of the cross-stream components of the mean velocity. 
This phenomenon is known as secondary flow of Prandtl's second kind \citep{prandtl26} and, despite the fact that it is of the order of magnitude of few percentage points of the streamwise velocity component, it has important effects on mean flow properties such as skin friction and effective diffusivity \citep{gess73}.
Contrary to the secondary flow of Prandtl's first kind, which consists of a usually skew-induced redistribution of the vorticity (\textit{e.g.}: in curved pipes, \cite{noor13}), the secondary flow of Prandtl's second kind is present in turbulent flows only. As intuitively discussed by \cite{eins58}, it is related to the correlation between the cross-stream components of velocity and to the anisotropy of the normal stress, which appear in the production terms of the mean streamwise vorticity transport equation. 
However, because of its weak intensity, it is challenging to identify the instantaneous phenomena in the turbulent flow which lead to the mean secondary flow and several hypothesis have been formulated to explain its dynamics. 
\par After the pioneering work by \cite{niku30}, \cite{gess65} performed one of the first experimental studies on square ducts and on rectangular ducts with aspect ratio $2$ (aspect ratio is defined as the total duct width divided by its total height). 
They considered bulk Reynolds numbers between $Re_b=25,000$ and $Re_b=150,000$ ($Re_b=h U_b / \nu$, where $h$ is the half-height of the duct, $U_b$ is the bulk velocity and $\nu$ is the kinematic viscosity) and they identified the balance between the Reynolds stresses and static-pressure gradients as the mechanism which generates the secondary flow. 
In a further work \cite{gess73} stated that the secondary flow is mainly due to gradients of the Reynolds-shear stress in the corner region and that the contribution from normal fluctuations is less relevant.
\par The first direct numerical simulation (DNS) of turbulent flow in square duct was carried out by \cite{gavr92} at a friction Reynolds number based on the friction velocity averaged over the perimeter of $Re^*_\tau=150$, corresponding to approximately $Re_b\approx2200$. 
Here the average friction Reynolds number $Re^*_\tau=h u^*_\tau / \nu$ is defined based on $h$, $\nu$ and the friction velocity averaged over the perimeter, which in the case of the squared duct is $u^*_\tau=\sqrt{- (h/2) ({\rm d} P / {\rm d} x)}$, where ${\rm d} P / {\rm d} x$ is the streamwise pressure gradient.
\cite{gavr92} recognised for the first time that two different velocity scalings are needed for the secondary motion in the viscous sub-layer and in the core region. 
\par \cite{huse93} carried out DNSs up to $Re^*_\tau=300$ and employed a generalised quadrant analysis \citep{wall72,sslu72} for the three non-diagonal components of the Reynolds-stress tensor.
These authors conjectured that the existence of the secondary flow is related to the fact that the turbulent production is stronger at the duct centre-plane than in channel flows, but lower along the corner bisectors, and that it can be the result of interaction between ejections from perpendicular walls. 
According to their findings, near the corner the couple of vortices induced by ejection events have a preferential direction of rotation, which leads to the momentum transport from the wall bisectors to the corners. 
\par \cite{uhlm07} and \cite{pine10} studied duct flow in marginally turbulent state ($Re^*_\tau=80$) and in the fully-turbulent range $Re^*_\tau=80-225$, respectively and clarified the effects of the geometrical constrains on the organisation of the instantaneous turbulent coherent structures in the near-wall region, together with the connection between those and the scaling properties of the secondary flow. 
\par The wall-shear stress in turbulent duct flows exhibits a characteristic behaviour, showing a certain number of local extrema: the position of the local maxima and minima correspond to the preferential location of high- and low-speed streaks, which have a spanwise spacing of $\approx50^+$ \citep[as in channel flow, see for instance][]{kim87}. 
\par Here the superscript $+$ denotes scaling with the viscous length $\ell^{*} = \nu / u^*_\tau$.
This pattern disappears far from the side walls at Reynolds numbers high enough to have the streaks uniformly distributed, but in square ducts a local maximum of the wall-shear stress remains visible near the corner at a spanwise distance of around than $50^+$, indicating the presence of a persistent streak at that location. The locations of the extrema of the streamwise vorticity have a behaviour similar to that of the streaks, thereby scaling in inner units. On the other hand, the maxima of the streamfunction scale in outer units, suggesting a relation with large-scale motion in the outer region \citep{pine10,piro18}. 
\par In the study by \cite{pine10}, the behaviour of the instantaneous streaks was described by considering the probability density functions (PDF) of the location of the minimum and maximum values of the wall-shear stress. Furthermore, it was shown that the secondary motion is associated with a discrepancy in the probability of occurrence of clockwise and anti-clockwise rotating vortices. The same fact was also reported for duct at marginally turbulent state by \cite{uhlm07}. 
The connection between coherent structures and secondary motion has been investigated by \cite{vinu16} as well, who considered the probability distribution of vortex clusters detected via the $\lambda_2$ criterion \citep{jeon95} in square and rectangular ducts at a friction Reynolds number at the centre-plane of $Re_\tau=180$ and $360$. 
Note that $Re_\tau=h u_\tau / \nu$ is defined based on $u_\tau=\sqrt{\tau_w/\rho}$, where $\tau_w$ is the wall-shear stress at the centre-plane and $\rho$ is the fluid density.  
These authors found that vortex clusters are more common in a region between the core of the duct and the wall, with the maximum probability located at a wall-normal distance of around $\approx 40$ viscous units. 
\par \cite{piro18} studied a wide Reynolds-number range with DNS, up to $Re^*_\tau=1000$ ($Re_b=40,000$), and confirmed that the secondary motion has two different contributions: a ``core circulation'', which scales in outer units, and a ``corner circulation'', the intensity of which also scales in outer units, but with size which scales in inner units. 
The latter is related to higher vorticity values but it has a lower contribution to the total circulation, which decreases for increasing Reynolds number, according to its length scales.
\par More recently, \cite{gavr19} studied in detail the evolution of the secondary motion with Reynolds number, considering five cases from $Re^*_\tau=157$ to $Re^*_\tau=861$. 
In this study the author identifies a vortex-generation process in the near-corner region as the primary cause of the secondary motion. 
This is due to the presence of an inflection point in the mean streamwise velocity along the corner bisector.
He also conjectured that the core circulation may be due to ``very long and long-living structures most evident in the form of oblique near-wall streaks'', which can propagate the corner effects to the core of the domain.
For cases at friction Reynolds number lower than $\approx300$ the corner circulation, generated by mean flow instability, is considered to be predominant. 
\par Turbulent ducts of other shapes have been also considered in the literature, showing how the geometry can affect the secondary flow. \cite{vinu18} studied rectangular ducts with aspect ratios from $1$ to $14$ at $Re_\tau=180$ and $360$. 
They found that the centres of the vortices attached to the shorter walls remain approximately constant for all the aspect ratios larger than one. 
On the other hand, the centres of the vortices on the longer walls move farther from the corner as the aspect ration increases and the secondary motion extends up to $4.3h$ far from the corner for the widest case. 
\cite{mari16} performed DNS and large-eddy simulation (LES) of hexagonal duct from $Re_\tau\approx180$ up to $550$, reporting that the intensity of the secondary motion reduces as $Re$ increases. 
This is due to the fact that interactions between events from the two walls become less likely, because of the angle of $120\degree$ between adjacent walls, reflected by the non-diagonal Reynolds stress term, which decreases with $Re$ as well. 
\par The present study is motivated by the lack of information in the literature regarding the instantaneous mechanism(s) which generates the secondary flow in turbulent duct. In most of the studies mentioned so far, the main emphasis was on the scaling properties of the secondary flow and the interpretation of the turbulence statistics, while in only few studies on ducts instantaneous turbulent structures have been examined. However, to the authors' knowledge, no quantitative evaluation of the contribution that such structures provide to the secondary motion has been performed before, nor their geometrical properties have been discussed in detail. These are the goals of the present work. 
\par The paper is organised as follows: in section 2 we describe the data-set and the detection technique; in section 3 we evaluate the fractional contribution of coherent structure to the secondary motion and their geometrical properties, performing a comparative analysis with turbulent channel flow at similar Reynolds number; in section 4 we discuss our conclusions. 
\section{Methodology}
\subsection{Numerical experiments}
We consider turbulent flow in square duct at $Re_\tau=180$ and $Re_\tau=360$ at the centre-plane, which is part of the data-set described by \cite{vinu18}, and a new data-set for turbulent channel flow at the same friction Reynolds number, similar to that described by \cite{alam06}.
\par The DNSs for the duct were performed with the spectral-element code \textit{Nek5000}, developed by \cite{fisc08}. The spatial derivatives in the incompressible Navier--Stokes equations are discretized employing a Garlerkin method, following the $P_N-P_{N-2}$ formulation by \cite{pate84} and the solution is expressed within each spectral element in terms of a nodal-base of Legendre polynomials on the Gauss--Lobatto--Legendre (GLL) quadrature points. The discretization of the time derivatives is explicit for the non-linear terms and implicit for the viscous term, employing an extrapolation and a backward differentiation scheme, respectively, both of the third order. In order to trigger transition to turbulence, we employed tripping through a volume force, implemented as proposed by \cite{schl12}, and active only during an initial transient. For a more complete description of the numerical setup we refer to \cite{vinu14}. 
\par The channel flow simulations are performed by integrating the incompressible Navier--Stokes equation in the form of evolution equations for the wall-normal vorticity and for the Laplacian of the wall-normal velocity, as in \cite{kim87}, and the spatial discretization is dealiased Fourier in the two wall-parallel directions and Chebychev polynomials in $y$. Time stepping is performed with the third-order semi-implicit Runge--Kutta as in \cite{mose99}. Both channels have computational domains in the streamwise and spanwise directions large enought to ensure that the largest structures of the flow are reasonably well represented.
The resolutions as well as the number of fields employed in the analysis for each case are reported in Table~\ref{tab:info}.
\begin{table}
    \centering
    \begin{tabular}{ccccccccccc}
     Case & $Re_b$  & $Re^*_\tau$ & $Re_\tau$ & No. grid points & $\Delta x^+$ & $\Delta y^+$ & $\Delta z^+$ & No. fields  \\
         \hline 
     \emph{D. 180} & $2500$ & 165 & 178 & $27.4\cdot10^6$ & $(1.98,\,9.80)$ & $(0.09,\,4.74)$ & $(0.09,\,4.74)$ & $879$ \\
         \hline 
    \emph{D. 360} & $5693$ & 342 & 356 & $122.4\cdot10^6$ & $(1.99,\,9.88)$ & $(0.15,\,4.65)$ & $(0.15,\,4.65)$ & $404$ \\
     \hline
     \emph{C. 180} & $3250$ & -- & 186 & $38.1\cdot10^6$ & $9.1$ & $(0.10,\,6.1)$ & $4.5$ & $76$  \\
     \hline
     \emph{C. 360} & $6739$ & -- & 354 & $151.8\cdot10^6$ & $8.7$ & $(0.05,\,5.8)$ & $4.3$ & $37$
    \end{tabular}
    \caption{Simulation parameters for the considered cases. The resolution is indicated in terms of the maximum and minimum grid spacing in inner units. The superscript $*$ indicates an averaged quantity over the complete duct perimeter.}
    \label{tab:info}
\end{table}
\par In the present work we adopt the following notation: $U$ denotes the streamwise mean velocity component, $V$ the vertical component and $W$ the horizontal component, for both channel and duct. 
The mean is intended as an average over time and the homogeneous directions (streamwise for the duct, streamwise and spanwise for the channel). 
The fluctuations with respect of the mean are denoted with lower-case variables, the components of the Reynolds-stress tensor with $\overline{u_i u_j}$, whereas root-mean-squared values and ensemble averages over the fields are indicated explicitly as $\xi_{\rm rms}$ and $\xi_{\rm ens}$, respectively.  Note that the differences between time and ensemble average are due only to statistical uncertainty, \emph{e.g.} $U = U_{\rm ens}$, for a number of samples large enough.
In the duct the origin of the reference frame is located at the bottom-left corner for an observer looking toward the streamwise direction, so that the two vertical walls are at $z=0$ and $z=2h$ and the two horizontal walls at $y=0$ and $y=2h$. Unless otherwise stated, outer scaling is employed, \textit{i.e.} the velocity components are scaled with the bulk velocity and the spatial coordinates with the half-height of the duct/channel.
\par The length of the computational domain is $L_x=25$ for the duct and the size in the streamwise and spanwise directions for the channel are, respectively, $L_x=12\pi$ and $L_z=4\pi$ for $Re_\tau=180$ and $L_x=8\pi$ and $L_z=3\pi$ for $Re_\tau=360$. 
\subsection{Identification of coherent structures}
The coherent structures are defined as connected regions of the domain where the instantaneous velocity fluctuations are higher than a threshold based on the value of the root-mean-square of the considered velocity components (the connected regions are defined through a six-points connectivity). This approach was introduced as a three-dimensional extension of the classic quadrant analysis by \cite{loza12}, who studied intense events of tangential Reynolds stress ($uv$) in turbulent channel flows. 
\par In channel flow the most relevant term of the Reynolds-stress tensor is considered to be $\overline{uv}$, since the only non-vanishing contribution to the production of the mean turbulent kinetic energy is $\overline{uv} \,\partial U/\partial y$. 
However, in duct flow the only terms among $\overline{u_i u_j} \,\partial U_i /\partial x_j$ which vanish are the ones with the spatial derivative in the streamwise direction. 
Furthermore, because of the symmetry, $uv$ and $uw$ events are equivalent to $uv$ in the channel near the horizontal and vertical walls, respectively, and $vw$ are equivalent to $vw$ in the channel in the proximity of the walls. 
Hence, in this study we consider intense events for all the off-diagonal components of the Reynolds-stress tensor \citep[as done by][]{huse93}, defined according to the conditions:
\begin{equation}
    |uv| > H_{uv} u_{\rm rms} v_{\rm rms}\,,\quad 
    |uw| > H_{uw} u_{\rm rms} w_{\rm rms}\quad {\rm and}\quad 
    |vw| > H_{vw} v_{\rm rms} w_{\rm rms}\,.
    \label{eqn:threshold}
\end{equation}
These conditions are consistent with those of \cite{sslu72} and \cite{loza12}.  
The actual thresholds $H_{\rm XX}$ (called ``hyperbolic hole'', referring to the hyperbola $uv=H_{uv}$ in the $uv$ plane) are single numerical values, while the normalisation terms $u_{\rm rms} v_{\rm rms}$, $u_{\rm rms} w_{\rm rms}$ and $v_{\rm rms} w_{\rm rms}$ are one-dimensional profiles for the channel and two-dimensional fields for the duct. 
Note that the normalised threshold for $uv$ in the duct is highest near the vertical walls, a region where $v$ is a fluctuation for the spanwise component of the velocity with respect to the side-wall boundary layers. 
\par
\begin{figure}
    \centering
    \includegraphics[width=0.45\textwidth]{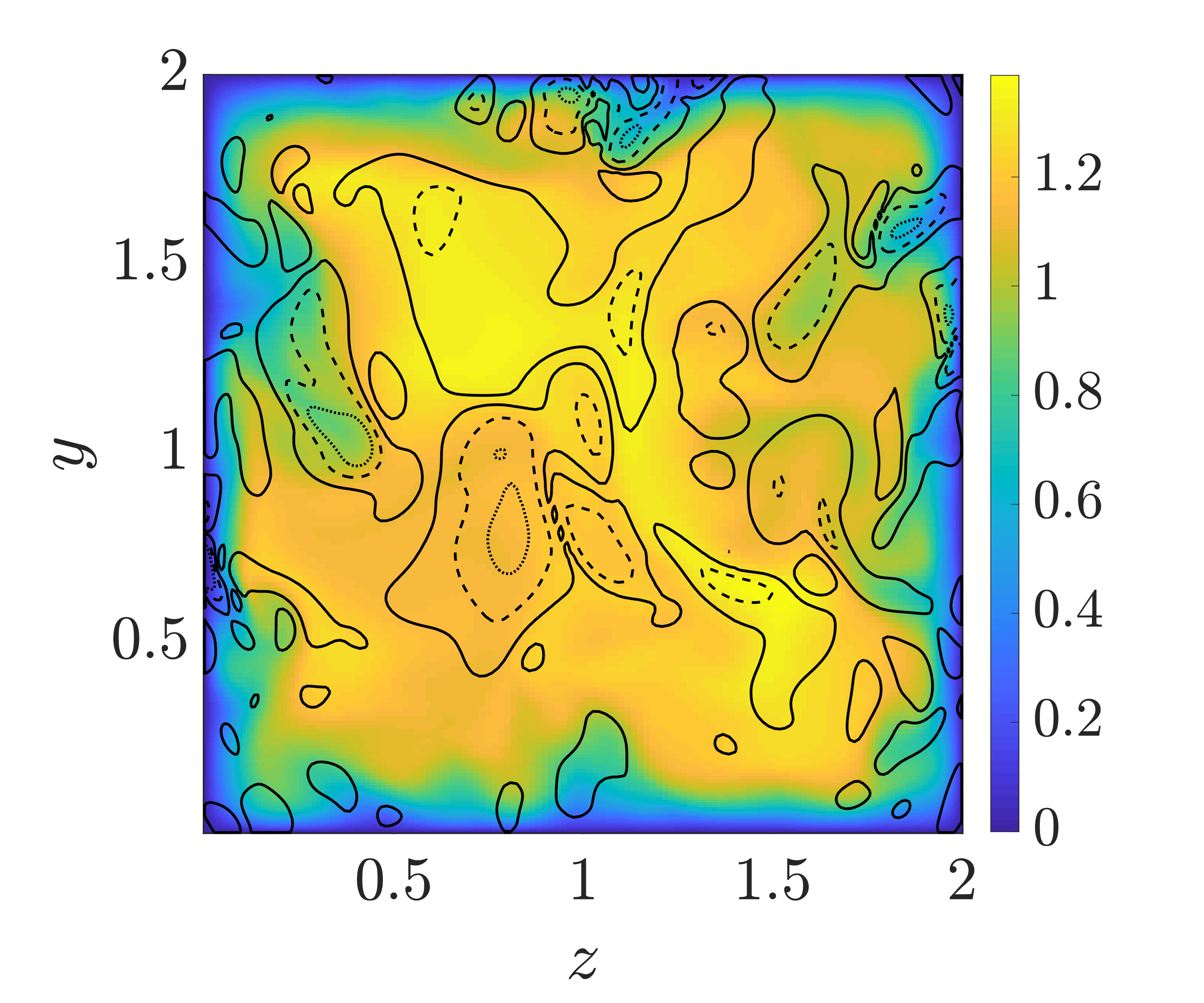}
    \includegraphics[width=0.375\textwidth]{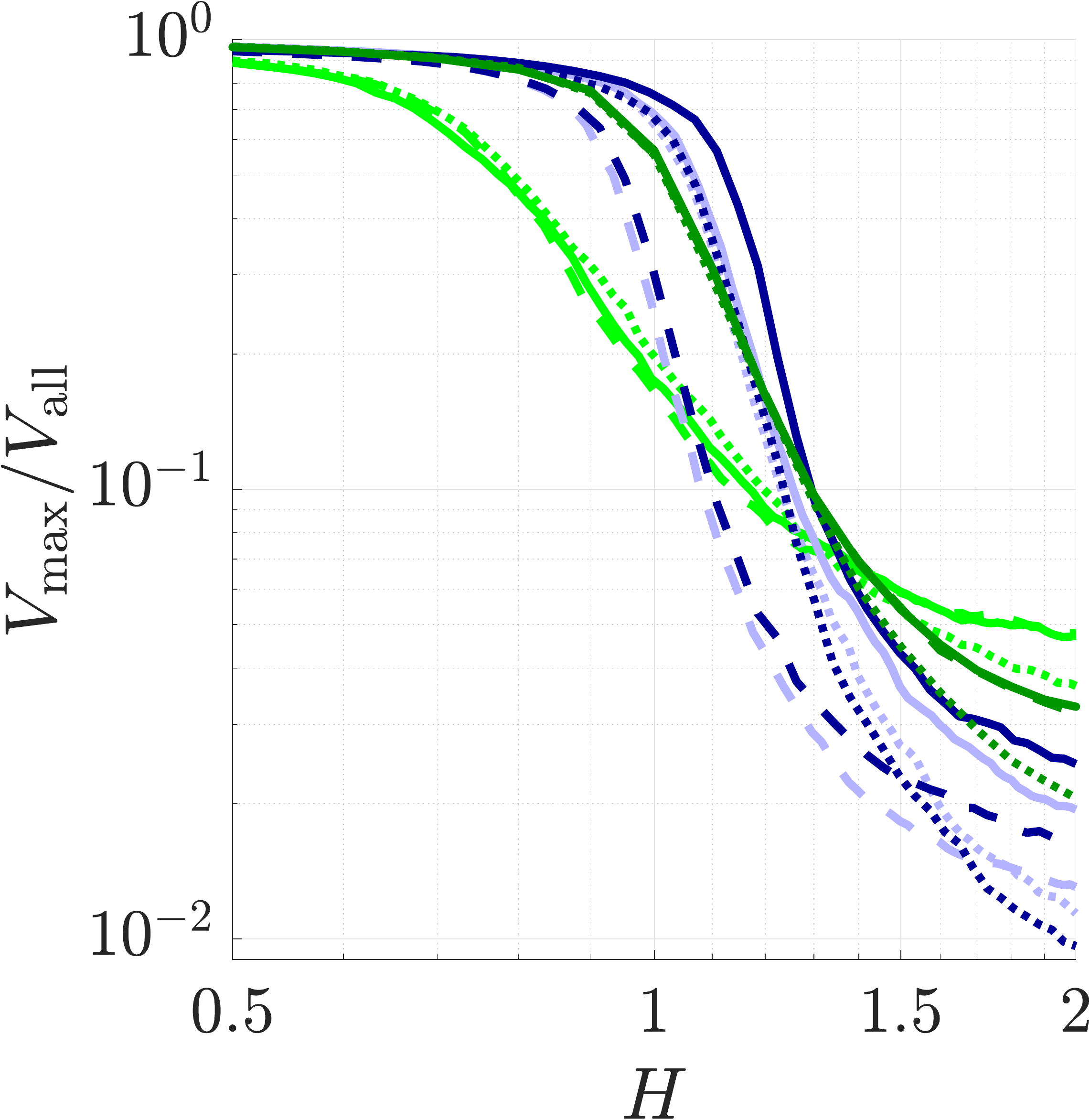}\\
    \includegraphics[width=0.375\textwidth]{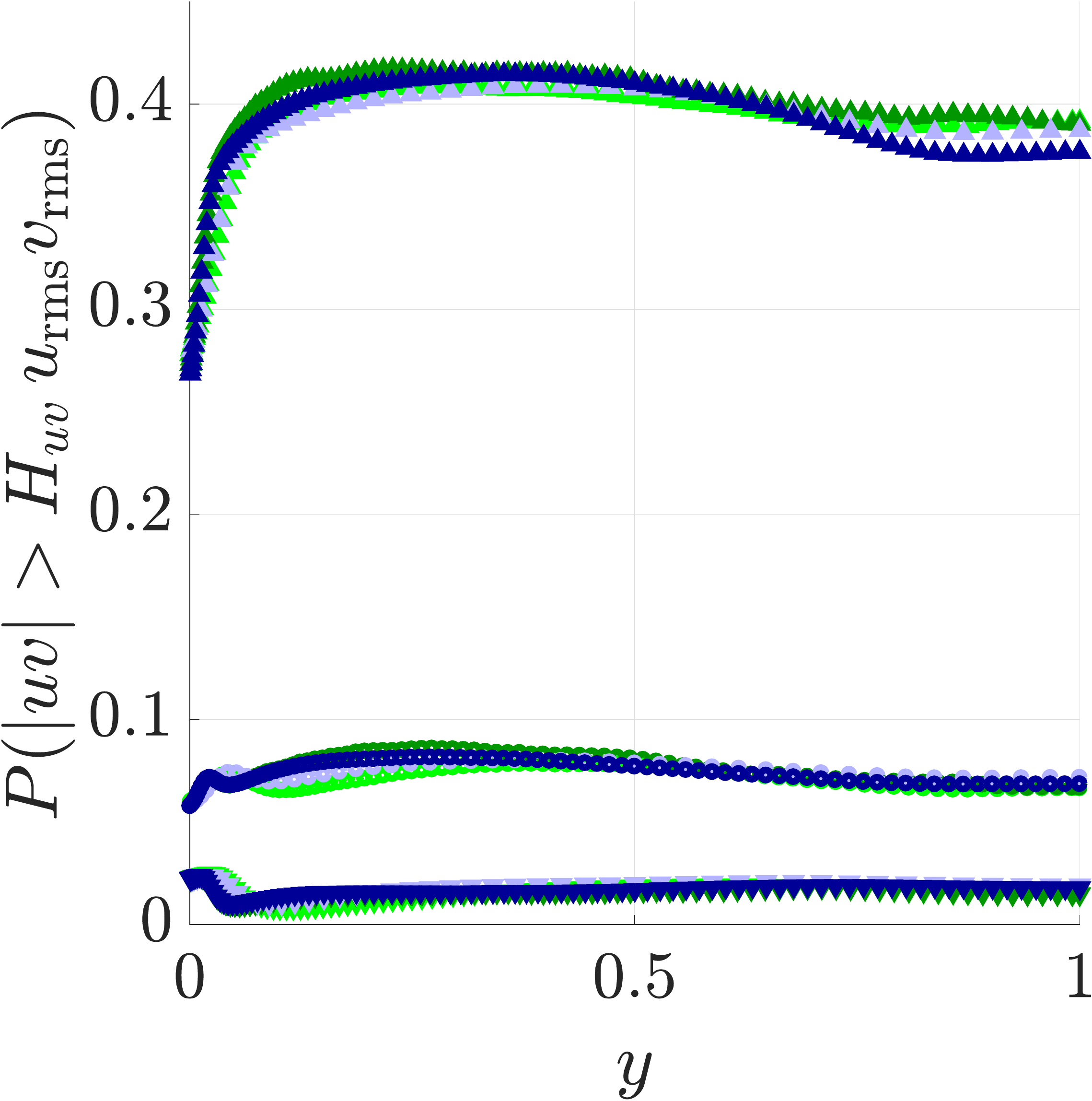}~
    \includegraphics[width=0.375\textwidth]{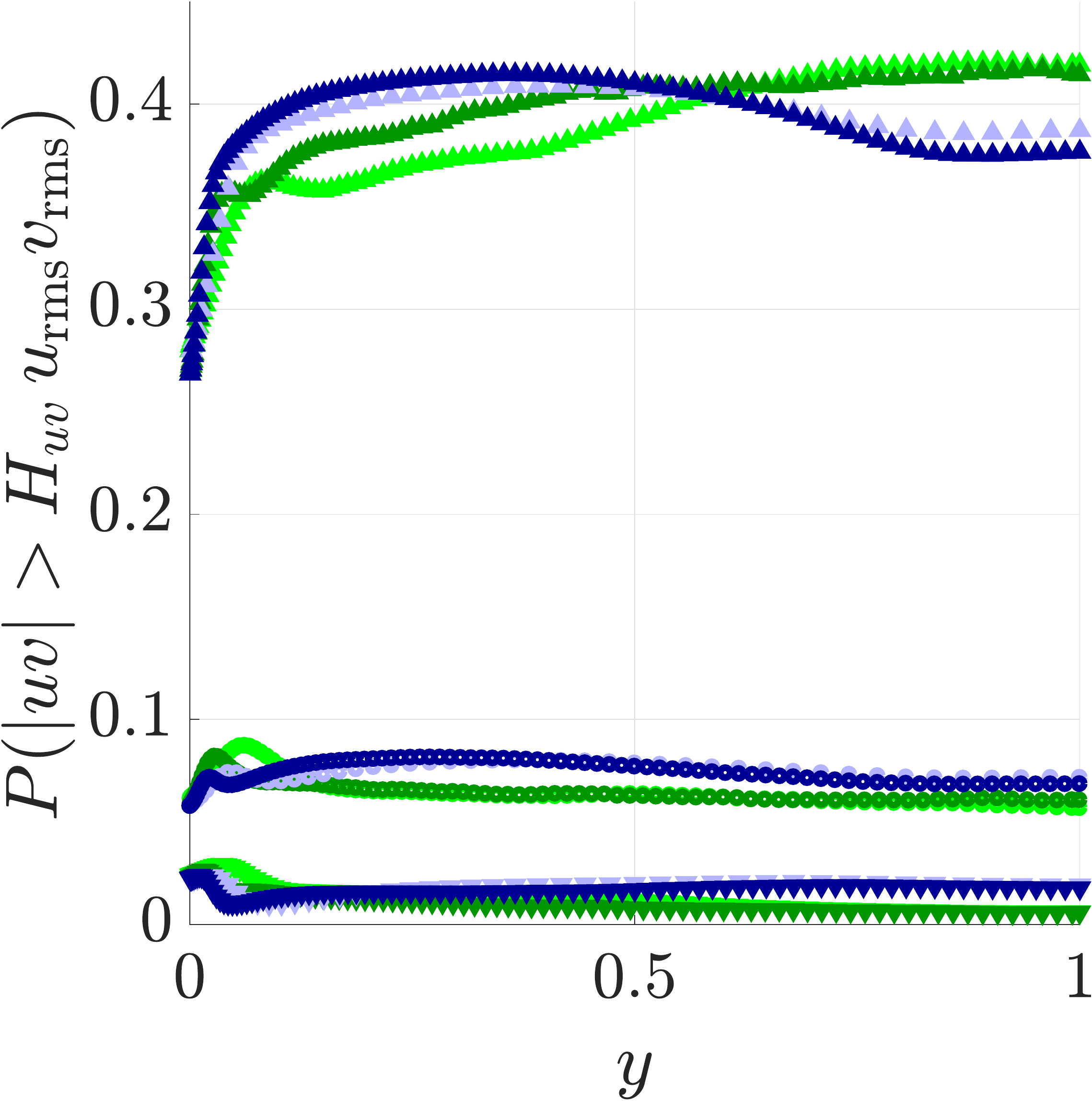}
    \caption{Top row: (left) slice orthogonal to the streamwise direction for duct flow at $Re_\tau=180$, coloured with the instantaneous streamwise component of the velocity. The black solid, dashed and dotted lines represent the boundaries of intense $uv$ events sampled for $H_{uv}=0.5$, $2.0$ and $4.0$, respectively. (Right) Percolation diagram for: $uv$ (solid lines), $uw$ (dashed lines) and $vw$ (dotted lines). Channel in blue, duct in green and dark and light colour for $Re_\tau=360$ and $Re_\tau=180$, respectively. Bottom row: probability of detection for $uv$ as a function of the distance from the wall for: $H_{uv}=0.5$ ($\blacktriangle$), $H_{uv}=2.0$ ($\bullet$) and $H_{uv}=4.0$ ($\blacktriangledown$). 
    Channel compared with (left) the centre-plane ($z=1$) and (right) a near-corner vertical profile ($z=0.1$) of the duct. Colours as for the top row.}
    \label{fig:perc}
\end{figure}

 The effect of increasing the threshold $H$ is illustrated in Figure~\ref{fig:perc} (top left) for $uv$ events in square duct at $Re_\tau=180$, which shows the contours of the identified structures for $H_{uv}=0.5$, $2$ and $4$. 
The use of the thresholds defined in equation~\ref{eqn:threshold} allows to identify flow regions of the velocity where two components of the velocity are highly correlated (or anti-correlated) in respect of the statistics. Additionally, such thresholds are chosen in order to detect the coherent structures homogeneously at the various wall-normal distances. 
This observation was first highlighted by \cite{naga03}, while studying vortex clusters with the $Q$ criterion \citep{hunt88}. 
\par More recently, \cite{alam06}, who adopted a similar threshold for the discriminant criterion \citep{chon90}, and \cite{loza12} introduced the percolation analysis in this context in order to qualitatively assess the effectiveness of the thresholds. Its usage is motivated by the fact that a percolation crisis is observed for both vortical and dissipative coherent structures in homogeneous isotropic turbulence \citep{mois04}. 
In these analyses the ratio between the volume of the largest structure, $\mathcal V_{\rm max}$, and the total volume occupied by all the structures, $\mathcal V_{\rm all}$, is computed for different values of $H$. 
For low values of $H$ most of the domain is occupied by few large objects and $\mathcal V_{\rm max}/\mathcal V_{\rm all}\approx1$, while for large enough values of $H$ only few and more isolated points fulfil the condition, therefore $\mathcal V_{\rm max}/\mathcal V_{\rm all}\ll1$. 
The fact that the percolation crisis occurs, \emph{i.e.} there is a relatively sharp transition between $\mathcal V_{\rm max}/\mathcal V_{\rm all}\approx1$ and $\mathcal V_{\rm max}/\mathcal V_{\rm all}\ll1$, is an evidence that the threshold properly captures the non-homogeneity of the flow. 
\par In the present work, in order to estimate $\mathcal V$ we assigned to each grid-point in the computational domain a characteristic volume based on the local grid spacing, and the volume of each structure is estimated as the sum of the characteristic volume of the grid points which belong to it. 
In the cases considered here the percolation analysis unveiled that the conditions (\ref{eqn:threshold}) are not fully appropriate when applied to the entire domain, since the normalisation factors become negligible near the wall.
Subsequently, a correction was introduced by excluding from the connectivity the region of the domain below $y^+=1$, computed with $\tau_w$ at the centre-plane. 
The wall distance $y^+=1$ was chosen for the correction because applying it in a wider region of the domain gives no appreciable differences in the results of the percolation analysis. 
\par The percolation diagram is shown in Figure~\ref{fig:perc} (top right) for all the cases and the considered structures, together with the probability for $uv$ events detected for different values of $H_{uv}$, in all the cases considered here (middle and right panels, respectively). 
The percolation diagram of $uv$ and $vw$ events is very similar in channel and in duct flow. However, it is different than that of $uw$ in the channel, while it is almost indistinguishable for $uv$ and $uw$ in the duct, as it is implied by the symmetry of the case. 
Small differences can be observed between the two different Reynolds numbers and in general between channel and duct. 
In particular, the percolation crisis is less evident for the duct at $Re_\tau=180$ and in both the Reynolds numbers is more sharp in the channel than in the duct. 
\par At the present state, it is not possible to discriminate whether such discrepancy is due to structural differences in both flows or rather to the different shape of the domain. 
This aspect will not be further investigated in this work, since our aim is to characterise the behaviour of the detected structures for a certain range of $H_{{\rm XX}}$ and we do not optimise the threshold. However, it is worth noting that the percolation crisis always occurs at a value of $H_{\rm}$ between 1.0 and 1.5 and that there are qualitative differences between the region of the domain sampled before and after the percolation crisis.
In the first case, $\mathcal V_{\rm all}$ can be a relatively small fraction of the domain, \emph{e.g.} $\approx40\%$ for $H_{uv}=0.5$, but despite this fact it is not possible to clearly identify the most intense events because they are incorporated into very large structures. To the contrary, in the second case, for which $\mathcal V_{\rm all}$ is even smaller, \emph{e.g.} $\approx7\%$ for $H_{uv}=2$, the intense events are isolated, since a further increase of $H_{uv}=2$ implies mostly a reduction of the volume for each event, rather than their splitting. 
\par The probability of belonging to a detected structure is in very good agreement between channel and duct at the centre-plane at both the Reynolds numbers and for all the values of $H_{uv}$ in the range considered. 
Its dependence on the wall-normal distance is weak, with the exception of the near-wall region ($y^+<15$), where very strong fluctuations are less likely. For the duct, it is also important to assess how the probability of detection differs at different spanwise locations. At $z=0.1$ ($z^+=18$ and $z^+=36$ for $Re_\tau=180$ and $Re_\tau=360$, respectively) the proximity of the vertical wall affects the probability of detection. However, the discrepancy between this region of the domain and the core region is less important for values of $H_{uv}$ higher than the one at which the percolation crisis occur. Similar results are obtained for $uw$ and $vw$ events, not shown here. 
These observations confirm that the normalisation factors are adequate and therefore it is possible to compare the coherent structures in the various cases. 
\par The turbulence statistics from the two duct cases under study here were computed by \cite{vinu18} using run-time statistics. The averaging periods were $7,158$ and $3,614$ convective time units (based on $U_{b}$ and $h$) for the $Re_{\tau}=180$ and 360 cases, respectively. On the other hand, the present study of coherent structures requires the analysis of instantaneous three-dimensional fields, and therefore it is important to assess the statistical convergence of the various quantities computed from such fields.
Due to the lack of periodicity in the spanwise direction and to the weak intensity of the secondary motion, duct flows require relatively long averaging times to obtain converged statistics \citep{vinu16}, and, since the coherent structure identification is performed as a post-processing step, the data-set needs to include a sufficiently large number of instantaneous fields.
\begin{figure}
    \centering
    \includegraphics[width=0.325\textwidth]{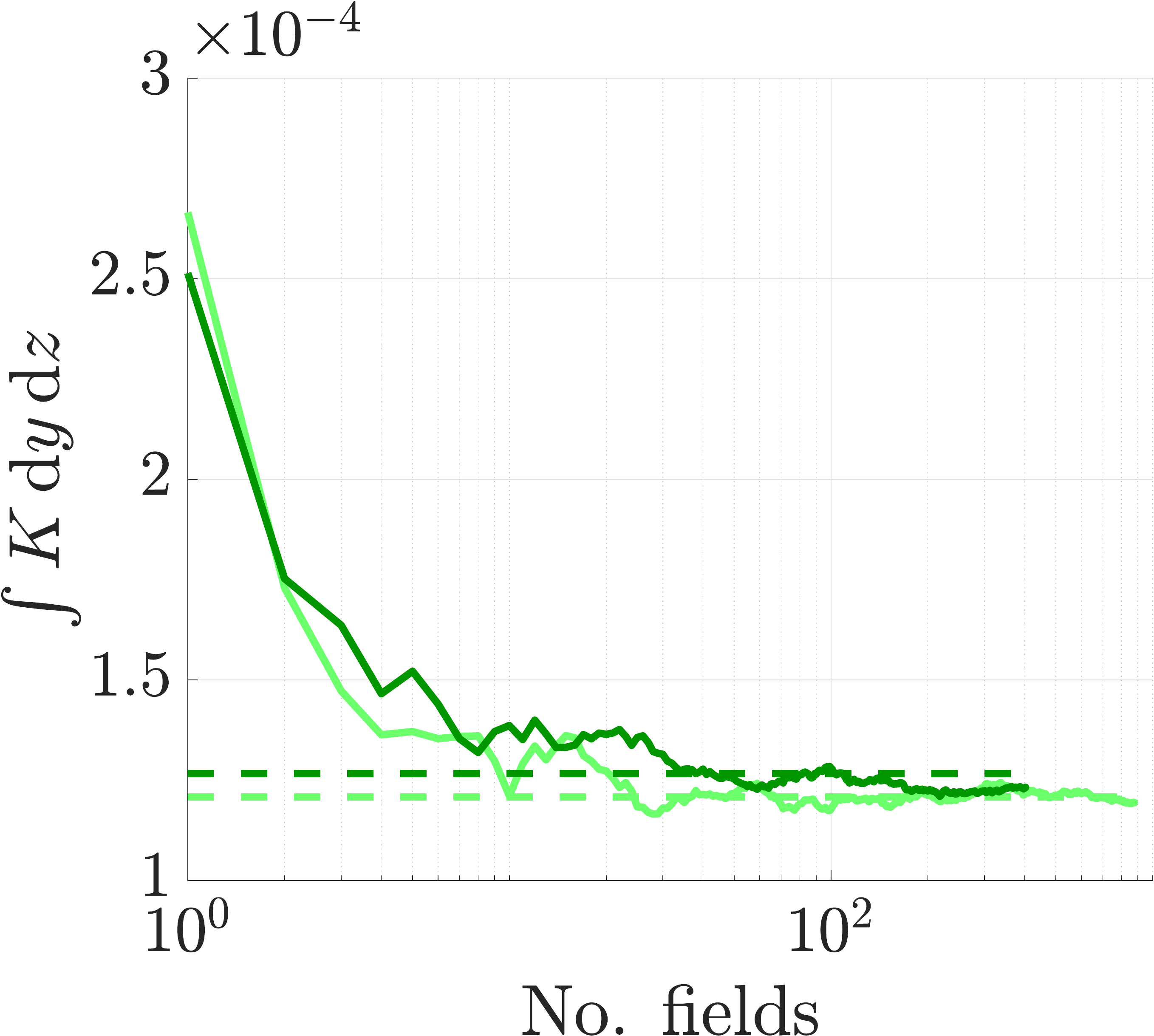}
    \includegraphics[width=0.325\textwidth]{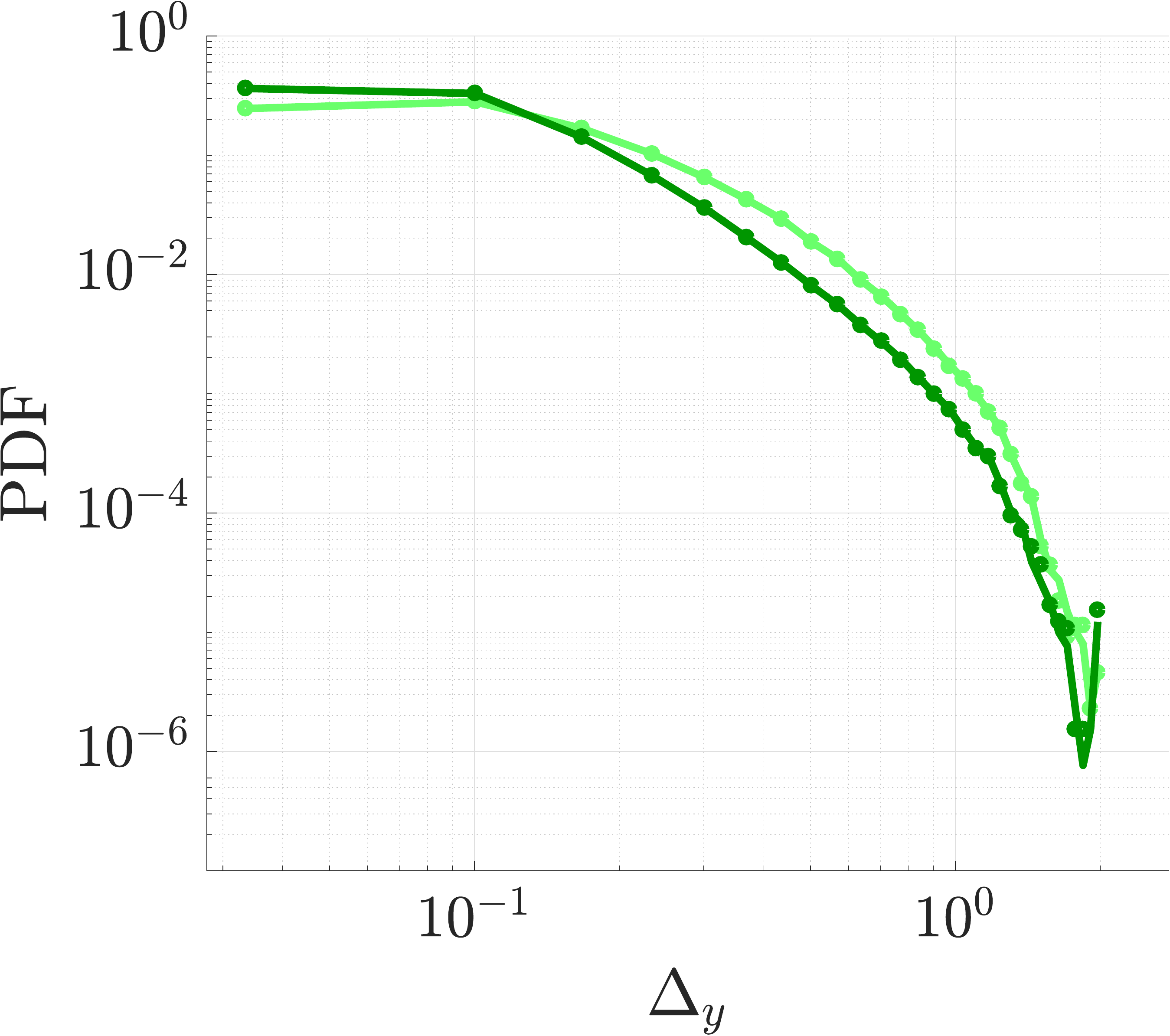}
    \includegraphics[width=0.325\textwidth]{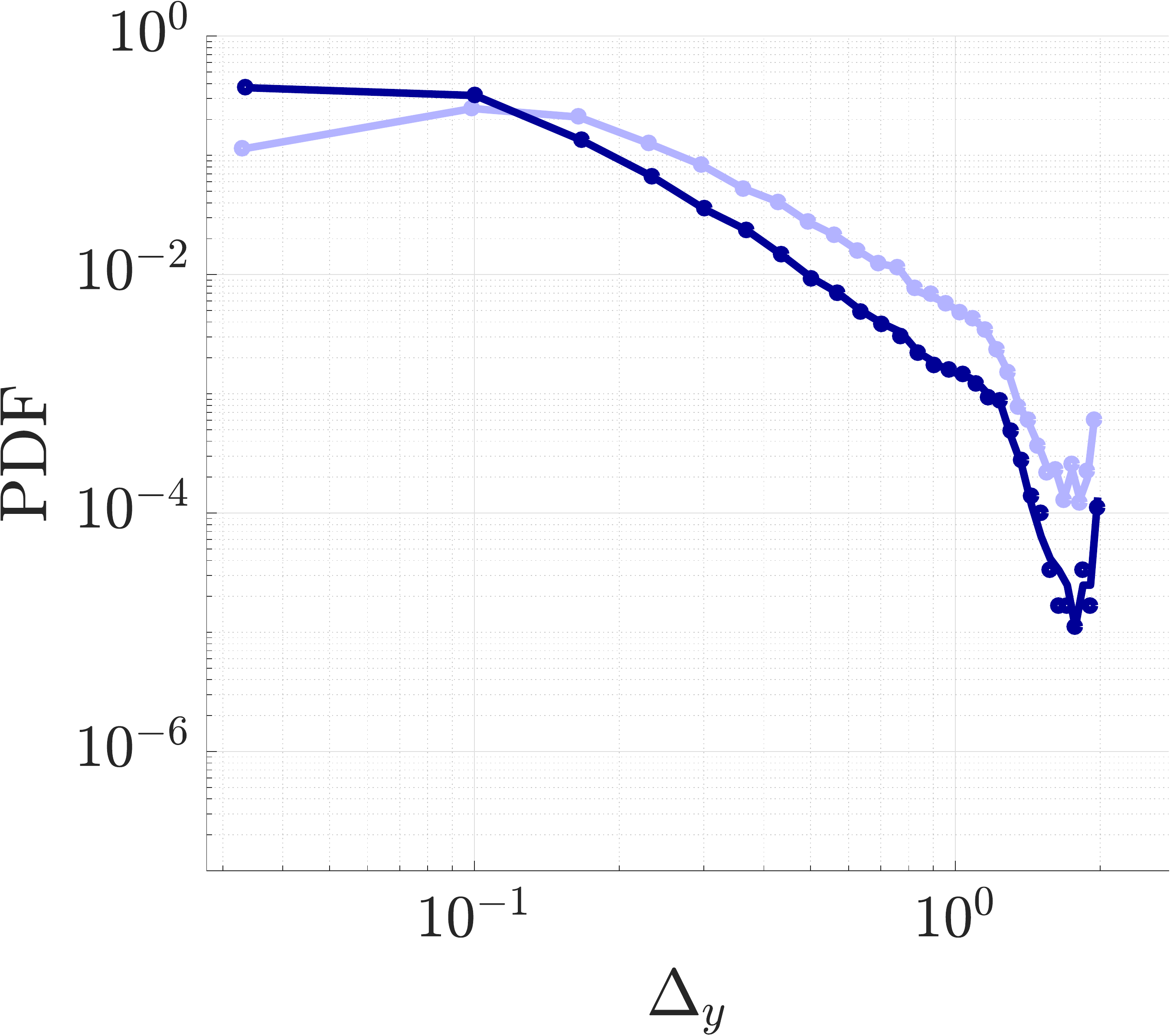}
    \caption{Statistical convergence: (left) integrated kinetic energy of the secondary motion computed via ensemble averaging (solid lines) as function of the number of fields and the (dashed lines) values reported by \cite{vinu18}. PDF of the length in the vertical direction of intense $uv$ events for (middle) duct and (right) channel, computed based on the entire data-set (solid lines) and (symbols) half of it. Colour code as in Figure~\ref{fig:perc}.}
    \label{fig:conv}
\end{figure}
\par In order to estimate the statistical uncertainty, we consider the kinetic energy associated with the secondary flow integrated over the section of the duct \cite[as in][]{vinu18}, denoted by $K=(V^2+W^2)/2$, and the probability density function (PDF) of the size of the detected structures in the vertical direction, $\Delta_y$ (Figure~\ref{fig:conv}). 
\par The integrated $K$ based on ensemble averaging over increasing number of fields is shown in Figure 2 (left), in which it can be observed that this quantity differs by $\approx4\%$ and $\approx1\%$ with respect to the statistics reported by \cite{vinu18} for the ducts at $Re_\tau=360$ and $Re_\tau=180$, respectively, when the entire data-set is considered. 
Note that the entire data-set consists of approximately $900$ and $400$ instantaneous fields for the ducts at $Re_\tau=180$ and $Re_\tau=360$, respectively, and that we study the fully developed turbulent flow (\emph{i.e.} the initial transient is excluded). The PDFs of $\Delta_y$ for the two ducts are shown in Figure~\ref{fig:conv} (middle), computed based on the entire data-set and for half of it; note the excellent agreement between both in the two duct cases.
\par Regarding the channel, we employ $76$ and $37$ fields for $Re_\tau=180$ and $Re_\tau=360$, respectively. The convergence of the ensemble average requires a significantly lower number of fields since both the streamwise and spanwise directions are periodic, and the discrepancy between the ensemble and the time average is negligible. The PDFs of $\Delta_y$ for the entire data-set and half of it for the channel are also in very good agreement, as shown in Figure 2 (right).
\subsection{Scaling properties of the secondary flow}
\begin{figure}
    \centering
    \includegraphics[width=0.45\textwidth]{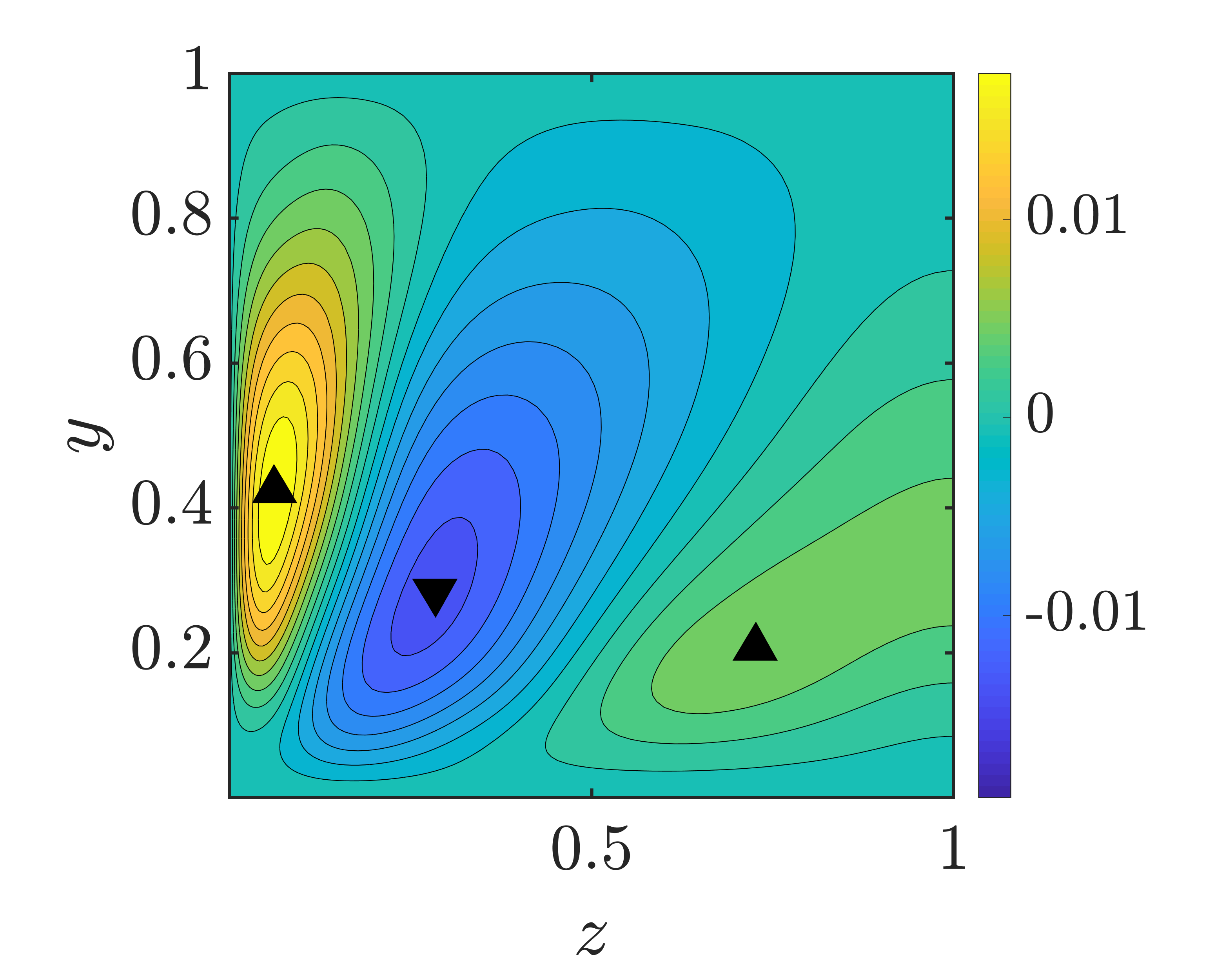}
    \includegraphics[width=0.45\textwidth]{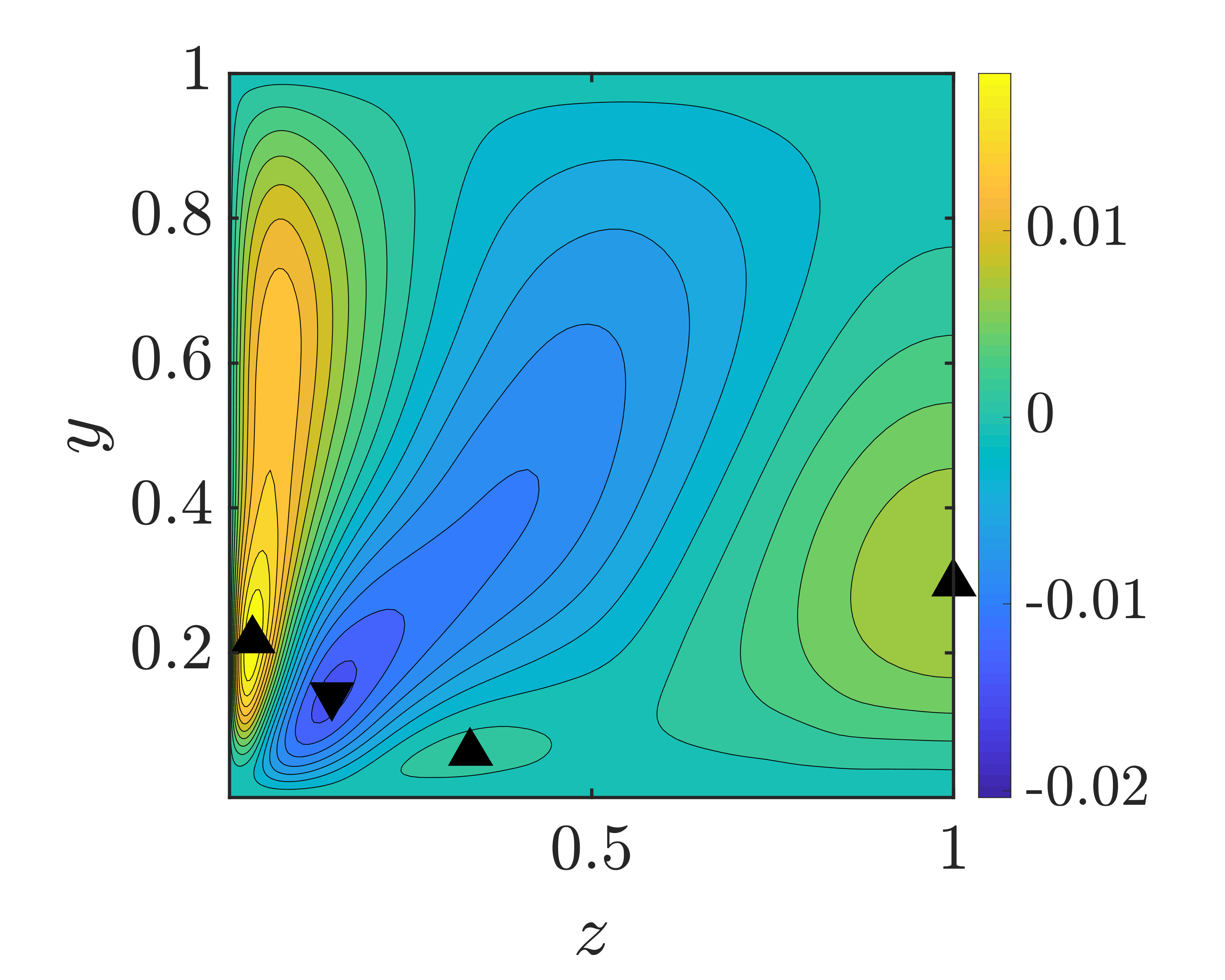}\\
    \includegraphics[width=0.325\textwidth]{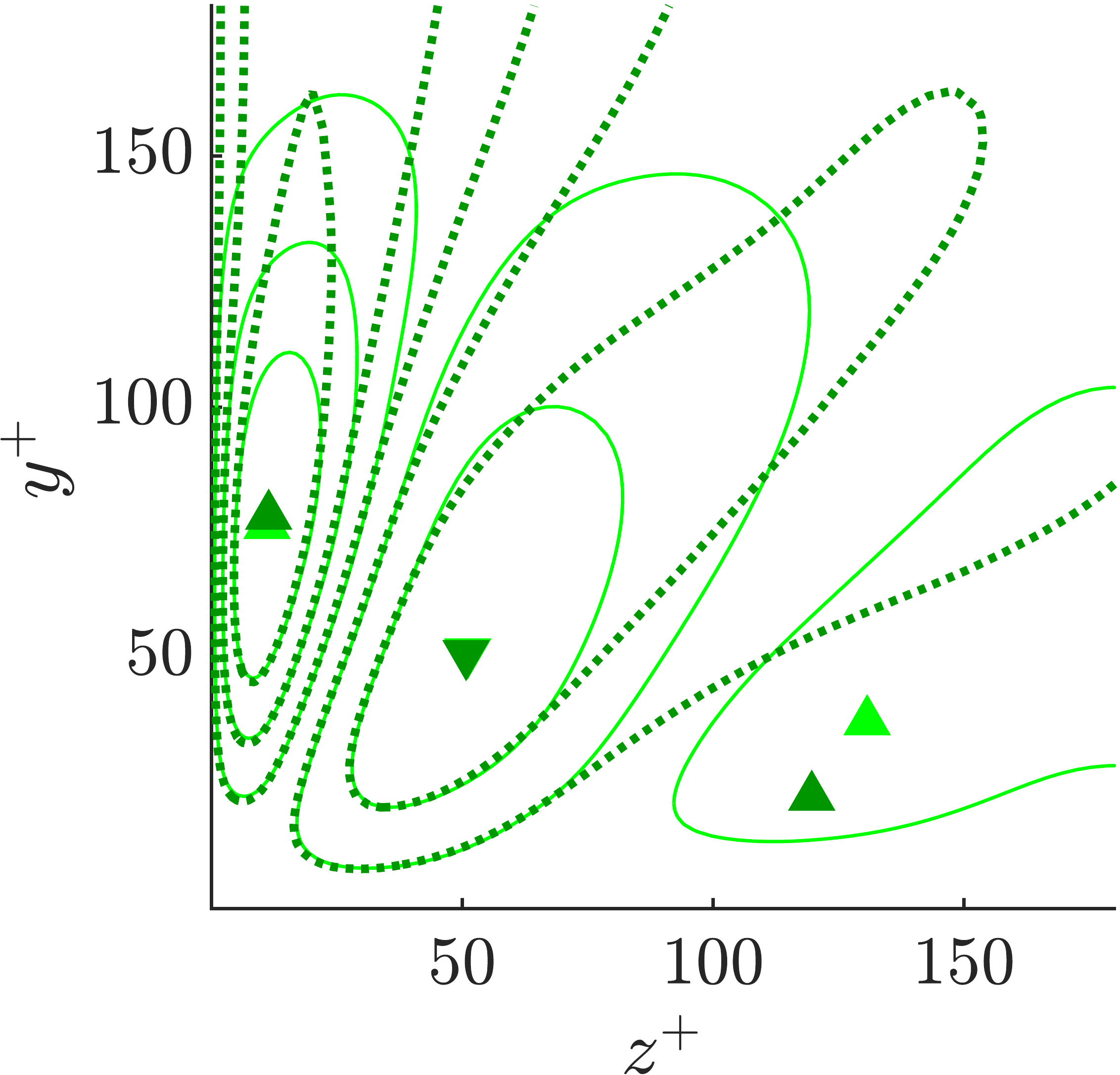}
    \includegraphics[width=0.325\textwidth]{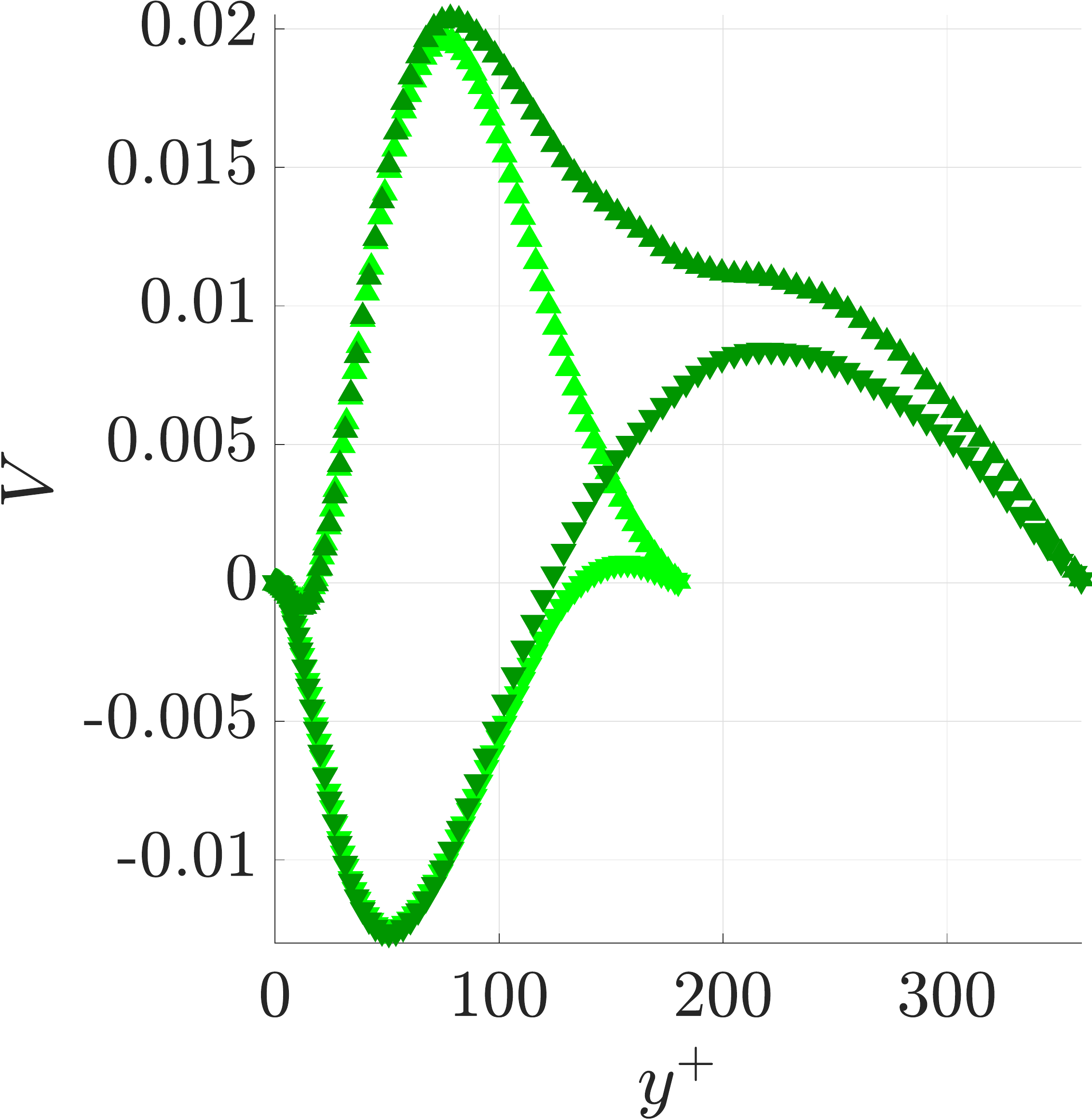}
    \includegraphics[width=0.325\textwidth]{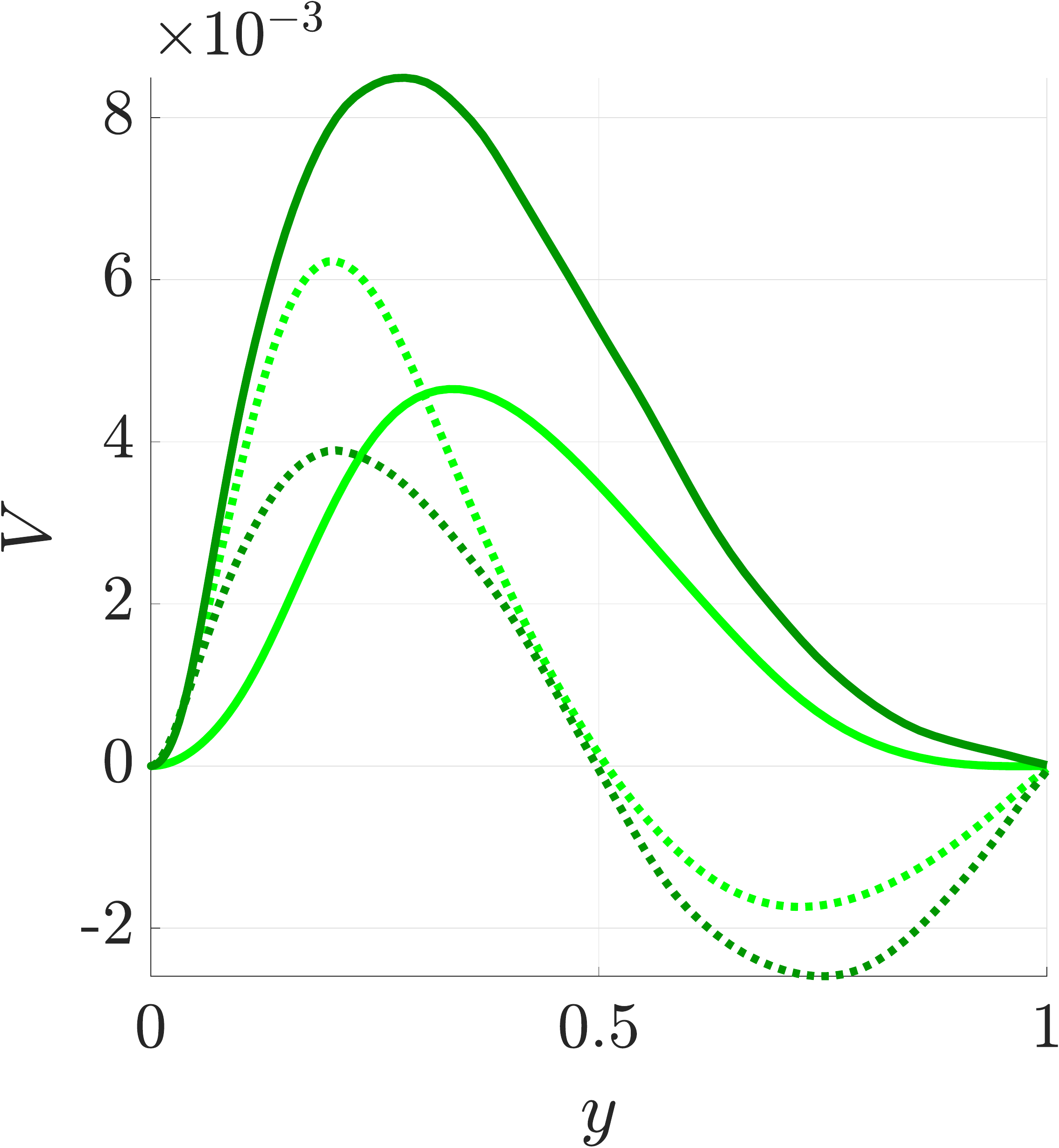}
    \caption{Top row: mean of the vertical component of the velocity $V$ for one quadrant of the duct at (left) $Re_\tau=180$ and (right) $Re_\tau=360$. Local maxima and local minima are indicated with $\blacktriangle$ and $\blacktriangledown$, respectively. Bottom row: scaling properties of $V$. Dark green and light green for $Re_\tau=360$ and $Re_\tau=180$, respectively. (Left) Contour and locations of maxima and local minima coordinates scaled in inner-units. (Middle) Vertical profile of $V$ as function of the inner-scaled vertical coordinate for the first maximum ($\blacktriangle$, $z^+\approx11$) and minimum ($\blacktriangledown$,  $z^+\approx50$) and (right) as function of the outer-scaled vertical coordinate for the second maximum (dotted line, $z^+\approx120$) and third maximum for $Re_\tau=360$ (solid line, $z=1$).}
\label{fig:V_extrema_top}
\end{figure}
As shown in Figure~\ref{fig:V_extrema_top}, in the lower-left quadrant of the duct at $Re_\tau=180$ the vertical mean velocity component $V$ exhibits one local maximum locate located at $(y^+=76;\,z^+=11)$, followed by a local minimum at $(y^+=49;\,z^+=49)$ and a second local maximum at $(y^+=35;\,z^+=125)$, which implies that a saddle point is located along the centre-line of the cross section. 
For $Re_\tau=360$ the same three local extrema are observed at approximately the same inner-scaled locations: $(78;\,11)$, $(50;\,50)$ and $(22;\,118)$, respectively. Furthermore, a new local maximum appears along the centre-line and the saddle point migrates to an intermediate position between the third and the forth local maxima.
\par As can be observed, despite the fact that the positions of the first three local extrema scale in inner units, the corresponding velocity profiles do so only for the first maximum and the minimum. 
The matching between $V$ at the two different Reynolds numbers is not valid for all $y^+$, which is expected because points such as $(y^+=150,\,z^+=11)$ in the duct at $Re_\tau=180$ belong to the region where the centre circulation takes place, \textit{i.e.} the outer-scaling region of the domain.
Nevertheless, the inner scaling remains valid far above the bisector, \textit{e.g.:} up to $y^+\approx50$ for $z^+=11$ and up to $y^+\approx100$ for $z^+=50$.
The profiles corresponding to the location of the third maximum do not exhibit a clear scaling, neither in inner or in outer units. 
In fact, if scaled in outer units, the velocity profile for $Re_\tau=180$ at the position of the second maximum overlaps with that at the centre-plane for $Re_\tau=360$ only in a region near the wall, and it is close to the that of the second maximum for $Re_\tau=360$ farther from the wall. 
The profile at the centre-plane of the profiles in the duct at $Re_\tau=180$ does not match with any of the profiles in duct at $Re_\tau=360$. 
\par Given the properties of the $V$ field that we have just discussed, we examined in detail the contributions of the coherent structures to the secondary flow at two spanwise locations: the local minimum of $V$ and the centre-plane ({\it i.e.} $z^{+} \simeq 11$ and $z=h$, respectively). The first choice is motivated by the fact that for this profile there is good agreement between the two Reynolds numbers if $V$ is scaled in outer units and the wall-normal coordinate in inner units. 
The latter because the behaviour for $Re_\tau=360$ at that profile is similar to the one observed by \cite{piro18} for the same location at higher Reynolds numbers and because the centre-plane of the duct has in common with the channel flow that $W=0$. 
%
%
\subsection{Definition of fractional contributions}
In the following we define $\Xi_{\rm XX}^>$ as the fractional contribution to the variable $\Xi$ sampled of over events ${\rm X X}$ detected according to conditions (\ref{eqn:threshold}) and $\Xi_{\rm XX}^<$ the fractional contribution over the portion of the domain which does not fulfil the same condition.
\par The fractional contributions are computed as ensemble averages and weighted with the sampled volume fraction, so that their sum recovers the ensemble average over the entire data-set, \emph{i.e.} $\Xi_{\rm ens} =  \Xi_{\rm XX}^> + \Xi_{\rm XX}^<$. 
For example, $U^>_{uv}$ is the mean streamwise velocity conditioned to the presence of strong $uv$ and weighted by the corresponding volume fraction and $U^<_{uv}$ is the mean streamwise velocity averaged over the complementary part of the domain. 
These can be expressed mathematically as:
\begin{equation}
    U^>_{uv} = \frac{\mathcal V_{uv}}{\mathcal V} \int U \delta_{uv}\, {\rm d} \mathcal V \qquad {\rm and} \qquad
    U^<_{uv} = \frac{\mathcal V - \mathcal V_{uv}}{\mathcal V} \int U (1-\delta_{uv})\, {\rm d} \mathcal V\,,
\end{equation}
where $\mathcal V$ is the volume of the entire domain, $\mathcal V_{uv}$ is the volume occupied by intense $uv$ events (for a certain value of $H_{uv}$) and $\delta_{uv}$ is $1$ if $|uv|>H_{uv} u_{\rm rms} v_{\rm rms}$ and $0$ elsewhere. It is important to note that $\Xi_{\rm XX}^>$ reflects both the intensity of the fluctuations and their probability of occurrence. For example, $|\Xi_{\rm XX}^>|$ can be higher for $H_{\rm XX}=2.0$ than for $H_{\rm XX}=4.0$ in a certain region, because a higher threshold $H_{\rm XX}$ corresponds to stronger fluctuations, but at the same time to a smaller portion of the domain.
The scope of the decomposition $\Xi_{\rm ens} =  \Xi_{\rm XX}^> + \Xi_{\rm XX}^<$ is two-fold: on the one hand, it allows to investigate how more and less intense events contribute to the realisation of $\Xi_{\rm ens}$; on the other hand, given that the probability of detection is in good agreement for a certain $H_{\rm XX}$, it allows a direct comparison between intense and weak events in the different cases.
\section{Results}
In this section we study the fractional contribution of intense events to the cross-stream component of the velocity. 
\subsection{Contributions to the secondary flow: core region}
\begin{figure}
    \centering
    \includegraphics[width=0.325\textwidth]{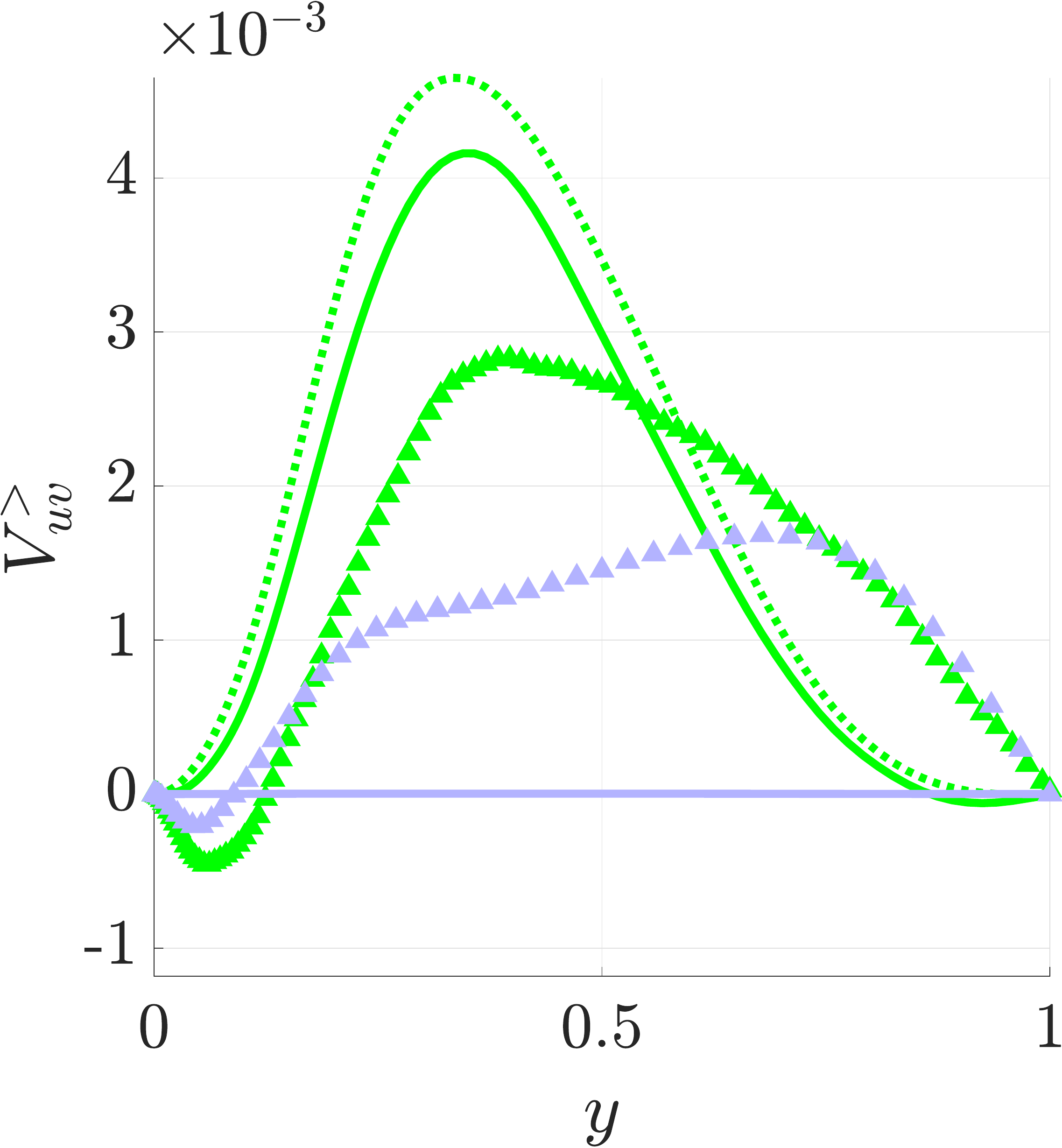}
    \includegraphics[width=0.325\textwidth]{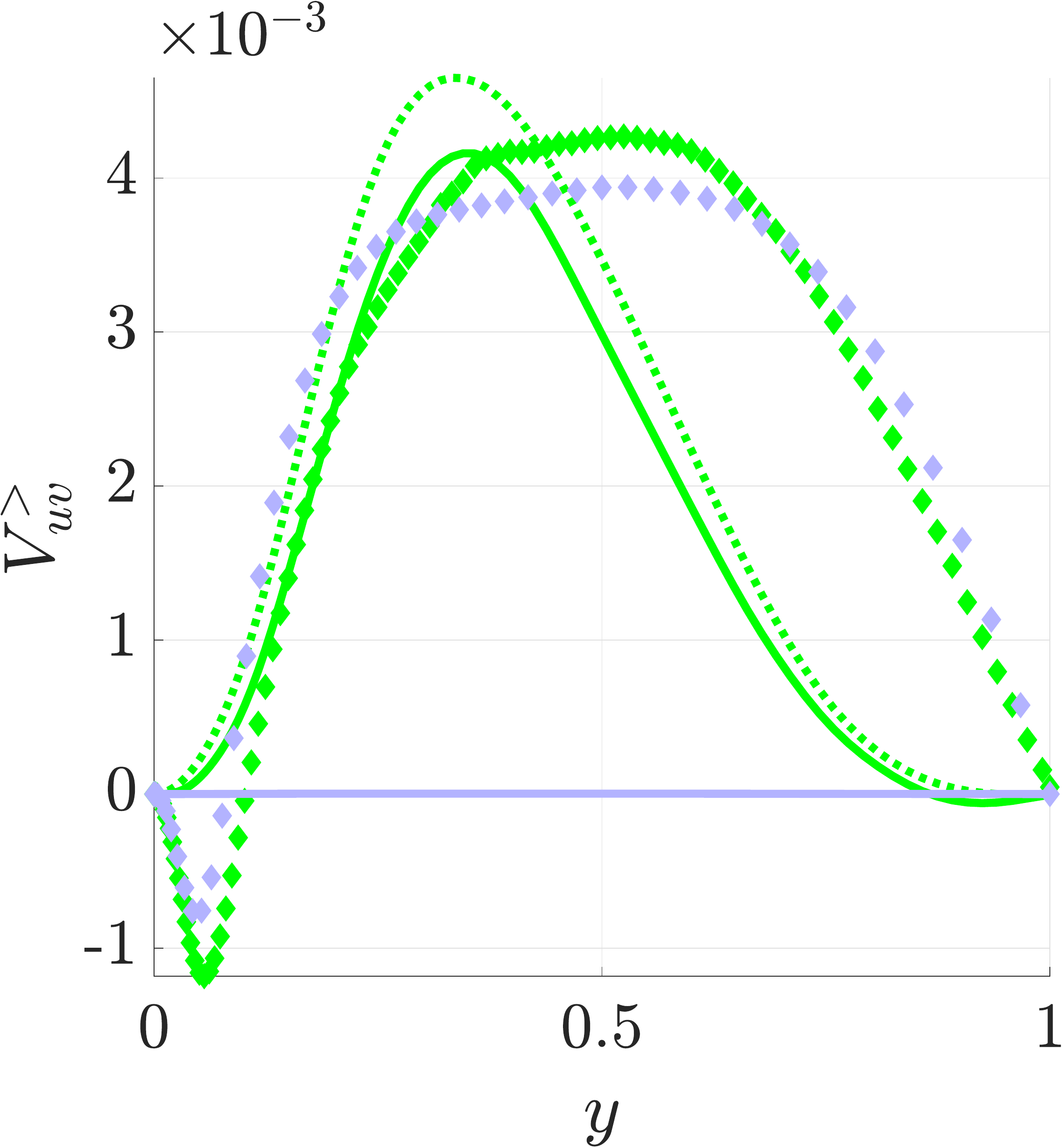}
    \includegraphics[width=0.325\textwidth]{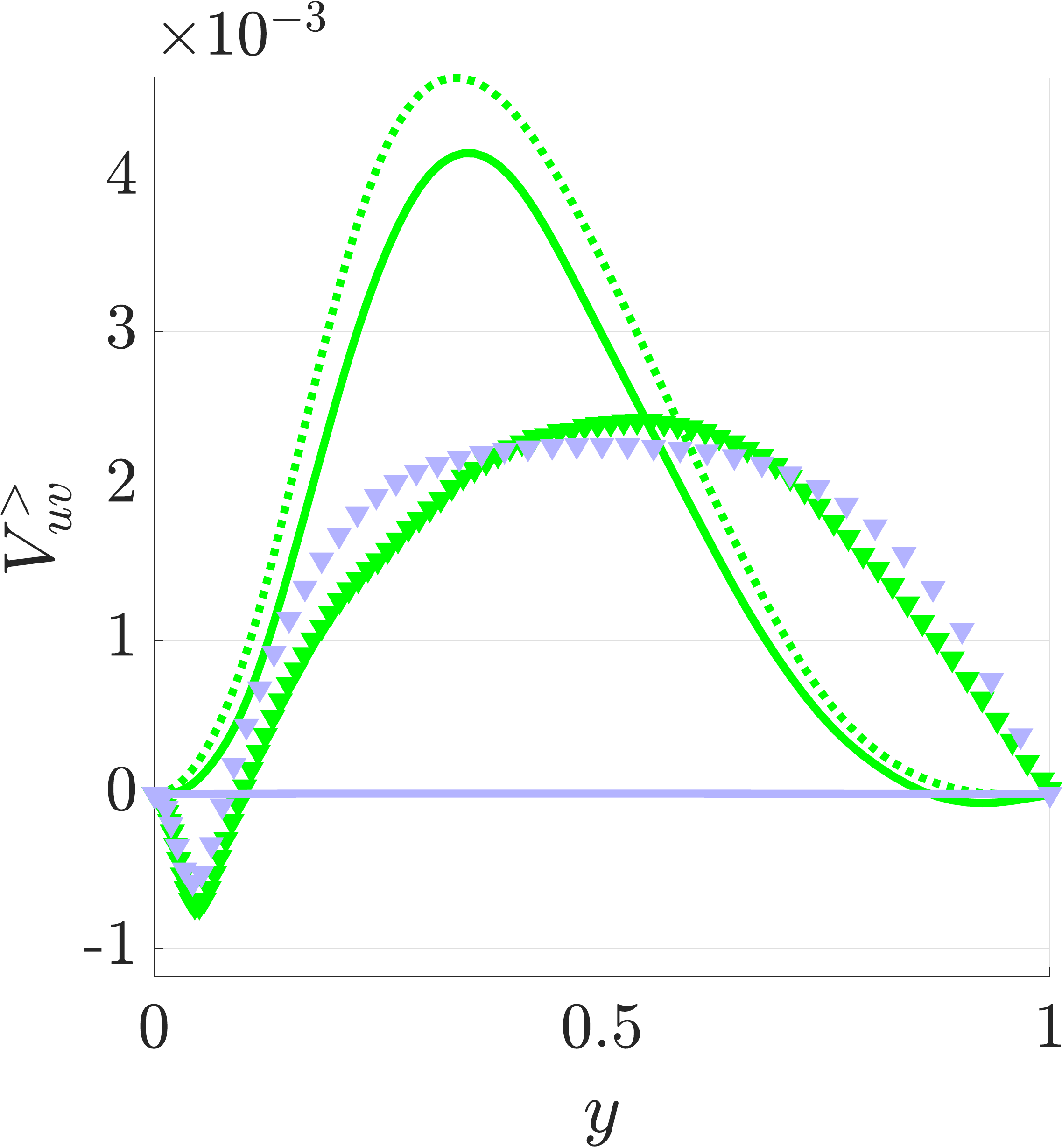}
    \includegraphics[width=0.325\textwidth]{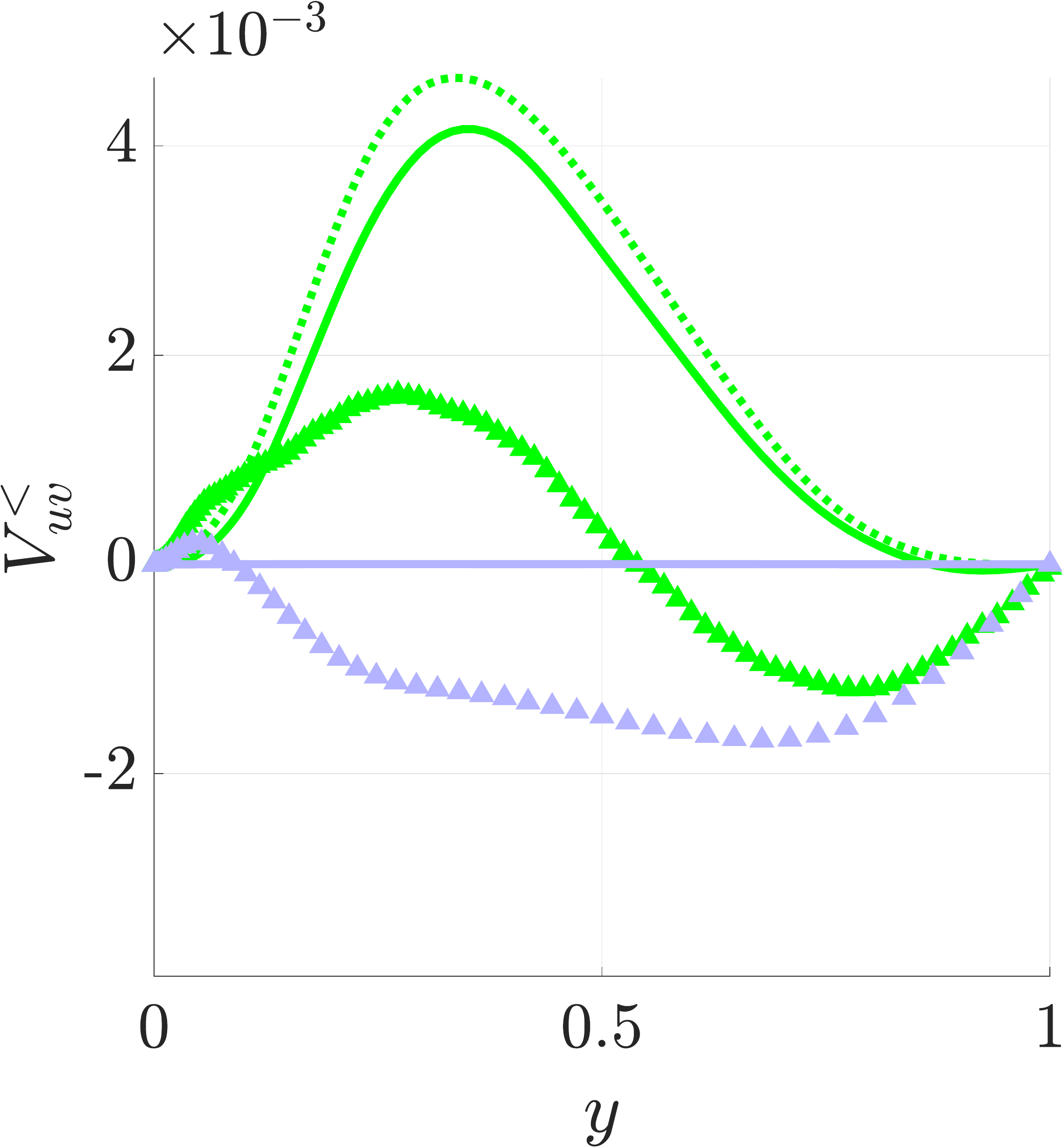}
    \includegraphics[width=0.325\textwidth]{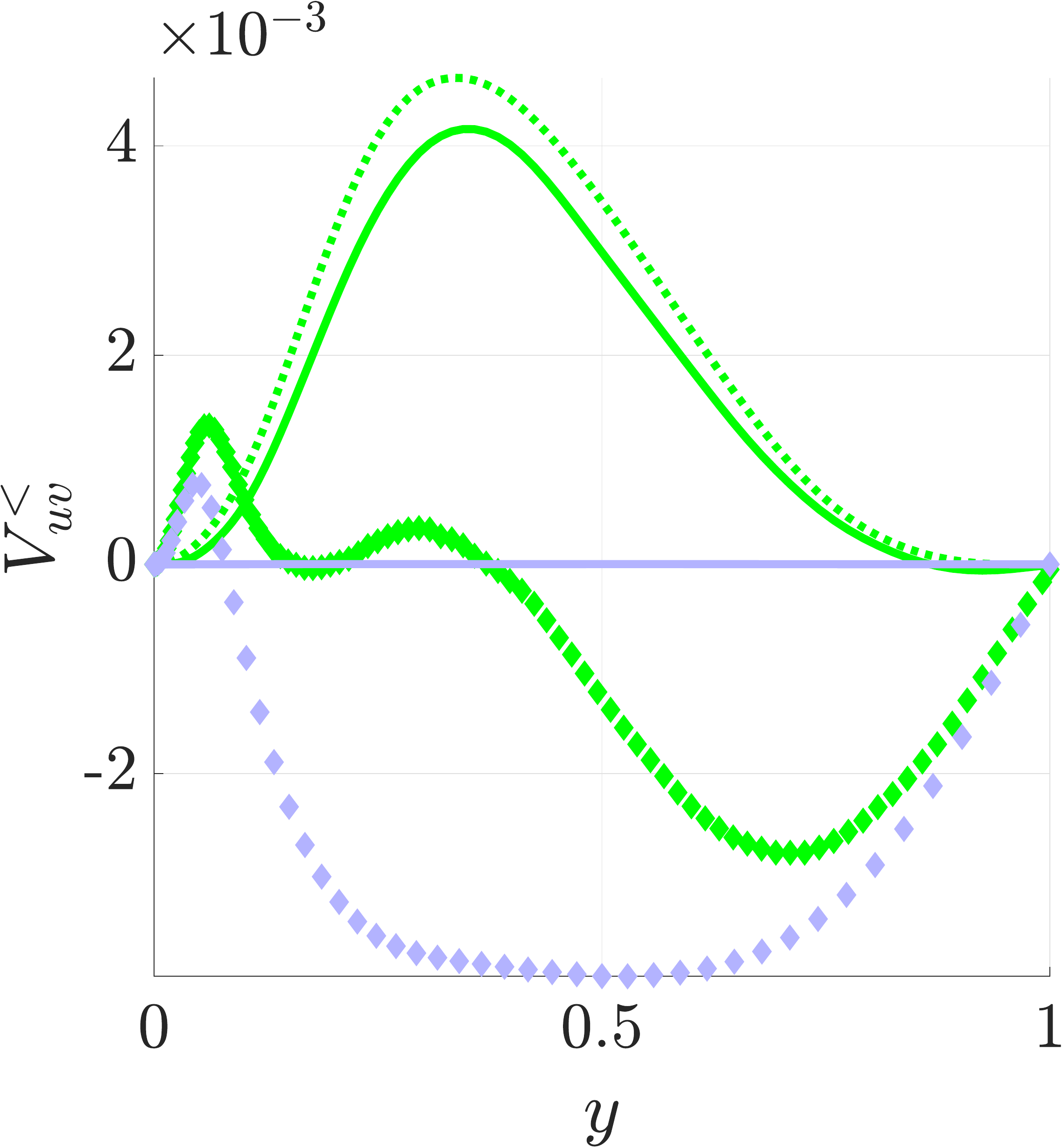}
    \includegraphics[width=0.325\textwidth]{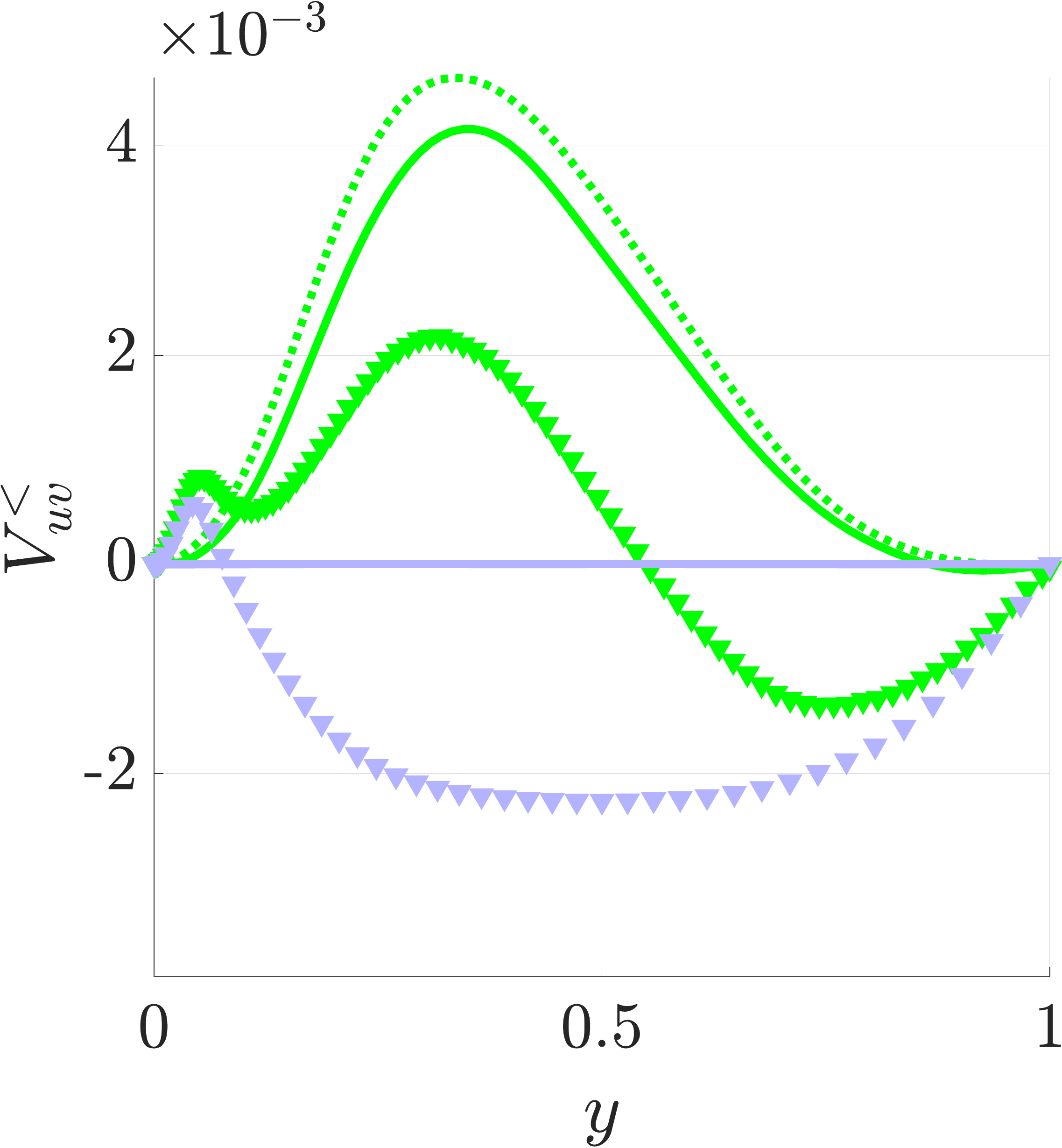}
    \caption{Contribution to the vertical mean component of the velocity from (top row) $uv$ structures and (bottom row) the complementary portion of the domain for (light green) duct at the centre-plane and (light blue) channel at $Re_\tau=180$. The fractional contributions are denoted with symbols and for increasing $H_{uv}$ from left to right: $H_{uv}=0.5$ ($\blacktriangle$),  $H_{uv}=2.0$ ($\vardiamond$) and $H_{uv}=4.0$ ($\blacktriangledown$). The ensemble average (solid lines) and the time average (dotted lines) are reported as reference.}
    \label{fig:Vuv_180}
\end{figure}

We firstly compare the core region of the duct and the channel. Figure~\ref{fig:Vuv_180} shows the contribution of intense $uv$ events to the vertical component of the mean velocity $V^>_{uv}$ and the complementary $V^<_{uv}$ in channel and duct for $Re_\tau=180$ at the centre-plane. 
In the channel, $V^>_{uv}$ is negative for $y^+<15$ and positive elsewhere, and $V^<_{uv}=-V^>_{uv}$, in agreement with the definitions above and the fact that in the channel $V=0$. The absolute values of $V^>_{uv}$ and $V^<_{uv}$ increase as $H_{uv}$ increases at all wall-distances up to $H_{uv}\approx2.0$, for which $\mathcal V_{\rm all}$ is $\approx 7\%$ of the total volume, and decreases for higher $H$, since too few events are sampled. 
This behaviour of $V^>_{uv}$ is in agreement with the observations by \cite{loza12}, who reported that intense ejections ($u<0$, $v>0$) are prevalent among intense $uv$ in the channel, with the exception of the near-wall region where such events are inhibited by the presence of the wall. 
\par In the duct a more complex behaviour is observed. At the centre-plane the relation between $V^>_{uv}$ and the mean $V_{\rm ens}$ varies at different wall-distances, and, since $V\neq0$, $V^>_{uv}\neq -V^<_{uv}$. 
Near the wall ($y^+<15$), $V^>_{uv}$ is negative and its intensity increases at higher $H_{uv}$, similarly to what was observed in channel flow. 
In the region between $y^+\approx20$ and $y\approx0.4$, where the secondary flow is more intense, $V^>_{uv}$ decreases as $H_{uv}$ increases up to $H\approx0.5$, increases for larger $H_{uv}$ between $H_{uv}\approx0.5$ and $H_{uv}\approx2.0$, and decreases again for $H_{uv}>2$.  
However, for $y>0.4$, $V^>_{uv}$ follows the same behaviour as $V^<_{uv}$ in channel flow, increasing for higher $H_{uv}$ up to $H\approx2$ and decreasing for larger $H_{uv}$. 
\par Interestingly, $V^>_{uv}$ in channel and duct are in good agreement for $H_{uv}\approx2.0$ and higher. 
Furthermore, in the intermediate region between $y^+=20$ and $y=0.4$, where the secondary motion is more intense, $V_{uv}^>$ of both duct and channel are in good agreement with $V_{\rm ens}$ in the duct, thus reaching in both cases the same intensity of the secondary motion. 
On the other hand, for $y>0.4$, where the secondary flow is less intense, there is no agreement between $V_{\rm ens}$ and $V^>_{uv}$.
\par The contribution to $V$ from the not sampled portion of the domain, $V^<_{uv}$, has a complementary trend. In the region between $y^+=20$ and $y=0.4$, since $V^<_{uv}\approx V_{\rm ens}$, $V^<_{uv}$ has an intensity lower than the same quantity in the channel. 
However, it becomes more relevant farther from the wall, where the weakening of the secondary motion is due to the balance between the positive $V^>_{uv}$ and the negative $V^<_{uv}$, which is the same phenomenon that in the channel leads to $V=0$. 
\par
\begin{figure}
    \centering
    \includegraphics[width=0.325\textwidth]{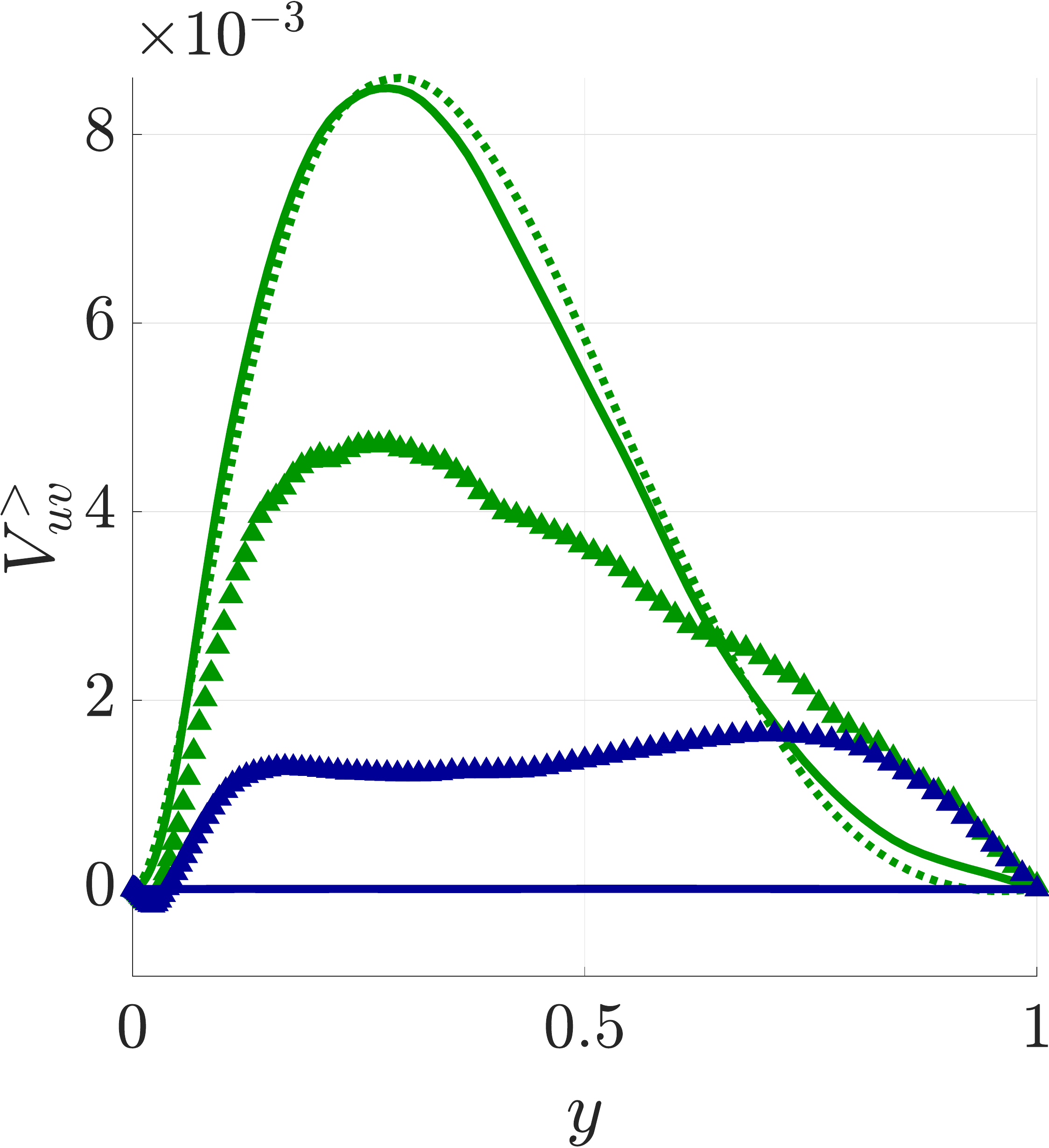}
    \includegraphics[width=0.325\textwidth]{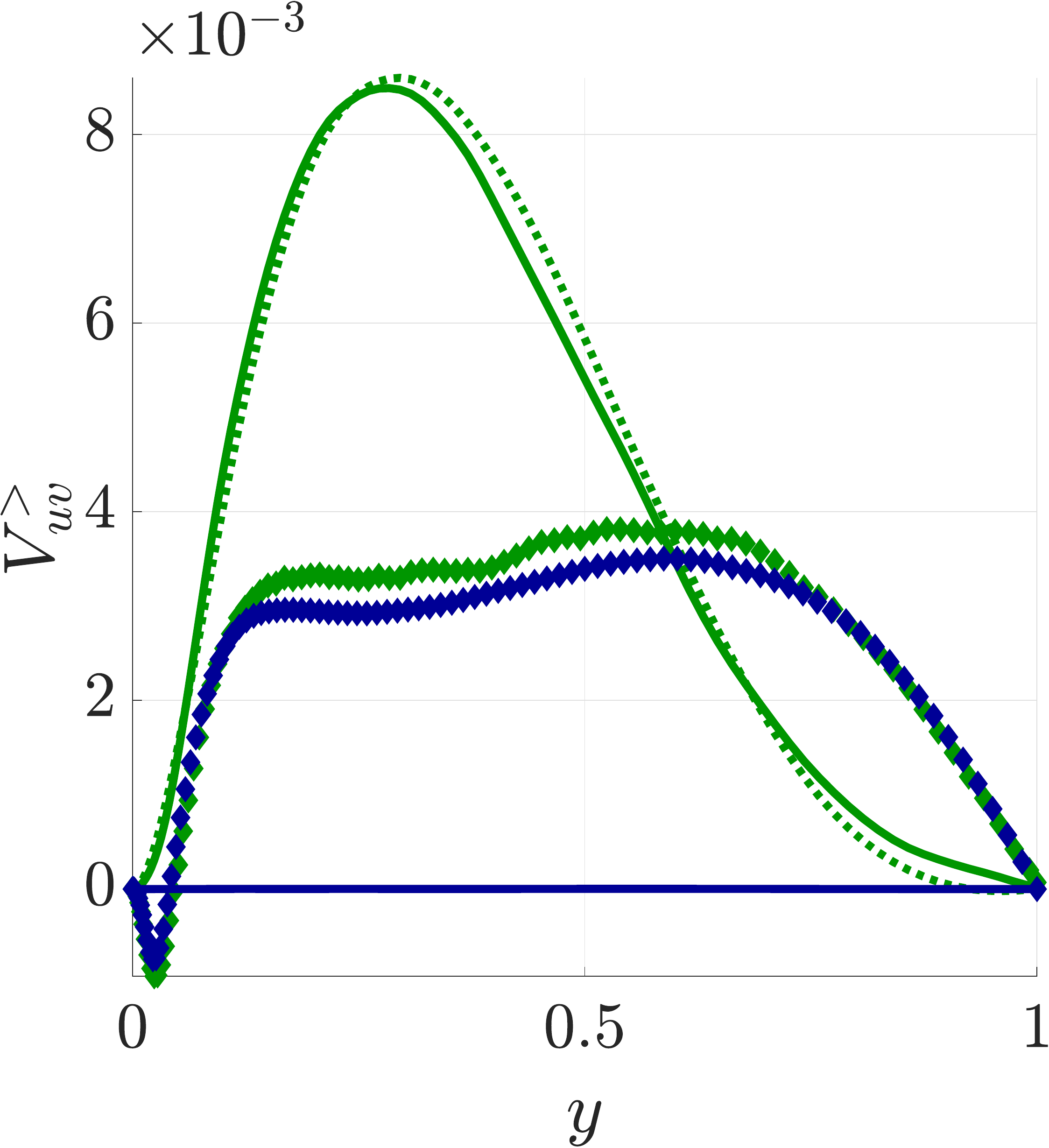}
    \includegraphics[width=0.325\textwidth]{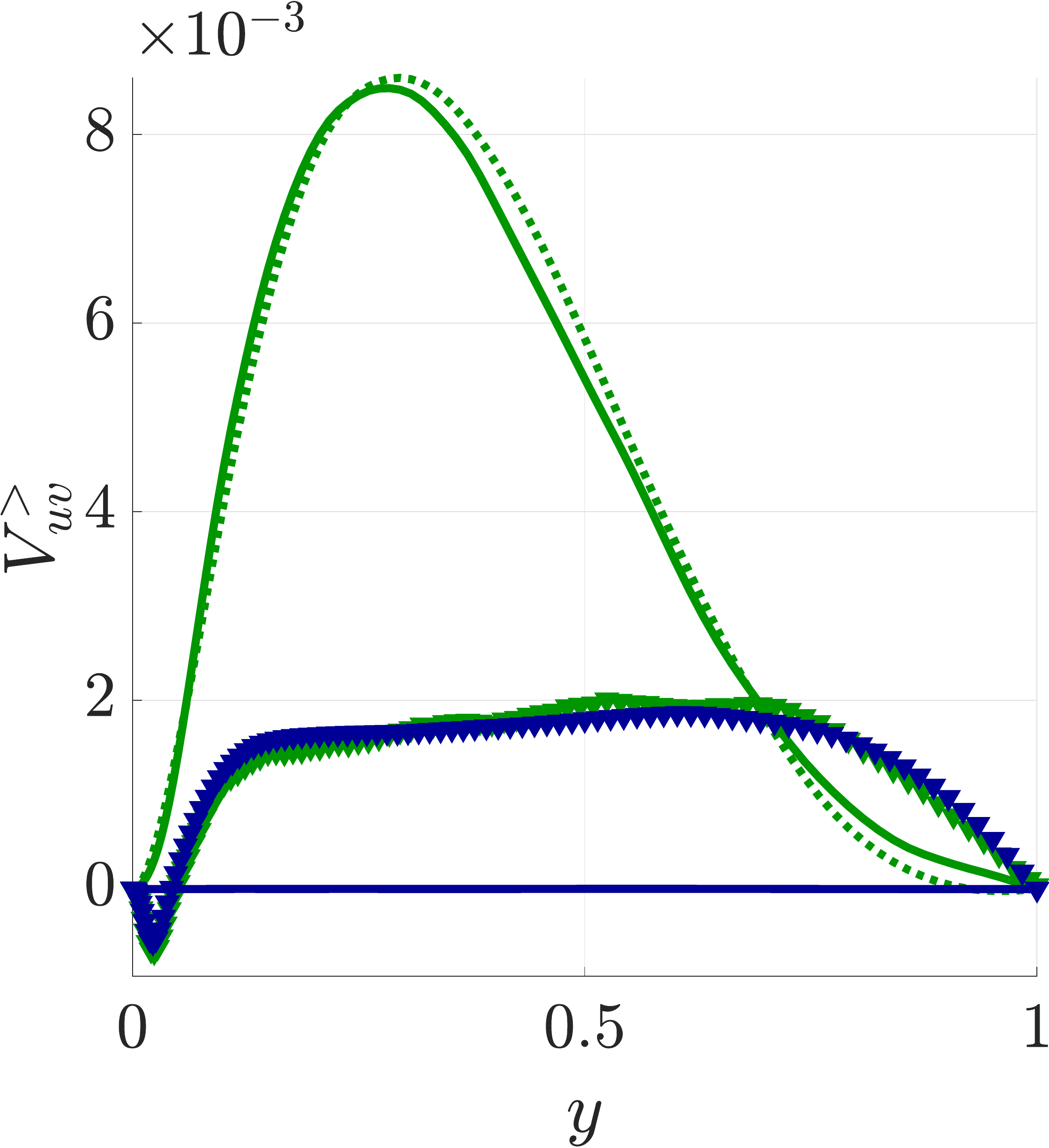}
    \includegraphics[width=0.325\textwidth]{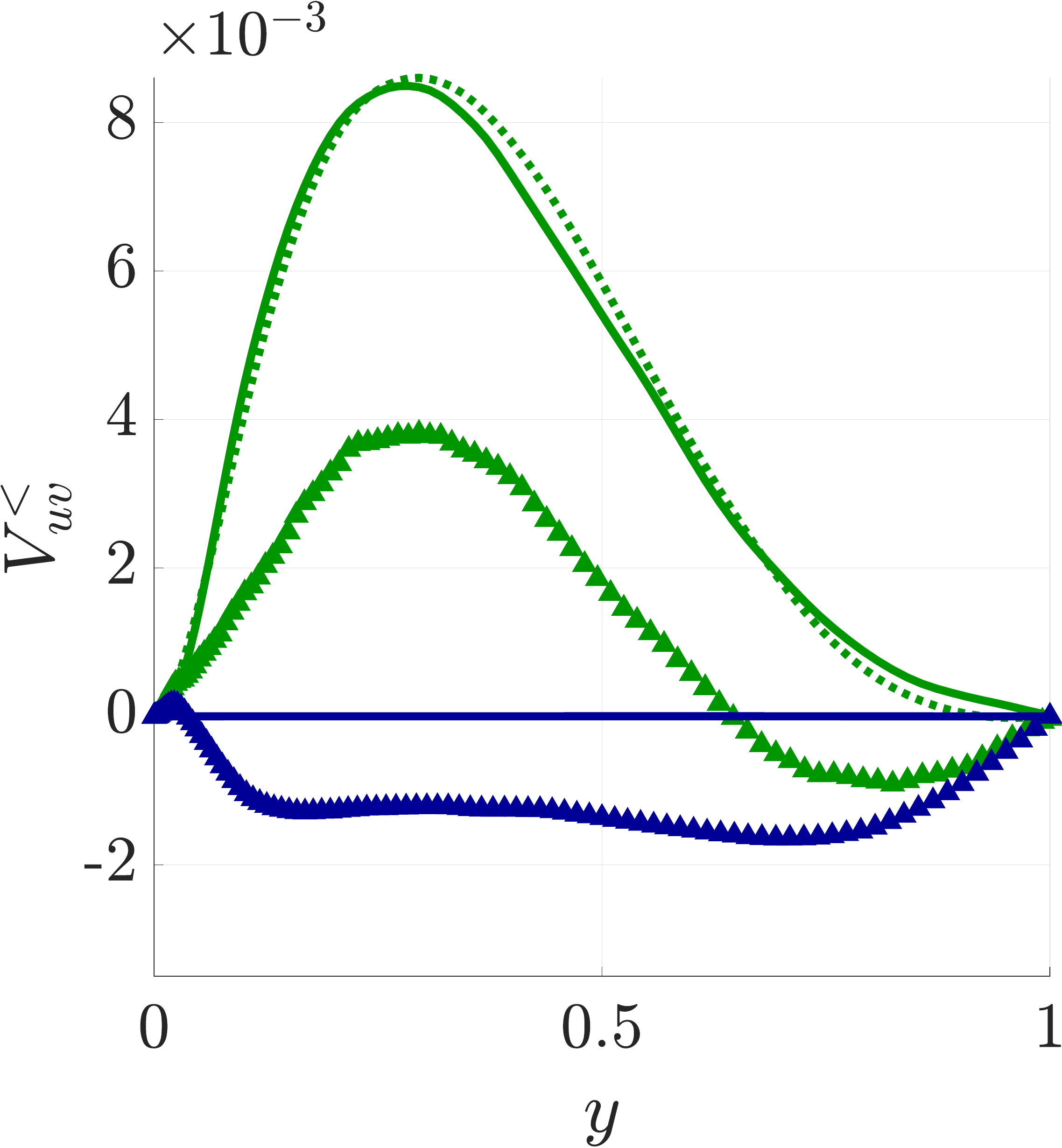}
    \includegraphics[width=0.325\textwidth]{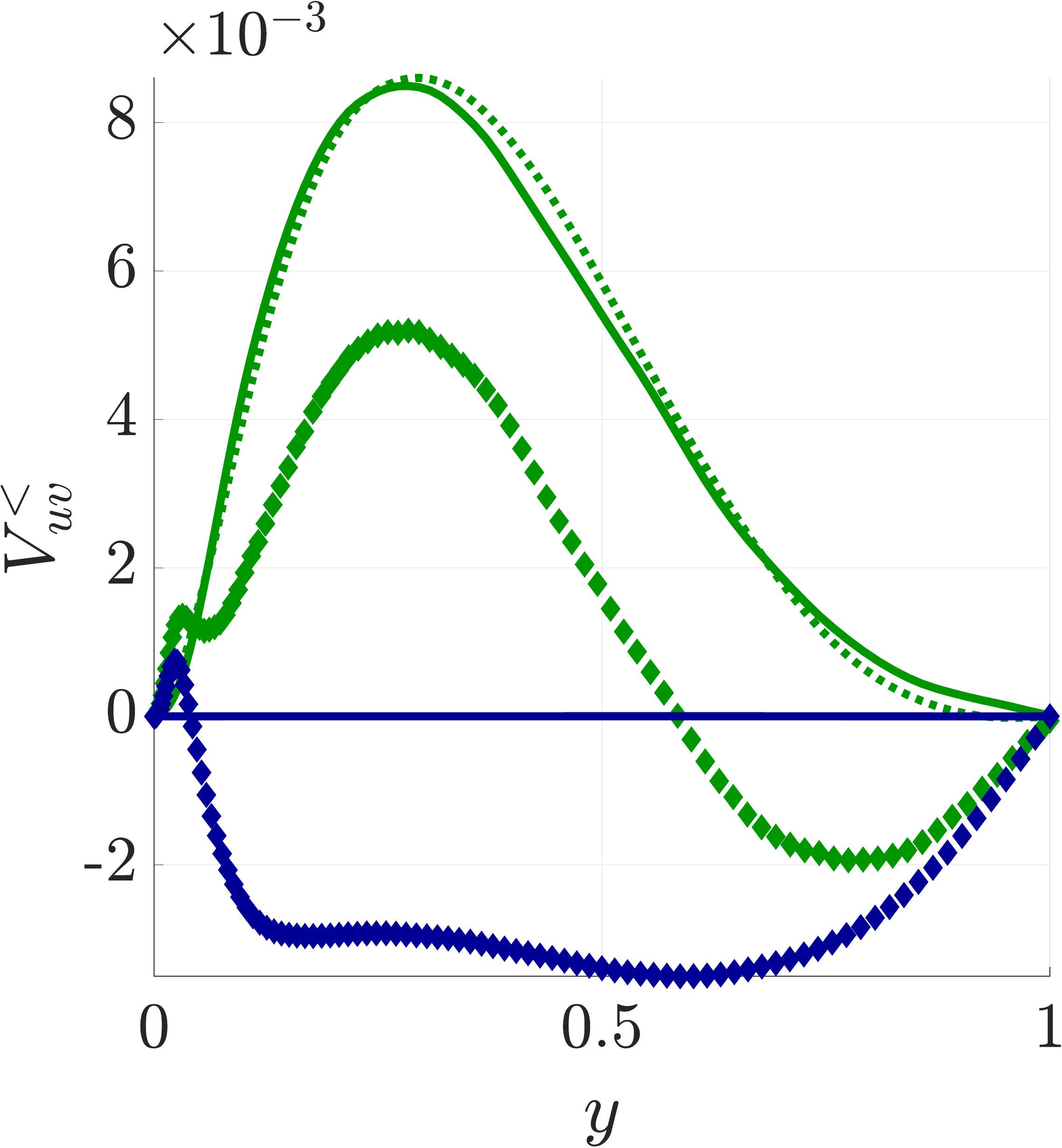}
    \includegraphics[width=0.325\textwidth]{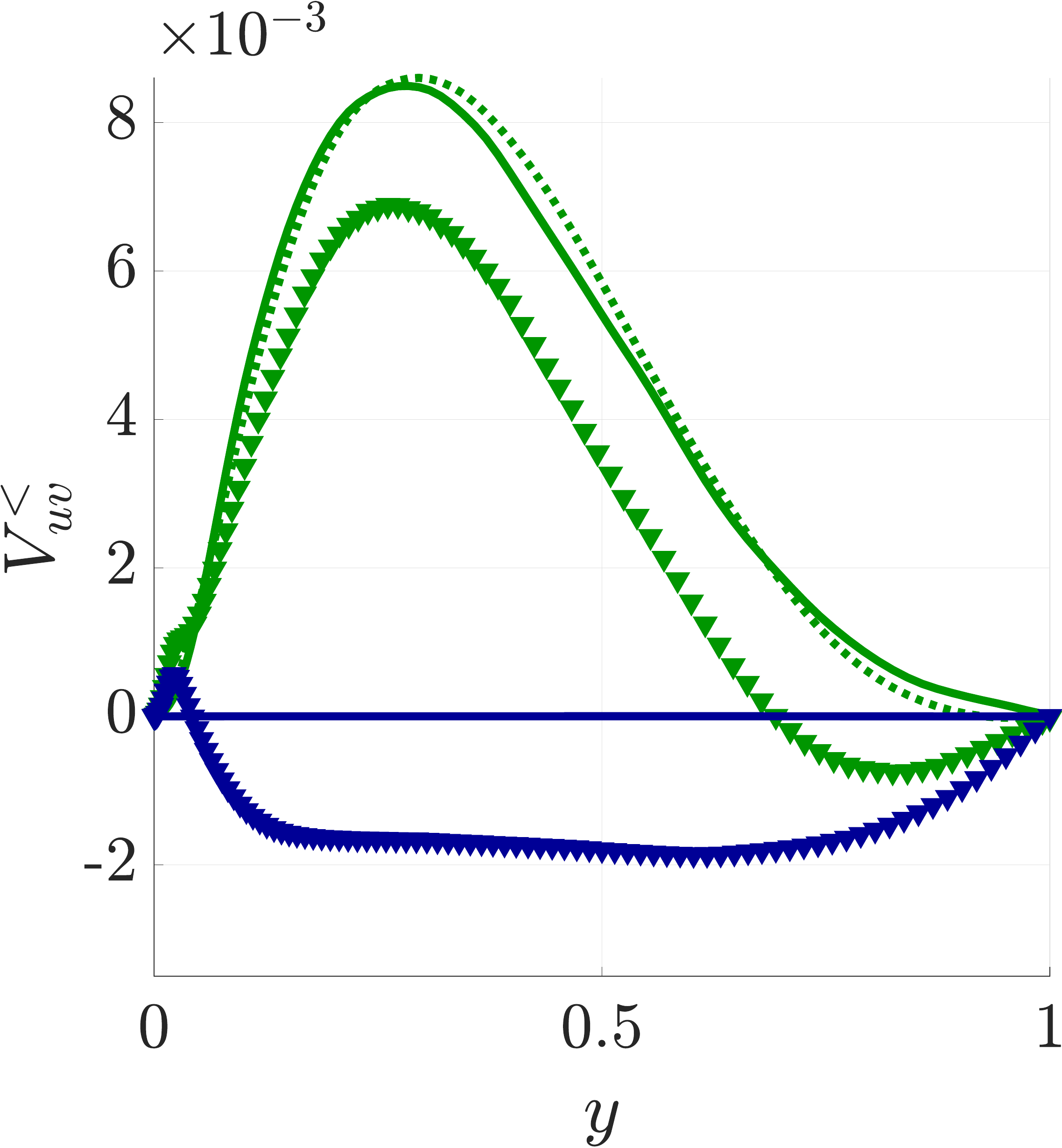}
    \caption{Contribution to the vertical mean component of the velocity from (top row) $uv$ structures and (bottom row) the complementary portion of the domain for (dark green) duct at the centre-plane and (dark blue) channel at $Re_\tau=360$. Symbols and lines as in Figure~\ref{fig:Vuv_180}.}
    \label{fig:Vuv_360}
\end{figure}

 The contributions for channel and duct at $Re_\tau=360$ are shown in Figure~\ref{fig:Vuv_360}.
In the channel, the trend for increasing $H_{uv}$ is as for $Re_\tau=180$: $V^>_{uv}$ (and $|V^<_{uv}|$) increases as $H_{uv}$ gets larger up to $H_{uv}\approx2.0$ and decreases for higher $H_{uv}$. 
In the duct, as opposed to the previous case, in the region where the secondary flow is particularly intense (approximately between $y^+\simeq20$ and $y\simeq0.4$), $V^>_{uv}$ monotonically decreases with $H_{uv}$. 
As for the cases at $Re_\tau=180$, for $H_{uv}\geq2$, the $V^>_{uv}$ distributions are in very good agreement between channel and duct. 
However, $V^>_{uv}$ accounts for less than half of the intensity of the secondary flow, thus the contribution from weak events is more relevant ($|V^<_{uv}|>|V^>_{uv}|$) and it significantly differs between duct and channel. 
\par These results can be summarised as follows: for every $H_{uv}$, at the wall-normal distances where $V^>_{uv}$ in the channel is higher than $V$ in the duct, $V^>_{uv}$ and $V^<_{uv}$ in channel and duct are in relatively good agreement; for $H_{uv}$ larger than the critical one identified by the percolation analysis, \emph{i.e.} when the intense events are isolated, $V^>_{uv}$ are always in good agreement, at every wall-normal distance between channel and ducts, and such agreement is better for more rare events. Furthermore, the dependence on the wall-normal distance of $V^>_{uv}$ for intense events in the duct does not resemble that of $V$ in the duct and it seems to be unaffected by the existence of the secondary flow. 
\par As previously mentioned, in duct flows all the off-diagonal terms of the Reynolds-stress tensor are different than zero, and the fact that the cross-stream term appears in the transport equation for the mean cross-stream vorticity suggests that $vw$ events may play an important role in the generation of the secondary motion. 
However, despite some quantitative differences, their contribution to $V$ turned out to be qualitatively similar to that of $uv$ events.
\begin{figure}
    \centering
    \includegraphics[width=0.325\textwidth]{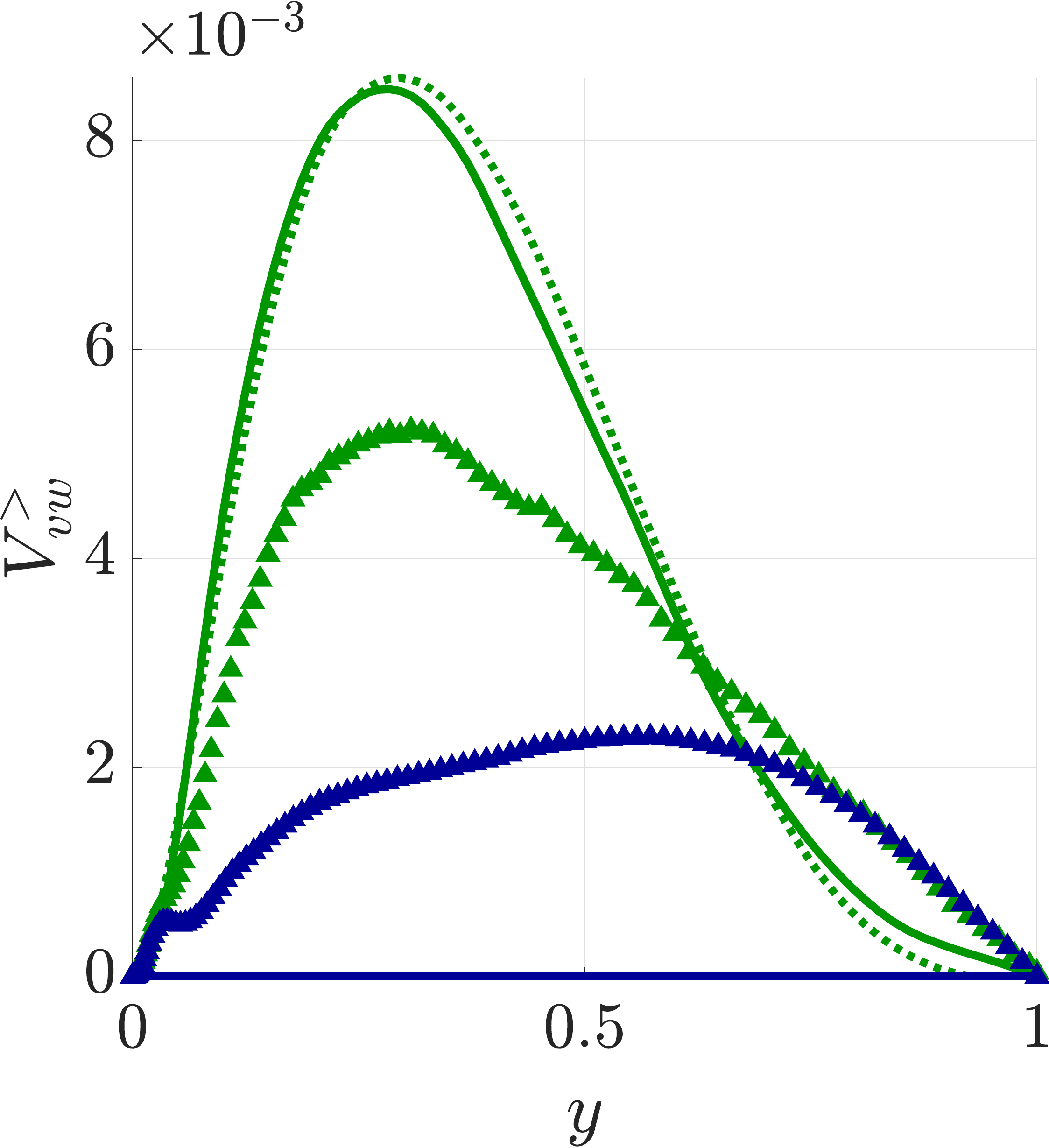}
    \includegraphics[width=0.325\textwidth]{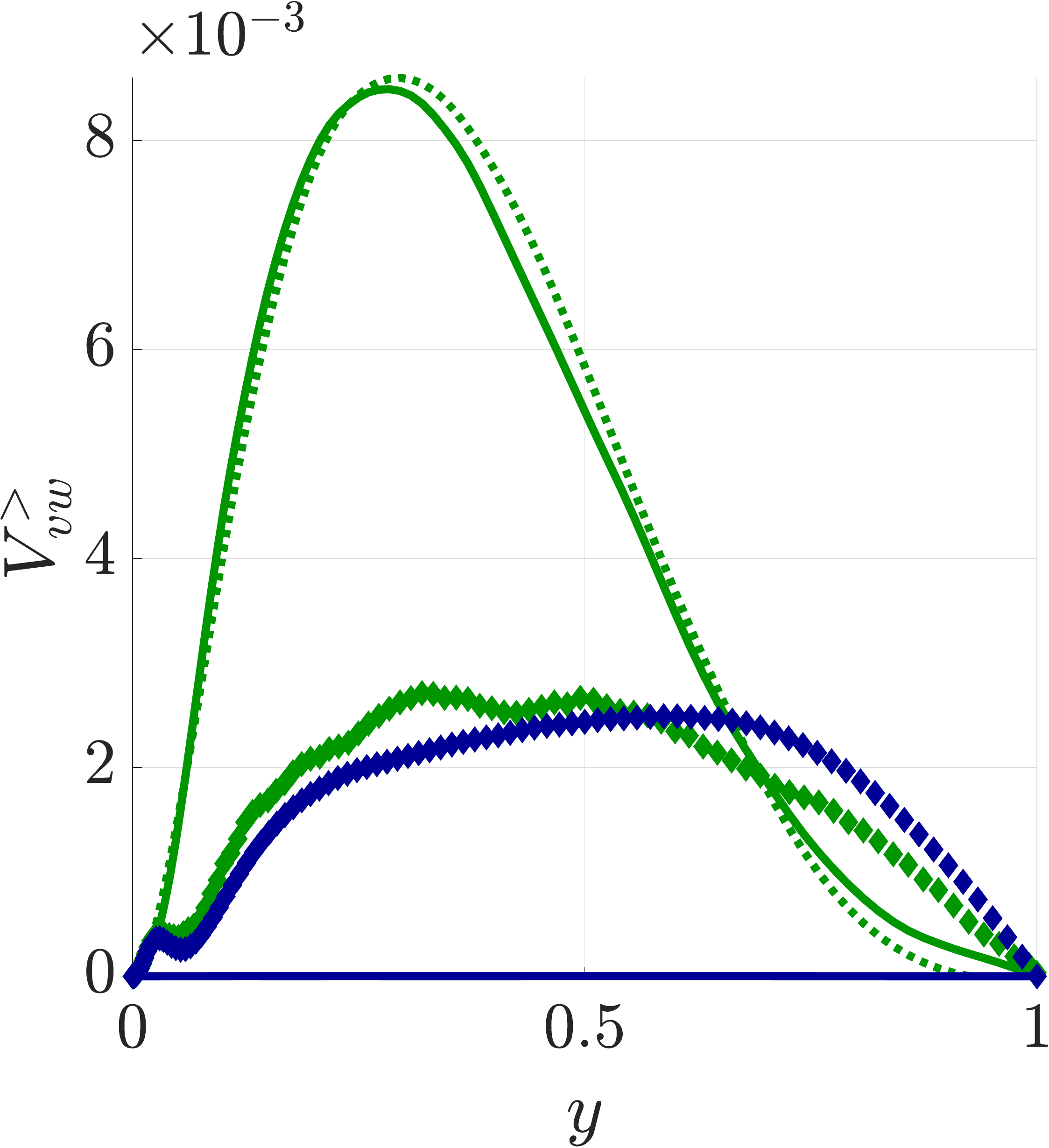}
    \includegraphics[width=0.325\textwidth]{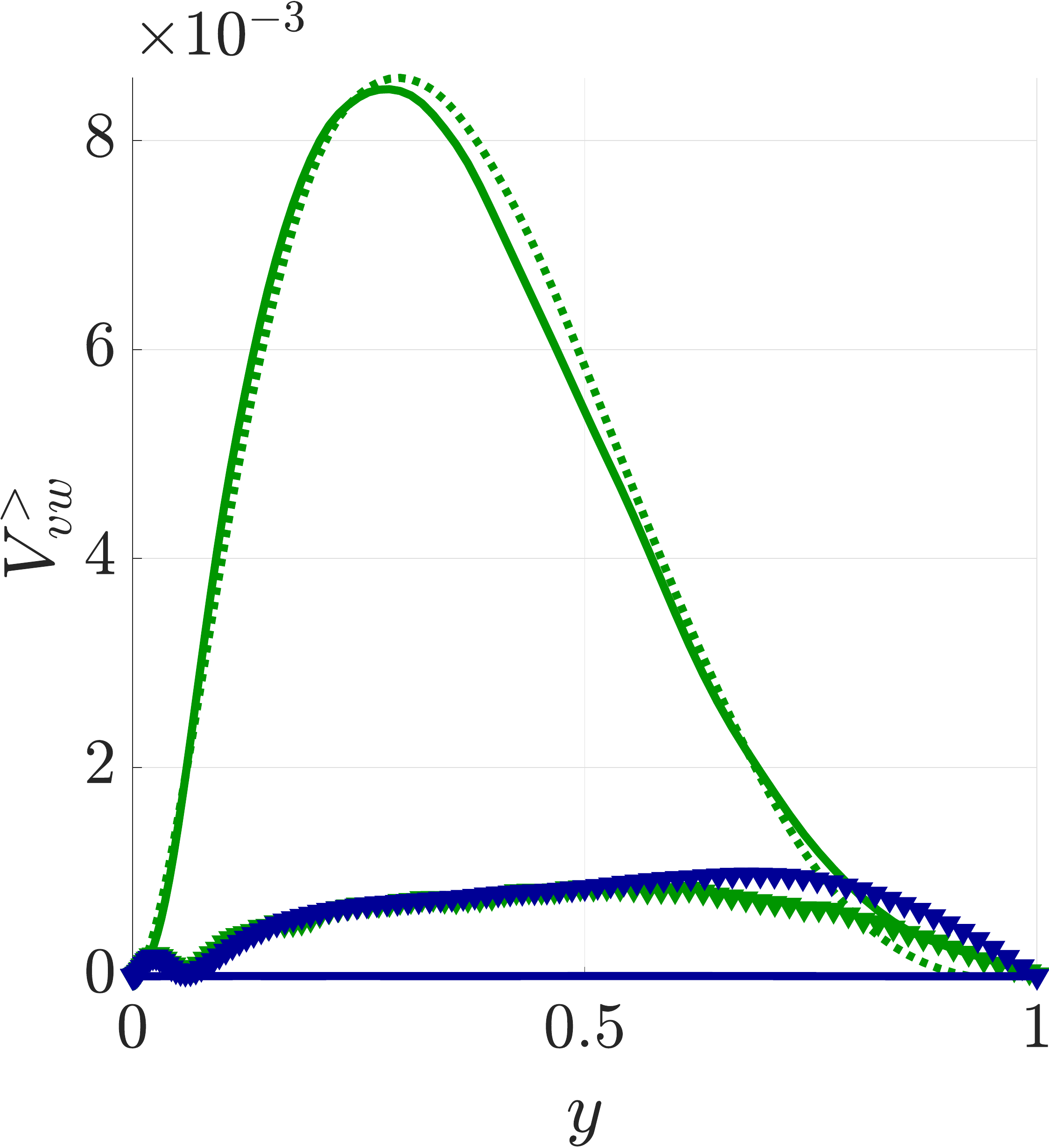}
    \caption{Contribution to the vertical component of the mean velocity $V$ from $vw$ events for (dark green) duct at the centre-plane and (dark blue) channel at $Re_\tau=360$. Symbols and lines as in Figure~\ref{fig:Vuv_180}.}
    \label{fig:Vvw}
\end{figure}
\par Figure~\ref{fig:Vvw} illustrates $V^>_{vw}$ at the centre-plane $V$ for all the considered cases. 
In general, $vw$ events have weaker fluctuations than $uv$, as can be argued by observing $\mathcal V_{\rm max}/\mathcal V_{\rm all}$ in the percolation diagram, which after the crisis is lower for $vw$ than for $uv$ in all cases, as observed in Figure~\ref{fig:perc}. 
Subsequently, in channel flow when $H_{vw}$ is such that the contribution $V^>_{vw}$ is maximised, $V^>_{vw}$ is lower than $V^>_{uv}$, and the optimal thresholds that maximised $V^>_{vw}$ is $H_{vw}\simeq1$, instead of $H_{uv}\simeq2$ for $uv$ (at both Reynolds numbers).
At the centre-plane in the duct, as for $V^>_{uv}$, the $V^>_{vw}$ profile from the duct is in good agreement with that of the channel for $H_{vw}>2$, but the two Reynolds numbers are more similar than the respective $V^>_{uv}$ profiles, since $V^>_{vw}$ in the region of more intense secondary flow is higher for $H_{vw}\approx0.5$ than for $H_{vw}\approx2$. 
This is due to the fact that $V^>_{vw}$ for $H_{vw}=2.0$ is lower than $V$, as it happens for $V^>_{uv}$ at $Re_\tau=360$.
\subsection{Contributions to the secondary flow: corner region}
\begin{figure}
    \centering
    \includegraphics[width=0.325\textwidth]{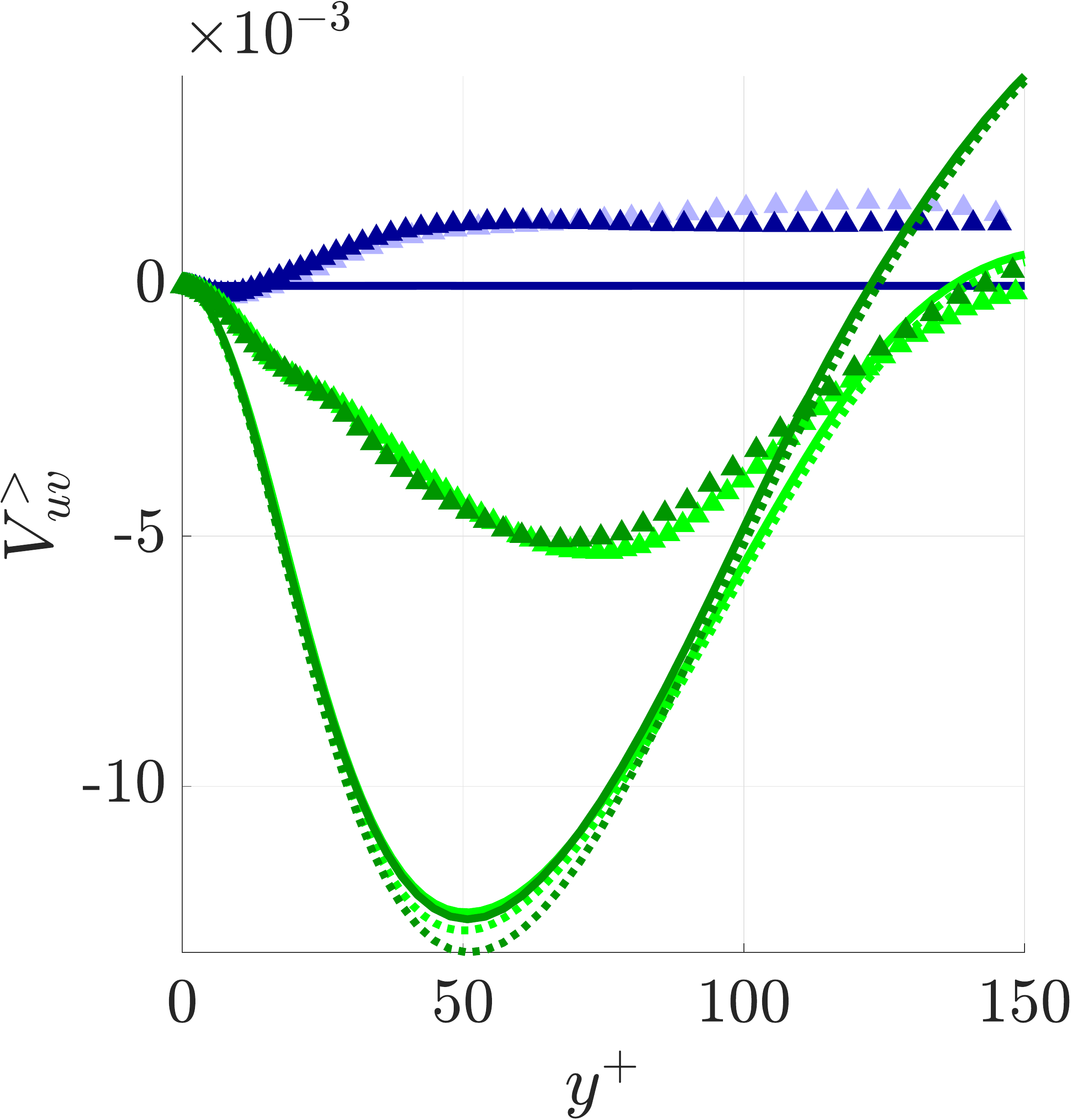}
    \includegraphics[width=0.325\textwidth]{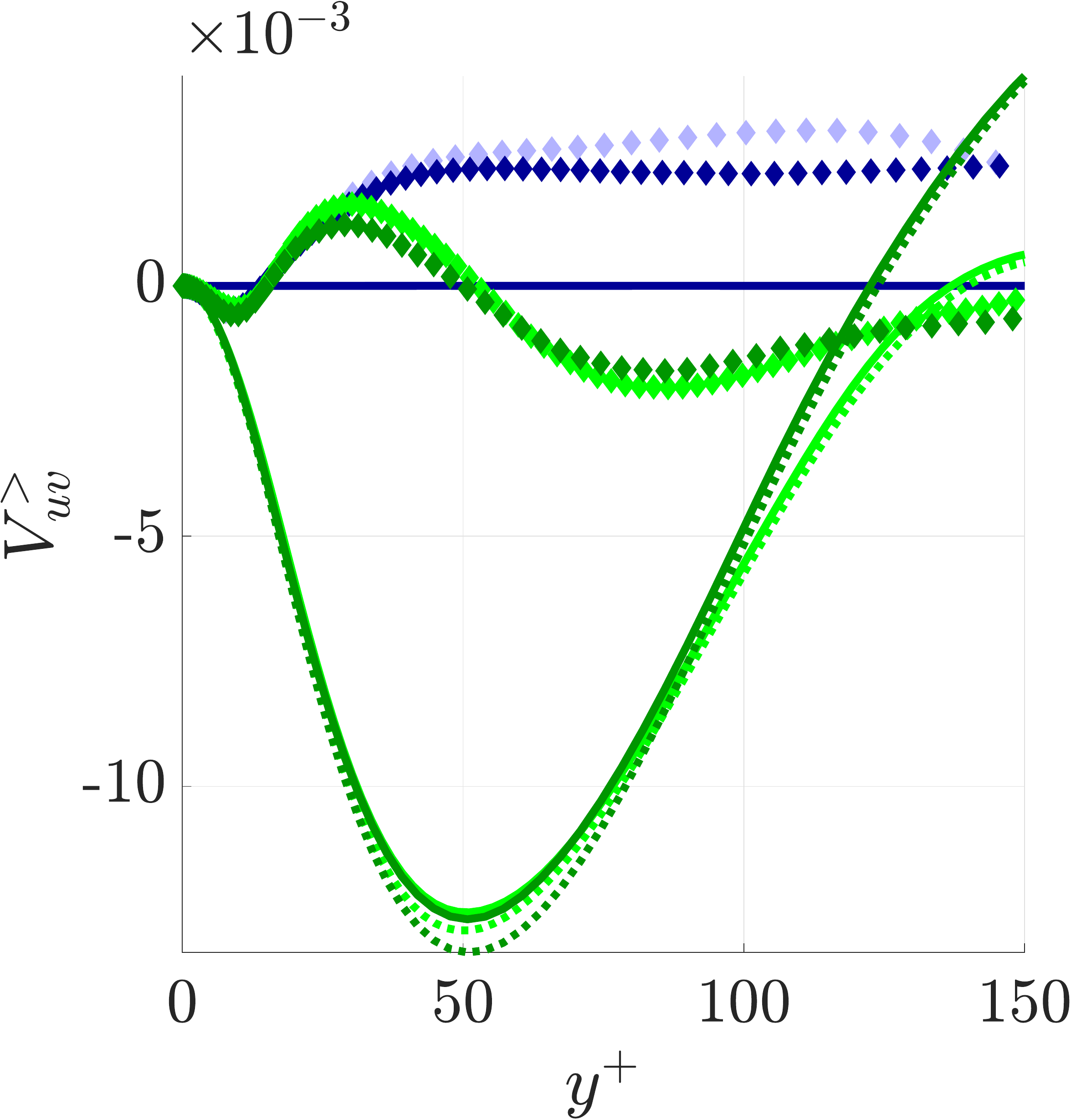}
    \includegraphics[width=0.325\textwidth]{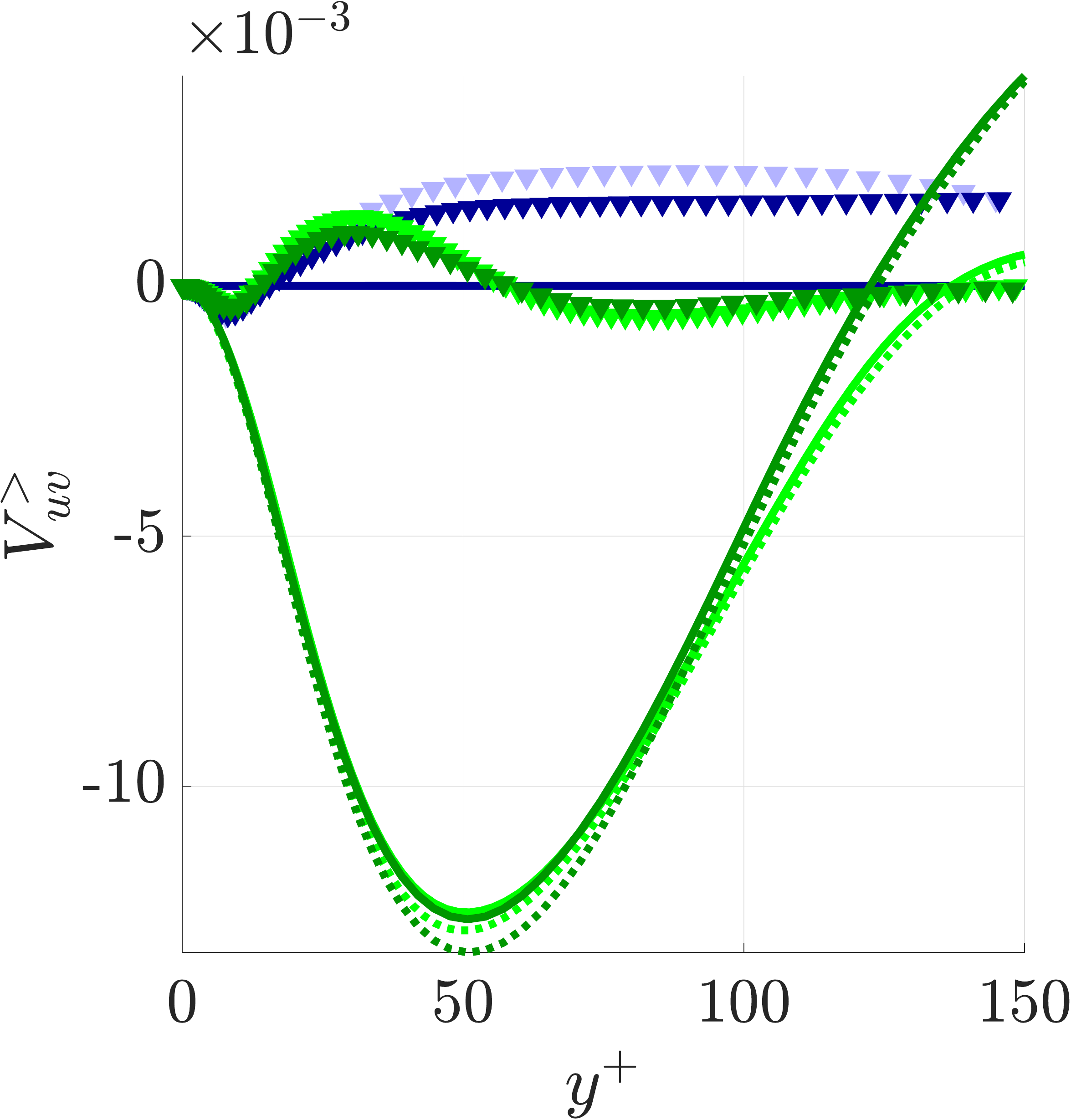}
    \includegraphics[width=0.325\textwidth]{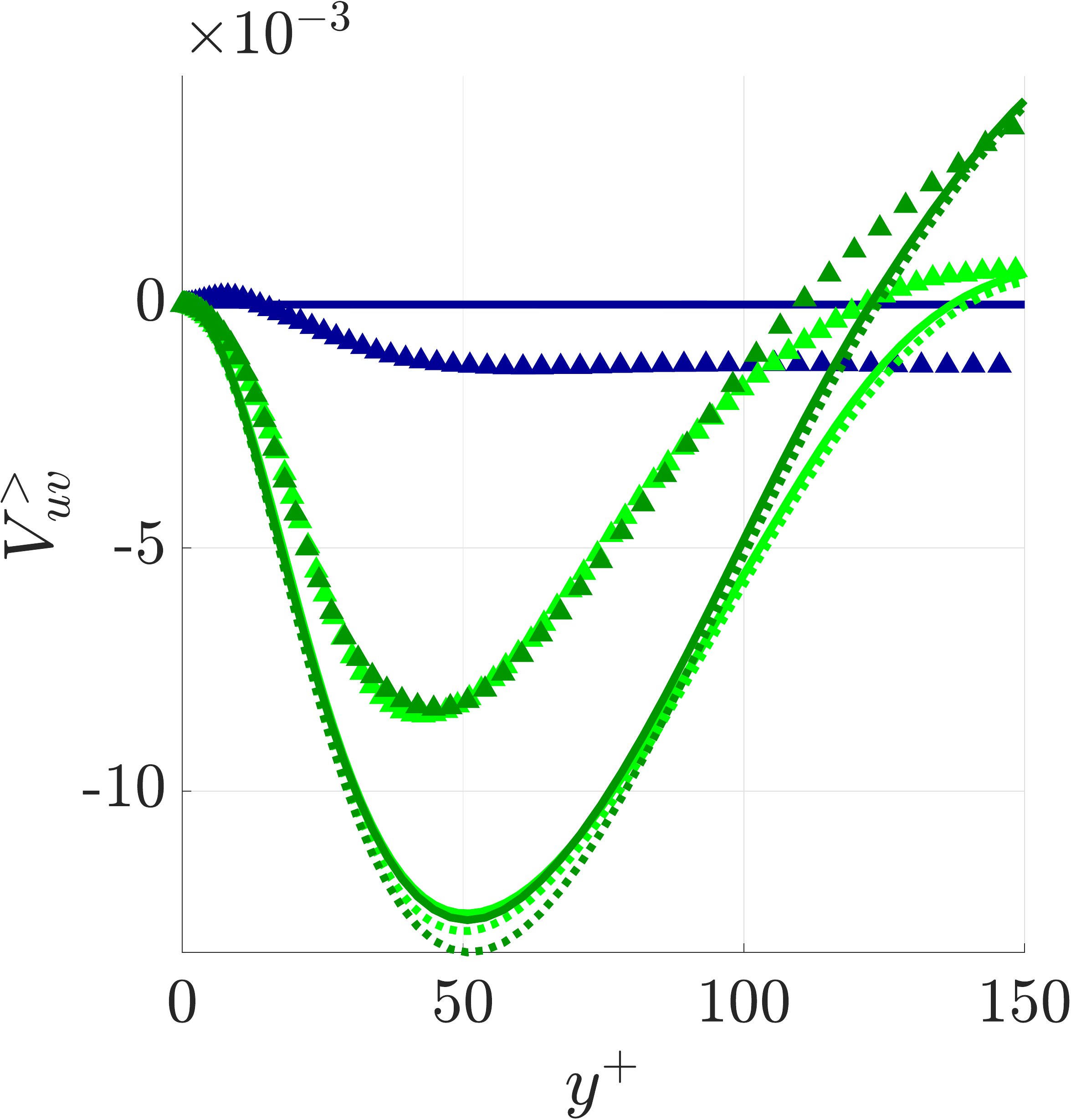}
    \includegraphics[width=0.325\textwidth]{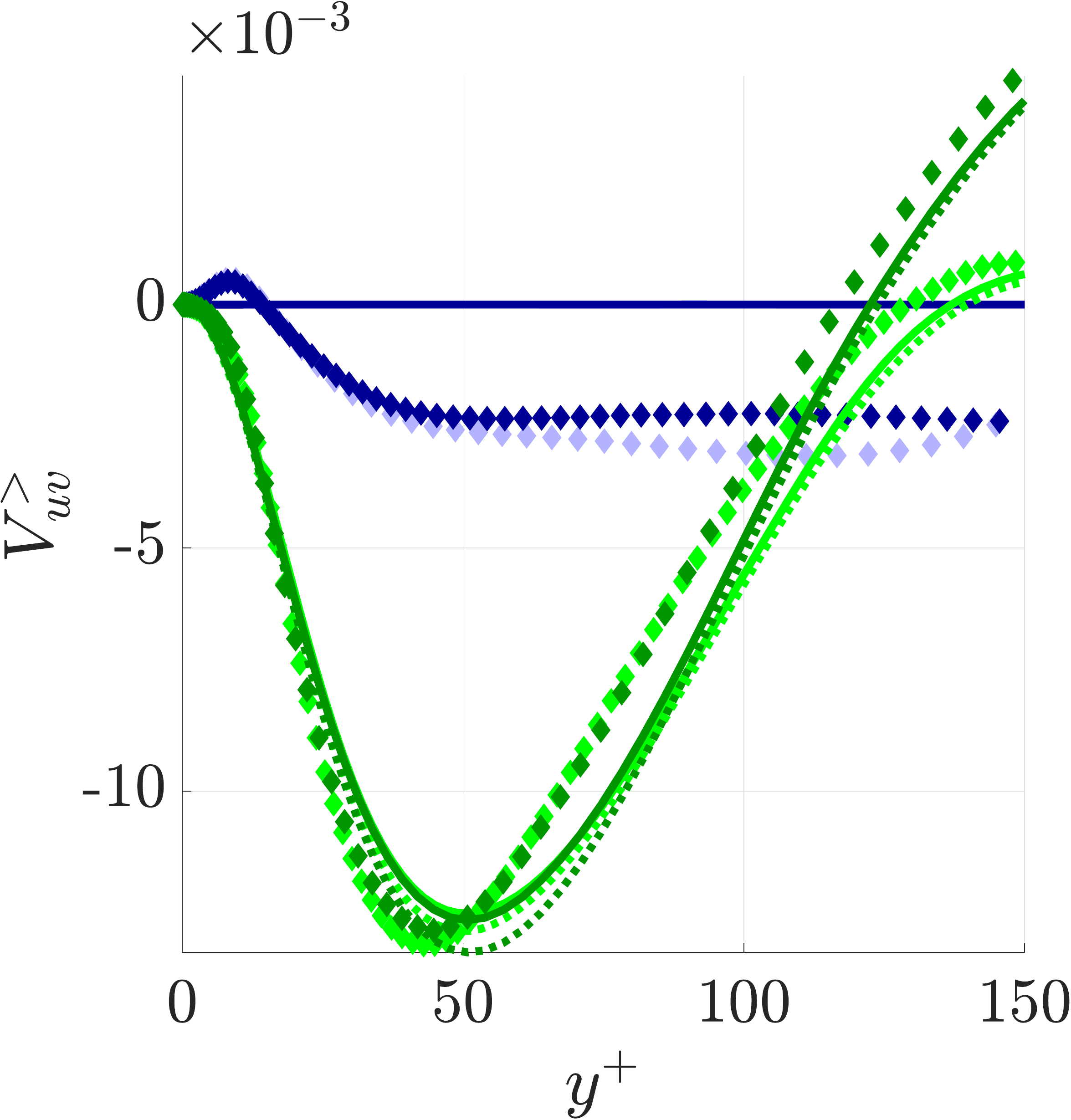}
    \includegraphics[width=0.325\textwidth]{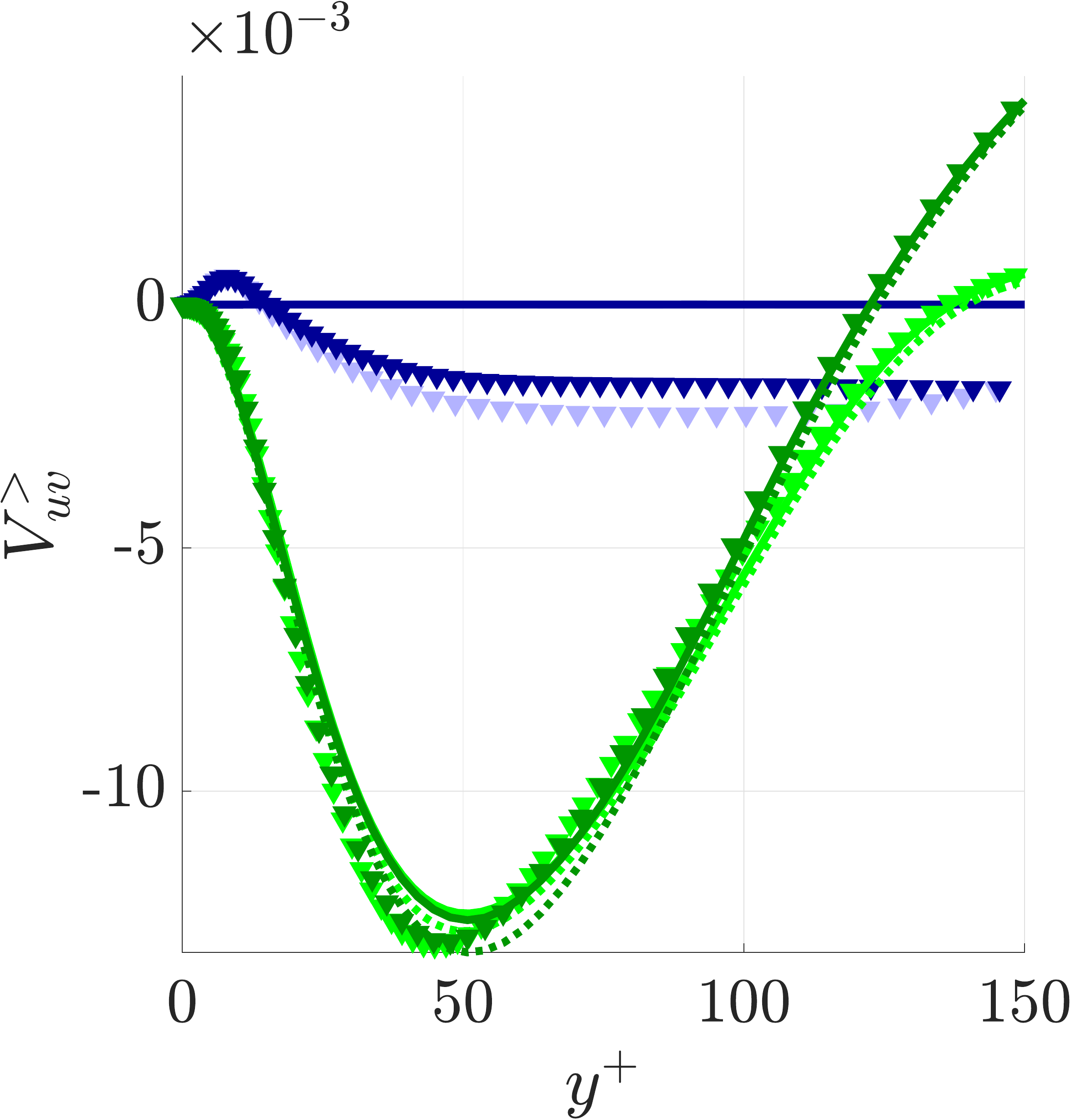}
    \caption{Contribution to the vertical mean component of the velocity from (top row) $uv$ structures and (bottom row) the complementary portion of the domain for (green) duct at the location of the first minimum ($z^{+} \approx 50$) and (blue) channel. Dark and light colours for $Re_\tau=360$ and $Re_\tau=180$, respectively. Symbols for the thresholds and the time and ensemble average as in Figure~\ref{fig:Vuv_180}.}
    \label{fig:Vuv_multi}
\end{figure}

We consider now the contribution to the secondary flow of intense events near the vertical walls. In Figure~\ref{fig:Vuv_multi} we show $V^>_{uv}$ and $V^<_{uv}$ at the location of the first minimum of $V$ ($z^+\approx50$) and $V^>_{uv}$ in the channel. Since in this region of the duct $V$ scales where it is expressed as a function of the inner-scaled wall distance, the data are reported for both Reynolds numbers.
\par It is worth noting that at this location $V$ of the duct is negative, which makes the behaviour of $V^>_{uv}$ not trivial to predict, given the predominance of ejection among intense events in the channel and in the core. In fact, $V^>_{uv}$ is qualitatively different than in the core region and in the channel.  
For $H_{uv}=0.5$, \emph{i.e.} before the percolation crisis, $V^>_{uv}$ in the duct is negative and it differs from $V^>_{uv}$ at every wall-distance. However, as $H_{uv}$ increases, the absolute value of the contribution $V^>_{uv}$ decreases monotonically and eventually $V^>_{uv}$ changes sign in the region below the corner bisector ($y^+\approx50$). For $H_{uv}$ high enough to isolate the intense events, in the region $y^+<25$ the contribution $V^>_{uv}$ is again in good agreement with the that of the channel. 
Therefore, it is possible to conclude that in this region ejections are still dominating within the intense events, despite the fact that the secondary flow has negative sign. 
On the other hand, for $y^+>25$, $V^>_{uv}$ does not match with the distribution of the channel in either of the two Reynolds numbers. 
\par Interestingly, the location of the bisector, $y^+=50$, is also the location where $V^>_{uv}$ changes sign where the most intense events are considered. The fact that for $y^+>50$ the contribution $V^>_{uv}$ for the duct is negative, is a clear sign of the influence of the vertical wall on the coherent structure near the corner. Such events are perceived as intense spanwise fluctuations by the flow on the vertical wall and they show a preferential orientation towards the nearest horizontal wall. 
\par It is worth noting that, although the contribution $V^>_{uv}$ remains relatively low, this is the only circumstance for which it is possible to observe a qualitative difference between intense events in channel and duct. 
\par At each wall-normal location, the absolute value of $V^>_{uv}$ is significantly lower than that of $V$, and for $H_{uv}=4.0$ $V^>_{uv}$ almost vanishes above the corner bisector ($y^+>50$), since in that region the probability of detection is also particularly low (Figure~\ref{fig:perc}). Therefore, $V^<_{uv}$ has a predominant role in the near-corner region, for both Reynolds numbers. In particular, for $H_{uv}=2.0$, $V^<_{uv}$ is already of the order of magnitude of $V$, and for $H_{uv}=4.0$ the agreement between $V$ and $V^<_{uv}$ is remarkably good. 
\par The relative importance of the fractional contributions to the mean velocity for $z^+=50$ is also observed for the entire near-corner region of the duct ($z^+<50$).
We also examined the contributions to $V$ from $vw$ events, $V^<_{vw}$ and $V^>_{vw}$, but they are similar to the ones previously described and they are not shown here. 
%
%

\subsection{Coherent structures: location and size}
In this section we examine the geometrical properties of $uv$ coherent structures in terms of their distance from the wall and the value of the thresholds $H_{uv}$. We focus on $uv$ events, because they have the highest contribution to the secondary motion, and we take into account the symmetries in the channel and the duct: in the former, we apply a change of reference system to the structures from which the centre of mass is located in the upper half of the domain; in the latter, all the structures are referred to the lower-left quadrant. 
Note that the centre of mass of the structures is defined as: 
\begin{equation}
    \mathbf x_{\rm cm} = \frac{1}{\mathcal{V}} \sum_{i=1}^{N} \mathcal{V}_i \mathbf{x}_i\,, 
\end{equation}
where $N$ is the number of connected grid points in the structure, $\mathcal{V}_i$ the characteristic volume assigned to each grid point, $\mathbf x_i$ is the position of the point and $\mathcal V$ is the total volume. 
\subsubsection{Classification by wall distance}
In turbulent channel flow, \cite{loza12} found that it is possible to identify three different families of $uv$ structures, which we will denote as follows: wall-attached objects (WA), whose centre of mass is located in a near-wall region below $y^+\approx20$; detached objects (D), located far from the wall; and tall wall-attached (TWA), which extend from the region near the wall up to the core of the channel, sometimes being connected with both walls. 
In the duct, the presence of two pairs of perpendicular walls leads to a more complex scenario. 
\par We classify $uv$ structures in the duct into seven different families making a distinction between objects attached to one wall or two perpendicular walls. 
The classification is based on the location of the points belonging to the structure which are closer to the walls, denoted $y_{\rm min}$ and $z_{\rm min}$ for the horizontal and the vertical walls, respectively, and the position of the centre of mass. 
\par The families under study in the duct are and defined as follows: detached structures (D) are entirely located far from the walls ($y_{\rm min}^+>20$ and $z_{\rm min}^+>20$); wall-attached structures (WA) have the centre of mass in the regions near horizontal walls ($y_{\rm cm}^+<20$); side-attached structures (SA) have the centre of mass in the regions near the vertical walls ($z_{\rm cm}^+<20$); tall-wall attached structures (TWA) have the centre of mass far from the horizontal walls but are partly contained in the corresponding near-wall region ($y_{\rm min}^+<20$); tall-side attached structures (TSA) are analogous to TWA for the vertical wall ($z_{\rm min}^+<20$); corner attached structures (CA) have the centre of mass in the near corner region ($y_{\rm cm}^+<20$ and $z_{\rm cm}^+<20$) and tall-corner attached structures (TCA), which are partially embedded in both a near-vertical wall and a near-horizontal wall ($y_{\rm min}^+<20$ and $z_{\rm min}^+<20$), but have the centre of mass far from the walls.
\par Note that the TCA structures do not necessary enter in the near-corner region, but are usually long and complex objects with some branches entering the near-wall regions of two adjacent perpendicular walls. 
\par In a preliminary study \citep{atzo18}, the authors compared the structures aspect ratio at three different spanwise locations and they employed the optimal value of $H_{uv}$ which maximises the number of detected structures. They concluded that the WA and SA structures in the duct exhibit features similar to those of the WA structures in the channel.
On the other hand, D objects closer to the vertical walls in the duct tend to be narrower than D objects in the channel, and the same is true for TWA structures closer to the vertical walls.  
It has also been observed that, in the duct, if TWA objects are considered together with TCA, they have similar geometrical properties as those of the tall wall-attached objects (TWA) in the channel; note however that certain TWA structures in the channel are wider than the spanwise size of the duct. 
\par In the present work, we focus on two different spanwise regions of duct, which are defined taking into account the scaling properties of the secondary motion: region A extends from the vertical wall to the location of the minimum of $V$ ($z^+<50$), and here the velocity profiles scale if the wall-normal vertical is expressed in inner units; region B, on the other hand, comprises the region where the core-circulation is dominant ($z>0.6$). 
The structures are assigned to region A or B based on the position of their centre of mass and those which are located neither in A nor B are discarded. 
\par
\begin{figure}
    \centering
    \includegraphics[width=0.49\textwidth]{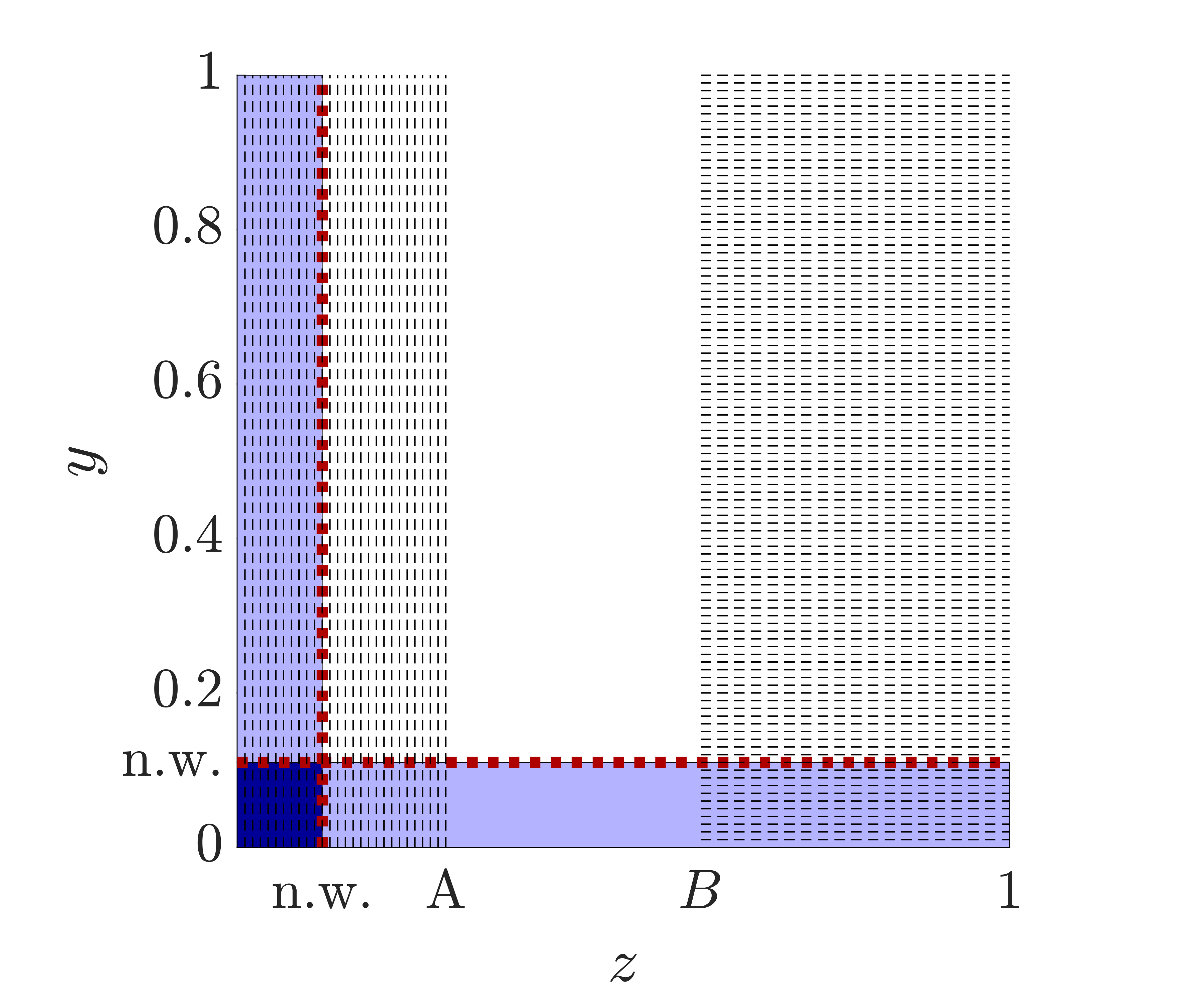}
    \includegraphics[width=0.49\textwidth]{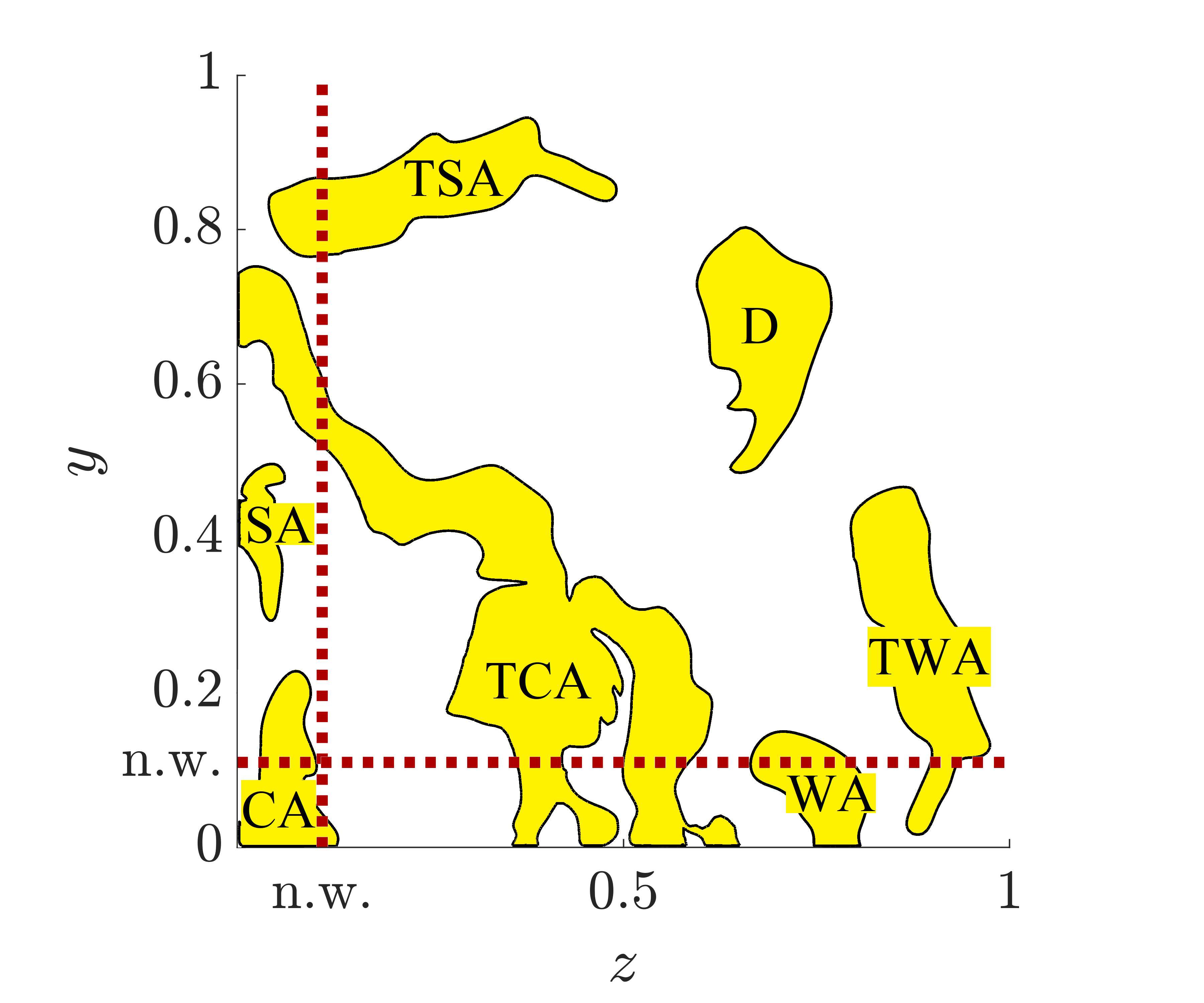}
    \caption{
    Summary of the structure classification in the duct. (Right) Regions of domain relevant for the structure analysis. The regions defined in inner units are illustrated for $Re_\tau=180$ only. Near-wall (n.w.) regions ($y^+<20$ or $z^+<20$) and the near-corner region ($y^+<20$ and $z^+<20$), spanwise regions (vertical dashed lines) A ($z^+<50$) and (horizontal dashed lines) B ($z>0.6$). (Left) Realistic representation of objects which belong to the different families: detached (D), wall attached (WA), side attached (SA), tall-wall attached (TWA), tall-side attached (TSA), corner attached and tall-corner attached (TCA). The figure does not represent an actual instantaneous cross-section. 
    }
    \label{fig:def_families}
\end{figure}

 This is motivated by the results reported the aforementioned preliminary study, which show consistent trends in how the distance from the vertical walls affect the geometrical properties of the structures. 
The boundaries of the classification regions are illustrated in Figure~\ref{fig:def_families}.  
\par
\begin{figure}
    \centering
    \includegraphics[width=0.35\textwidth]{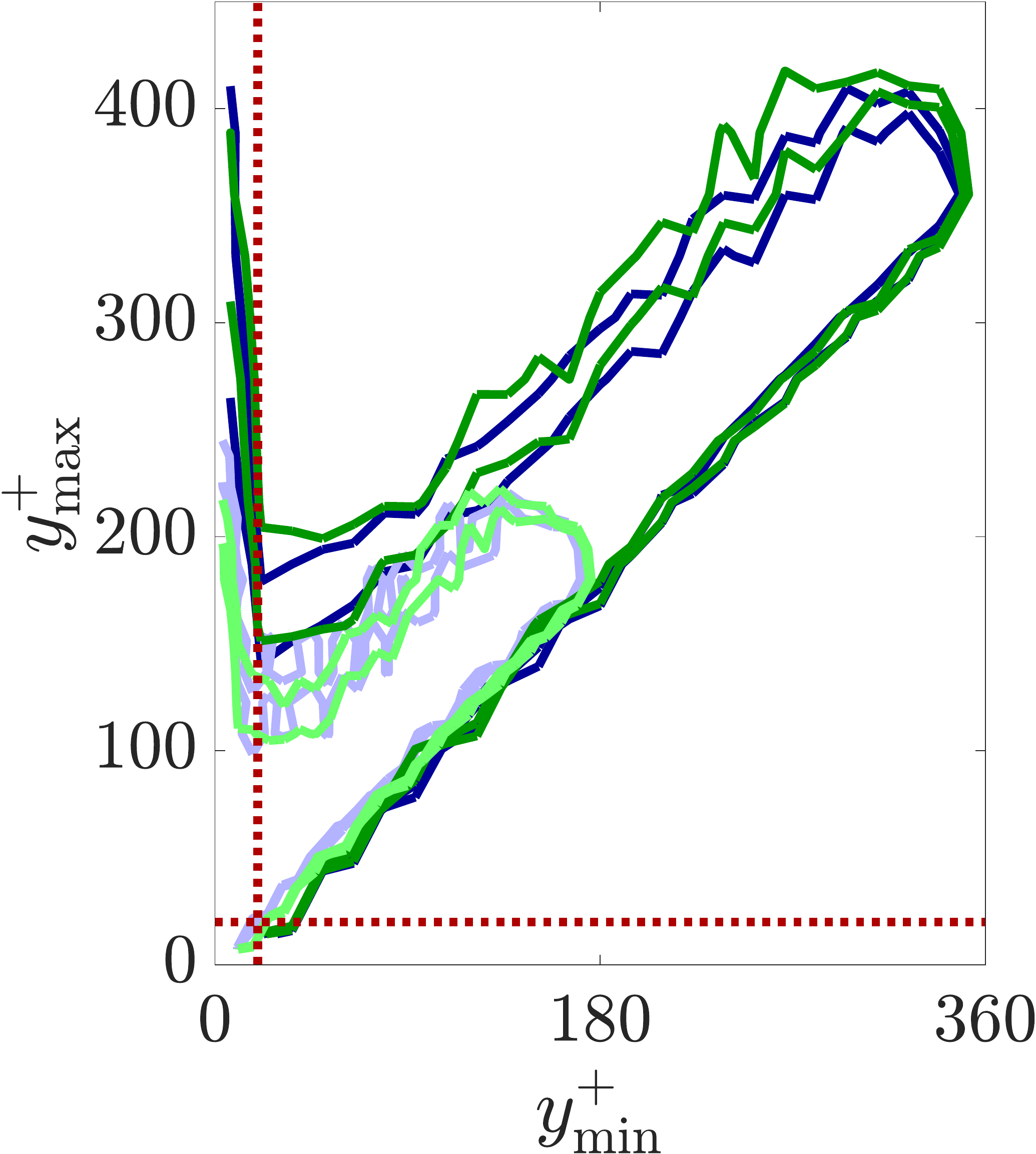}
    \includegraphics[width=0.35\textwidth]{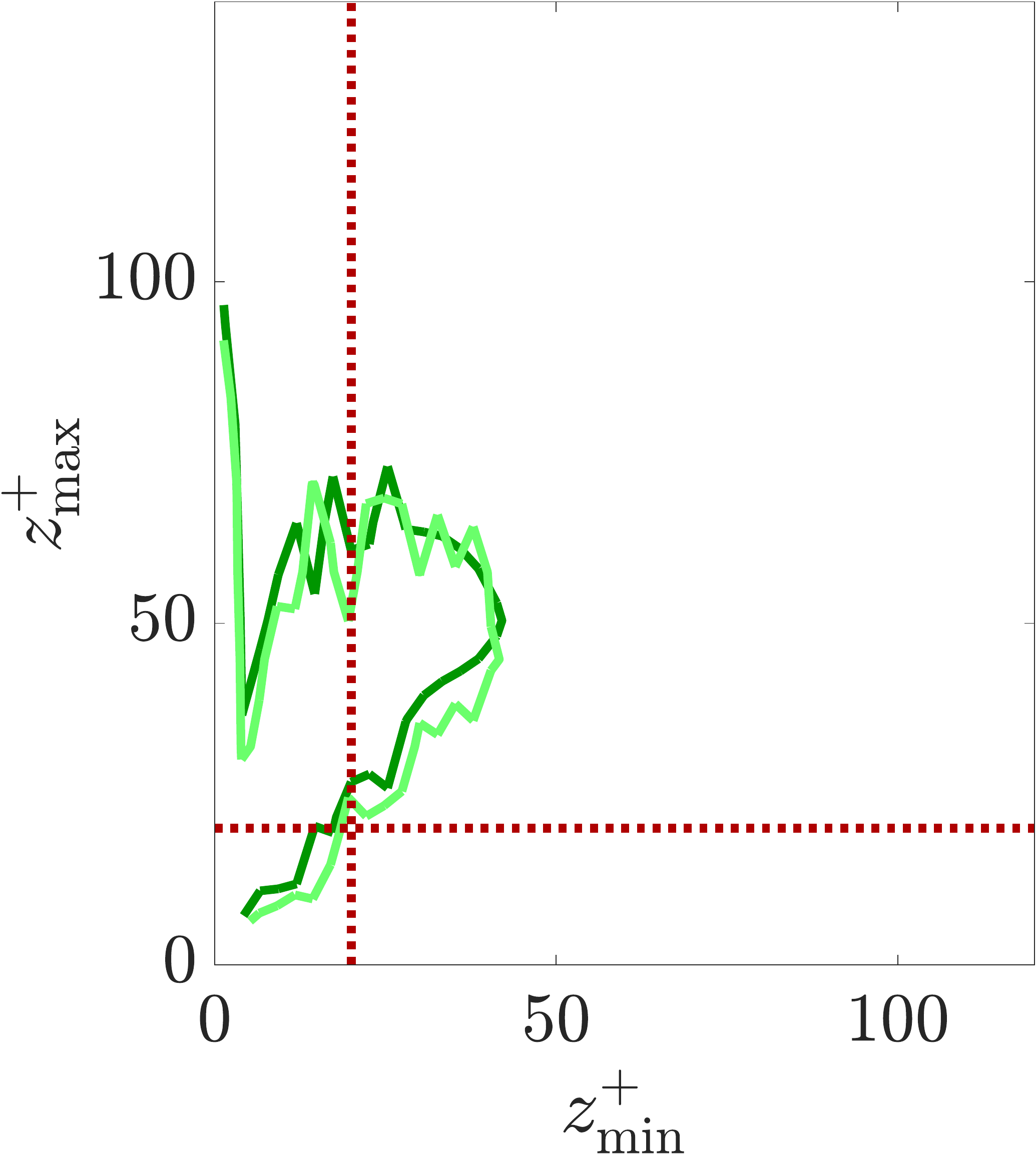}
    \caption{(Left) JPDF of the minimum and maximum distance of the identified objects to the horizontal walls in the duct region B and in the channel. (Right) JPDF of the minimum and maximum distance of the identified objects to the vertical walls in region A. The contours represent $95\%$ of the sampled objects. The dotted red lines are $y^+=20$ and $z^+=20$. Blue for channel and green for duct flow. Dark and light colour for $Re_\tau=360$ and $Re_\tau=180$, respectively.}
    \label{fig:show_families}
\end{figure}

 In order to emphasise the suitability of such a classification, we consider the joint-probability density functions (JPDF) of the minimum and maximum distances from the wall of points within each structure, denoted by $p(y^+_{\rm min},y^+_{\rm max})$ for the vertical coordinate and by $p(z^+_{\rm min},z^+_{\rm max})$ for the spanwise. 
We focus on $p(y^+_{\rm min},y^+_{\rm max})$ for structures within region B in the duct and the channel, and on $p(z^+_{\rm min},z^+_{\rm max})$ for structures within region A in the duct, for both Reynolds numbers (Figure~\ref{fig:show_families}).  
\par It is possible to note a number of similarities in $p(y^+_{\rm min},y^+_{\rm max})$ between channel and duct at the same Reynolds number: the existence of objects extending from the near-wall region to the core region in both flows ($y^+_{\rm min}<20$ and $y^+_{\rm max}>20$) and, in particular, of very large objects which extend trough the near-wall region to the opposite half of the domain ($y^+_{\rm min}<20$ and $y^+_{\rm max}>180$ for $Re_\tau=180$ and $y^+_{\rm min}<20$ and $y^+_{\rm max}>360$ for $Re_\tau=360$). 
\par Interestingly, despite the fact that the geometrical constraints prevent a proper matching of $p(y^+_{\rm min},y^+_{\rm max})$ in inner-scaling, $y^+<20$ is recognisable at both Reynolds numbers as the region from where the TWA objects originate, since for higher $y^+$ there are no structures with an extension in the vertical direction comparable to that of TWA structures. On the other hand, $p(z^+_{\rm min},z^+_{\rm max})$ for region B exhibits a very good agreement between both Reynolds numbers. 
Note that such agreement is not limited to the extreme values of $z^+_{\rm min}$ and $z^+_{\rm max}$, which is just a consequence of the sampling region being defined in inner units. 
\par An aspect which has received little attention in similar studies is how the classification to different families is affected by the choice of different thresholds $H$. 
This is in part due to the fact that the percolation analysis provides natural choices of $H_{uv}$ based either on fractions of the critical value for which the percolation crisis occurs, or on the optimal value at which the highest number of structured are detected. 
Furthermore, the contributions of the structures to the statistics generally have a relatively smooth dependency on $H_{uv}$ \citep{loza12}.  
However, $V^>_{uv}$ changes qualitatively in the duct for different $H_{uv}$, and in particular in the core region it is in better agreement with the channel for higher $H_{uv}$ at both Reynolds numbers. 
\par We therefore examined how the volume fraction occupied by different families evolves with respect volume of all the identified structures $\mathcal{V}_{\rm all}$ for different values of $H_{uv}$ (including those for which we compared $V^>_{uv}$), as shown in Table~\ref{tab:frac_V}.
\begin{table}
    \centering
    \begin{tabular}{lccccccc}
&    \multicolumn{3}{c}{$Re_\tau=180$}&~& \multicolumn{3}{c}{$Re_\tau=360$}\\
\hline
$H_{uv}$ &    $1.0$ & $2.0$ & $4.0$  &~& $1.0$ & $2.0$ & $4.0$  \\
\hline
$N_{\rm struct}$ & \underline{$343\cdot10^3$} &\underline{$313\cdot10^3$} &\underline{$154\cdot10^3$} &~& \underline{$448\cdot10^3$} &\underline{$362\cdot10^3$} &\underline{$174\cdot10^3$}  \\
$\mathcal{V}_{\rm all}/\mathcal{V}_{\rm domain}$ & \underline{$0.20$} &\underline{$0.07$} &\underline{$0.02$} &~& \underline{$0.20$} &\underline{$0.07$} &\underline{$0.02$}  \\
D({c})   & 0.60 (\textbf{0.10}) & 0.52 (\textbf{0.30}) & 0.50 (\textbf{0.68}) &~& 0.65 (\textbf{0.13}) & 0.58 (\textbf{0.36}) & 0.52 (\textbf{0.79}) \\
WA({c})  & 0.25 (\textbf{0.01}) & 0.26 (\textbf{0.05}) & 0.33 (\textbf{0.07}) &~& 0.23 (\textbf{0.01}) & 0.24 (\textbf{0.02}) & 0.33 (\textbf{0.04}) \\
TWA({c}) & 0.15 (\textbf{0.89}) & 0.22 (\textbf{0.65}) & 0.17 (\textbf{0.25}) &~& 0.12 (\textbf{0.86}) & 0.18 (\textbf{0.62}) & 0.15 (\textbf{0.17}) \\
\hline 
$N_{\rm struct}$ & \underline{$1.16\cdot10^6$} & \underline{$877\cdot10^3$} & \underline{$390\cdot10^3$} &~& \underline{$1.46\cdot10^6$} &\underline{$1.31\cdot10^6$} &\underline{$565\cdot10^3$}  \\
$\mathcal{V}_{\rm all}/\mathcal{V}_{\rm domain}$ & \underline{$0.20$} & \underline{$0.07$} & \underline{$0.01$} &~& \underline{$0.34$} &\underline{$0.11$} &\underline{$0.02$}  \\
D({d})   & 0.45 (\textbf{0.13}) & 0.46 (\textbf{0.31}) & 0.44 (\textbf{0.52}) &~& 0.52 (\textbf{0.11}) & 0.50 (\textbf{0.42}) & 0.48 (\textbf{0.72}) \\
WA({d})  & 0.16 (\textbf{0.02}) & 0.16 (\textbf{0.06}) & 0.22 (\textbf{0.12}) &~& 0.14 (\textbf{0.01}) & 0.14 (\textbf{0.03}) & 0.21 (\textbf{0.06}) \\
SA({d})  & 0.14 (\textbf{0.04}) & 0.15 (\textbf{0.08}) & 0.16 (\textbf{0.09}) &~& 0.15 (\textbf{0.01}) & 0.14 (\textbf{0.03}) & 0.17 (\textbf{0.05}) \\
TWA({d}) & 0.07 (\textbf{0.14}) & 0.09 (\textbf{0.32}) & 0.08 (\textbf{0.19}) &~& 0.06 (\textbf{0.05}) & 0.10 (\textbf{0.34}) & 0.08 (\textbf{0.14}) \\
TSA({d}) & 0.09 (\textbf{0.17}) & 0.08 (\textbf{0.15}) & 0.04 (\textbf{0.05}) &~& 0.09 (\textbf{0.08}) & 0.09 (\textbf{0.13}) & 0.03 (\textbf{0.03}) \\
CA({d})  & 0.07 (\textbf{ $\approx$0 }) & 0.05 (\textbf{0.01}) & 0.06 (\textbf{0.02}) &~& 0.03 (\textbf{0.01}) & 0.02 (\textbf{ $\approx$0 }) & 0.03 (\textbf{ $\approx$0 }) \\
TCA({d}) & 0.02 (\textbf{0.50}) & 0.01 (\textbf{0.07}) & ~$\,\approx$0  (\textbf{0.01}) &~& 0.01 (\textbf{0.73}) & 0.01 (\textbf{0.05}) & ~$\,\approx$0  (\textbf{ $\approx$0 }) \\         
\hline 
\end{tabular}
    \caption{Number of structures in the entire data-set $N_{\rm struct}$, ratio between the overall volume occupied by structures and the computational domain $\mathcal{V}_{\rm all}/\mathcal{V}_{\rm domain}$, fraction in terms of number of objects volume (the latter in bold font) for $uv$ structures of different families in channel (c) and duct (d).}
    \label{tab:frac_V}
\end{table}
\par In channel flow, where the D, WA and TWA families are considered, the effect of increasing $H_{uv}$ is to reduce the relative space occupied by TWA structures, to the advantage of both WA and D objects. 
For $H_{uv}=0.5$, \emph{i.e.} a threshold lower than the critical one but still high enough to have a high number of individual structures, most of the points fulfilling the condition are gathered in large TWA objects which together provide $\approx90\%$ of $\mathcal{V}_{\rm all}$ (results not shown here). 
Despite this, $\mathcal{V}_{\rm all}$ is between $\approx30\%$ and $\approx40\%$ of the entire computational domain. 
For $H_{uv}=2.0$ the TWA objects still occupy the largest portion of volume among all the types of structures, but their relative importance is reduced to $\approx60\%$ of $\mathcal{V}_{\rm all}$. For even higher values of the threshold, \textit{e.g.} for $H_{uv}=4.0$, only the stronger fluctuations are still detected and therefore most of the TWA are broken down into D structures, which then become the most representative in terms of volume ($\approx70\%$ of $\mathcal{V}_{\rm all}$). Note that at such high values of $H_{uv}$, only rare and small very intense events are sampled and $\mathcal{V}_{\rm all}$ is $\approx2\%$ of the domain. 
\par In duct flow, where seven different families are defined, the effect of increasing the threshold is more complex. 
Similarly to the channel, the largest structures with $H_{uv}=0.5$, which are TCA connected to at least two contiguous walls, account for more than the $\approx90\%$ of $\mathcal{V}_{\rm all}$. 
However, as $H_{uv}$ increases, the contribution of TCA to the total volume drops and, for $H_{uv}=2.0$, it is reduced to less than $10\%$ of $\mathcal{V}_{\rm all}$. 
Interestingly, at this $H_{uv}$ the combined volume fraction from tall-attached objects (TWA, TSA and TCA) is lower in the duct than in the channel, and, subsequently, detached and wall-attached objects (WA and SA) have a higher share of $\mathcal{V}_{\rm all}$. 
For $H_{uv}=4.0$, following the same trend previously observed in channel flow, large attached structures become less important than detached ones in terms of occupied volume. 
\par Reynold-number effects can also be observed in the distribution of structures among families for both channel and duct.
Since the near-wall region is a relatively smaller portion of the domain (in outer scaling) for higher Reynolds number, the volume fraction of D objects increases and that of SA and WA decreases as the Reynolds number increases. 
On the other hand, the volume fraction of tall attached objects has a weaker Reynolds-number dependence, since the bigger objects in these families tend to extend through a large portion of the cross section of the domain, a fact that is in agreement with what was reported for channel at much higher Reynolds number by \cite{loza12}.
\subsubsection{Size of structures in different families}
In the following we describe the geometrical properties of structures belonging to different families, with the main aim of identifying features which distinguish coherent structures in duct and channel flow. 
\par The structure sizes are defined based on the bounding box, \emph{i.e.} the length in the streamwise direction is defined for each structure as $\Delta_x=x_{\rm max}-x_{\rm min}$. 
The sizes are scaled in viscous units, employing in the duct the wall-shear stress at the centre plane, despite the fact that it is not uniform in the spanwise direction. 
Note that scaling with the local friction velocity was also considered, but since most of the coherent structures are relatively large the results do not differ significantly. 
We observed that the choice of $H_{uv}$ has a lower impact on the probability density functions (PDF) of the sizes then on the volume fraction for the different families, therefore we report the PDF only for  $H_{uv}=2.0$ in most cases. 
\par We firstly examine the size of small objects attached to the wall.
\begin{figure} 
    \centering
    \includegraphics[width=0.325\textwidth]{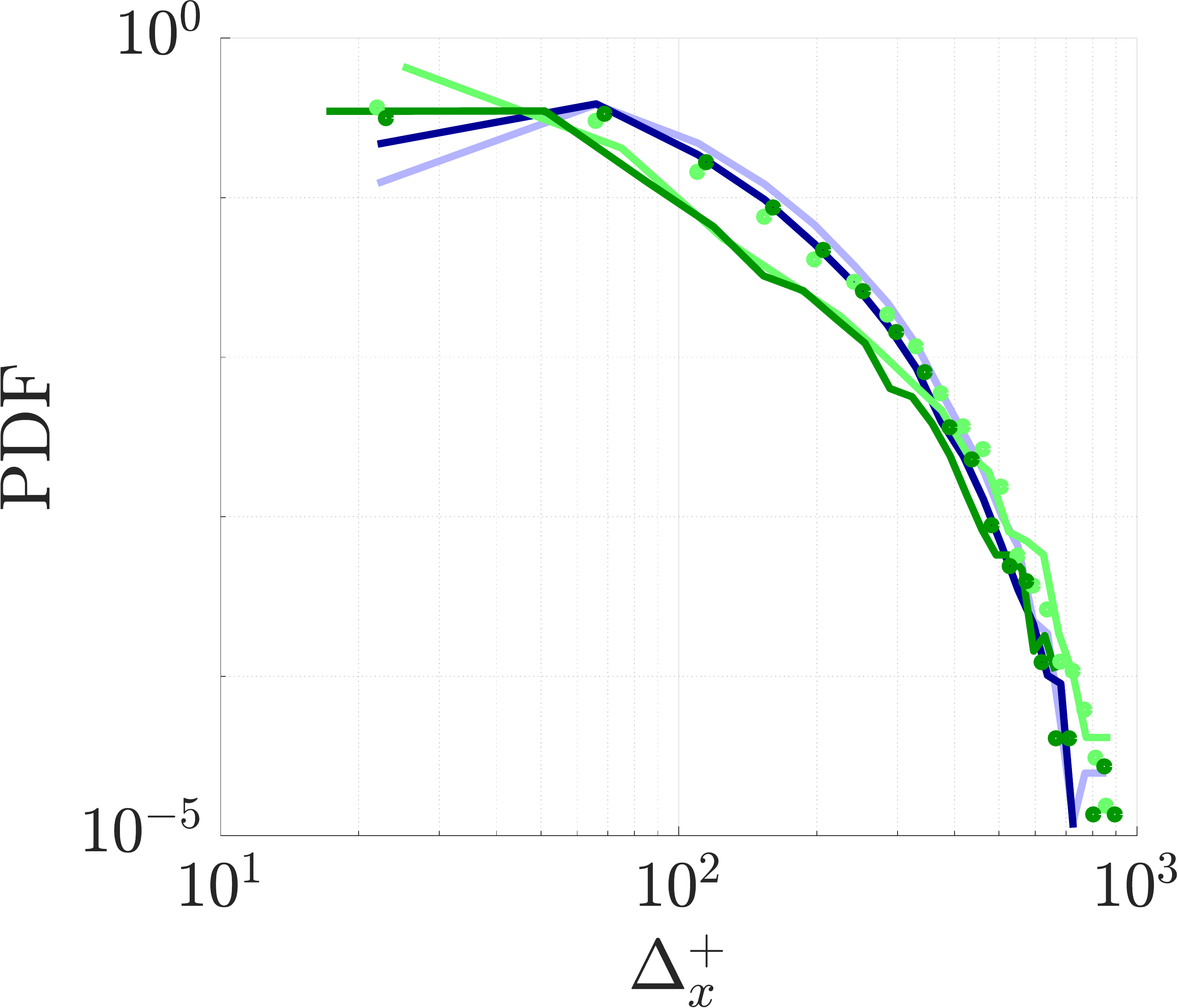}
    \includegraphics[width=0.325\textwidth]{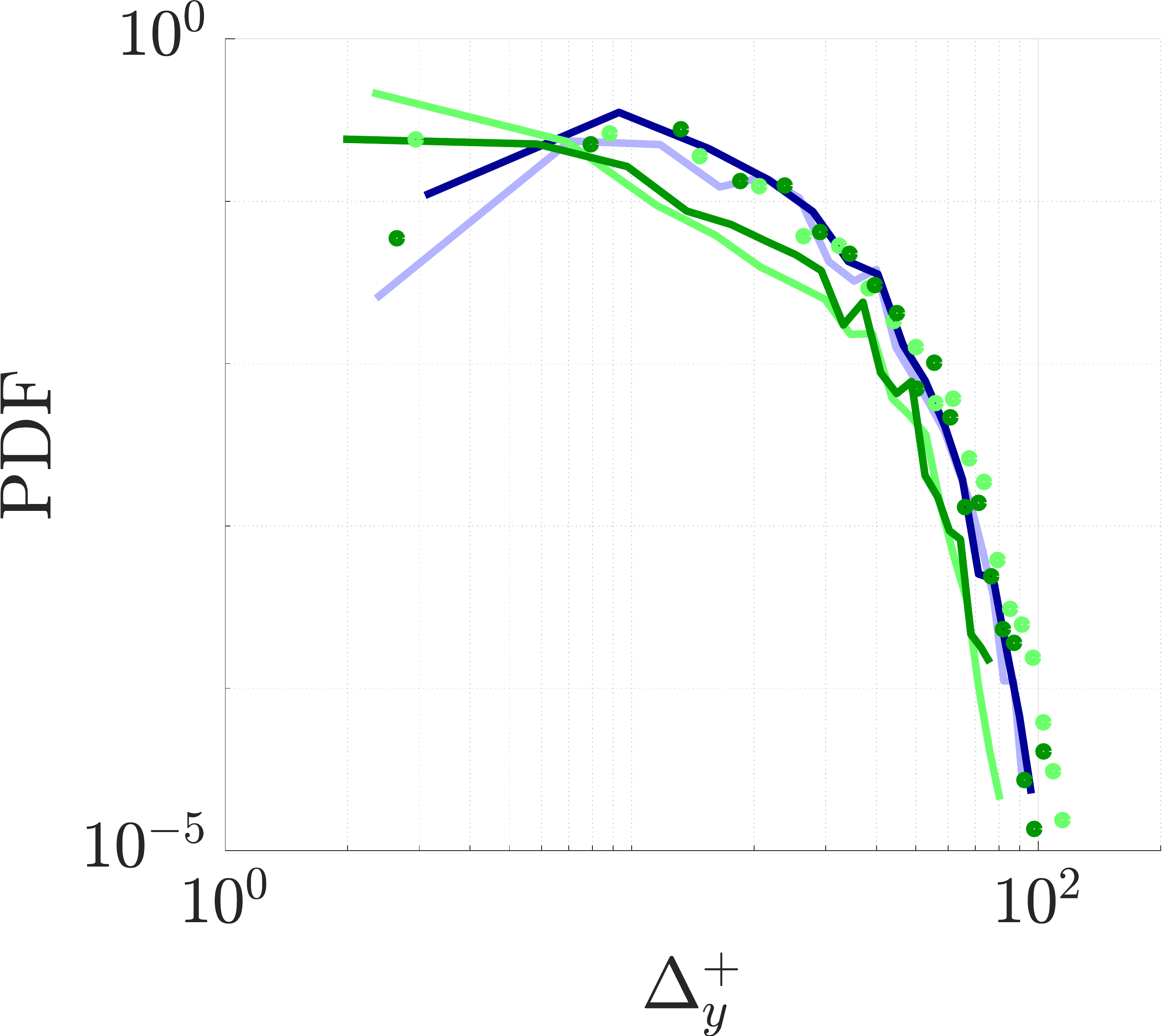}
    \includegraphics[width=0.325\textwidth]{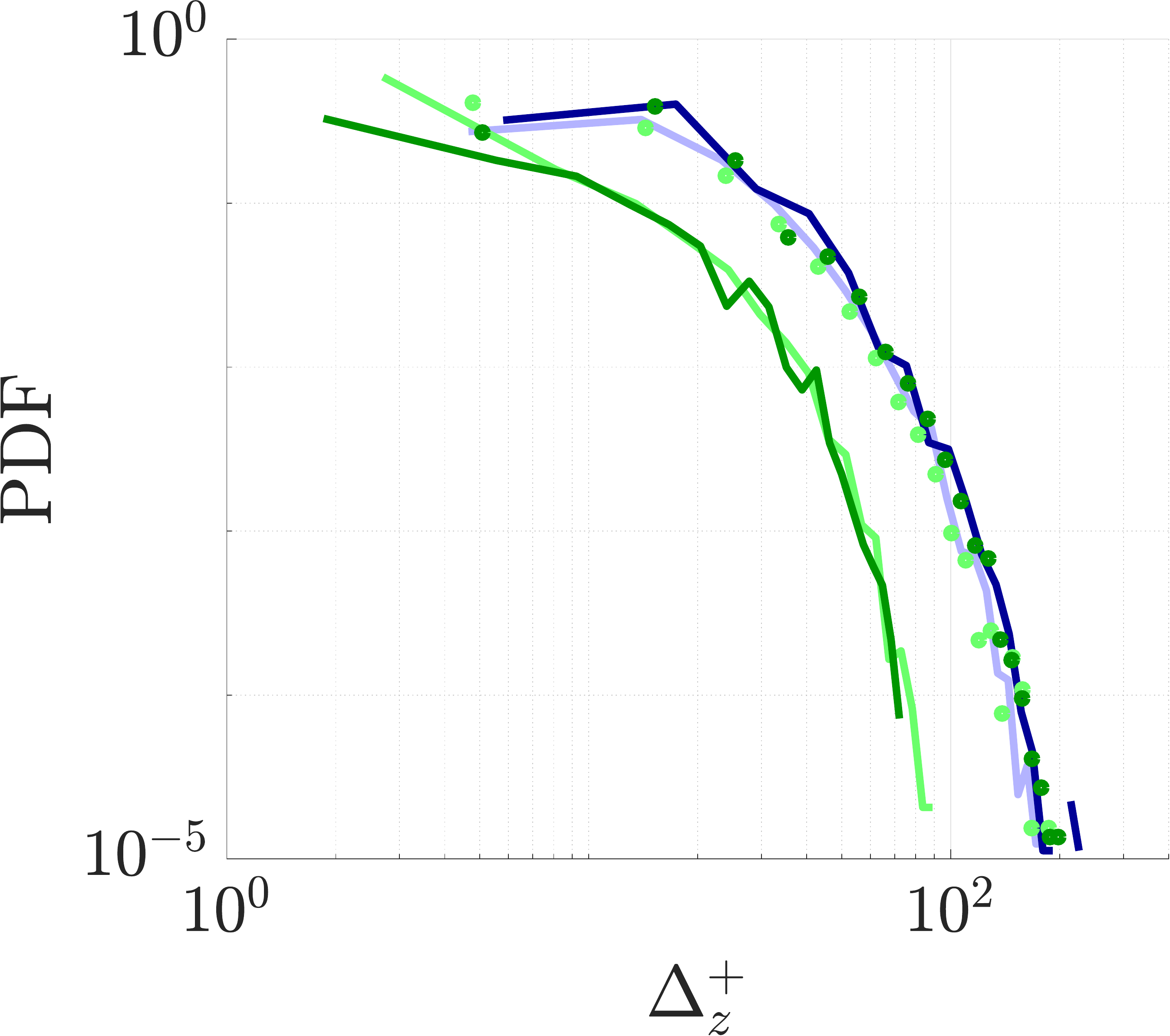}
    \includegraphics[width=0.325\textwidth]{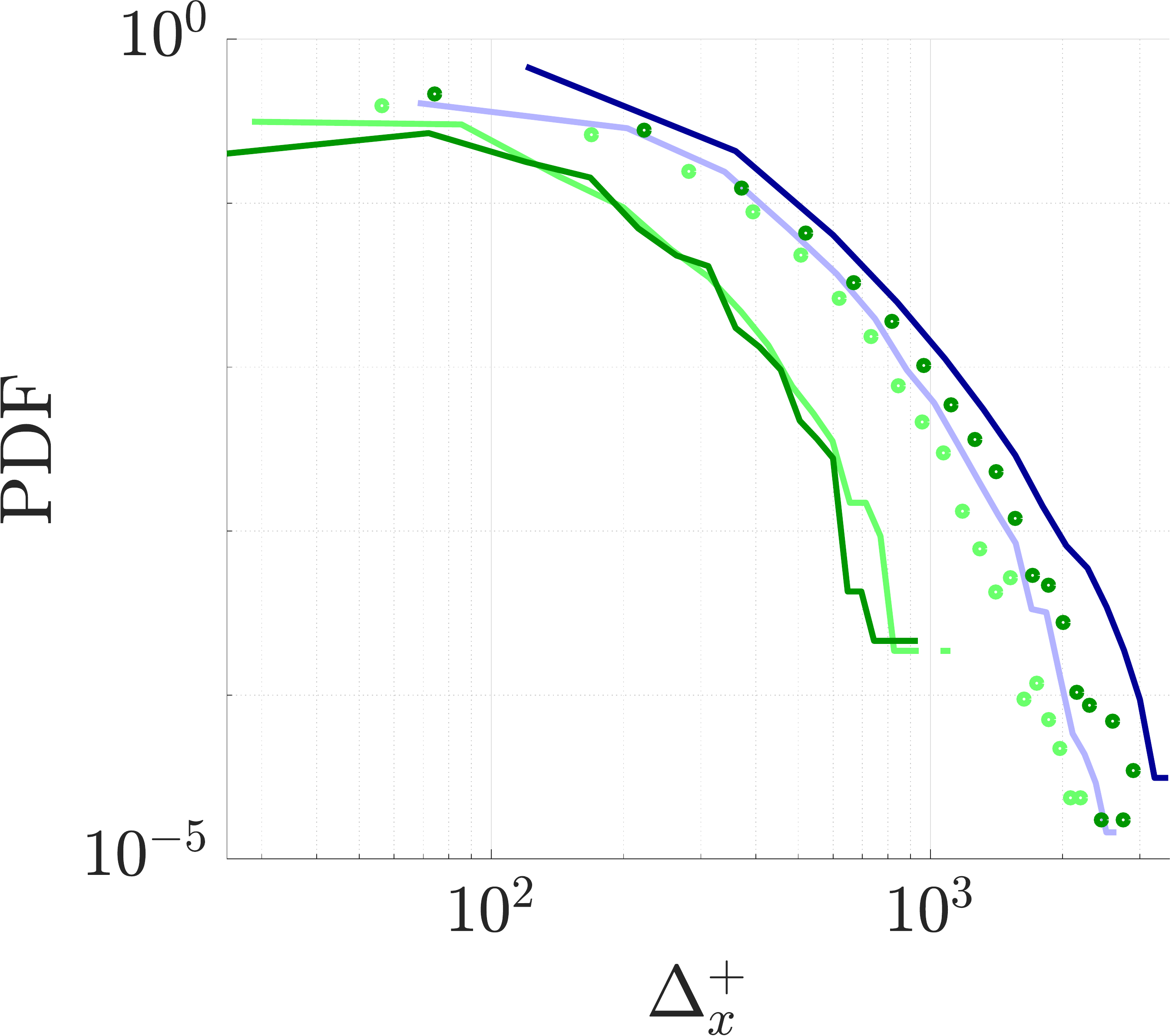}
    \includegraphics[width=0.325\textwidth]{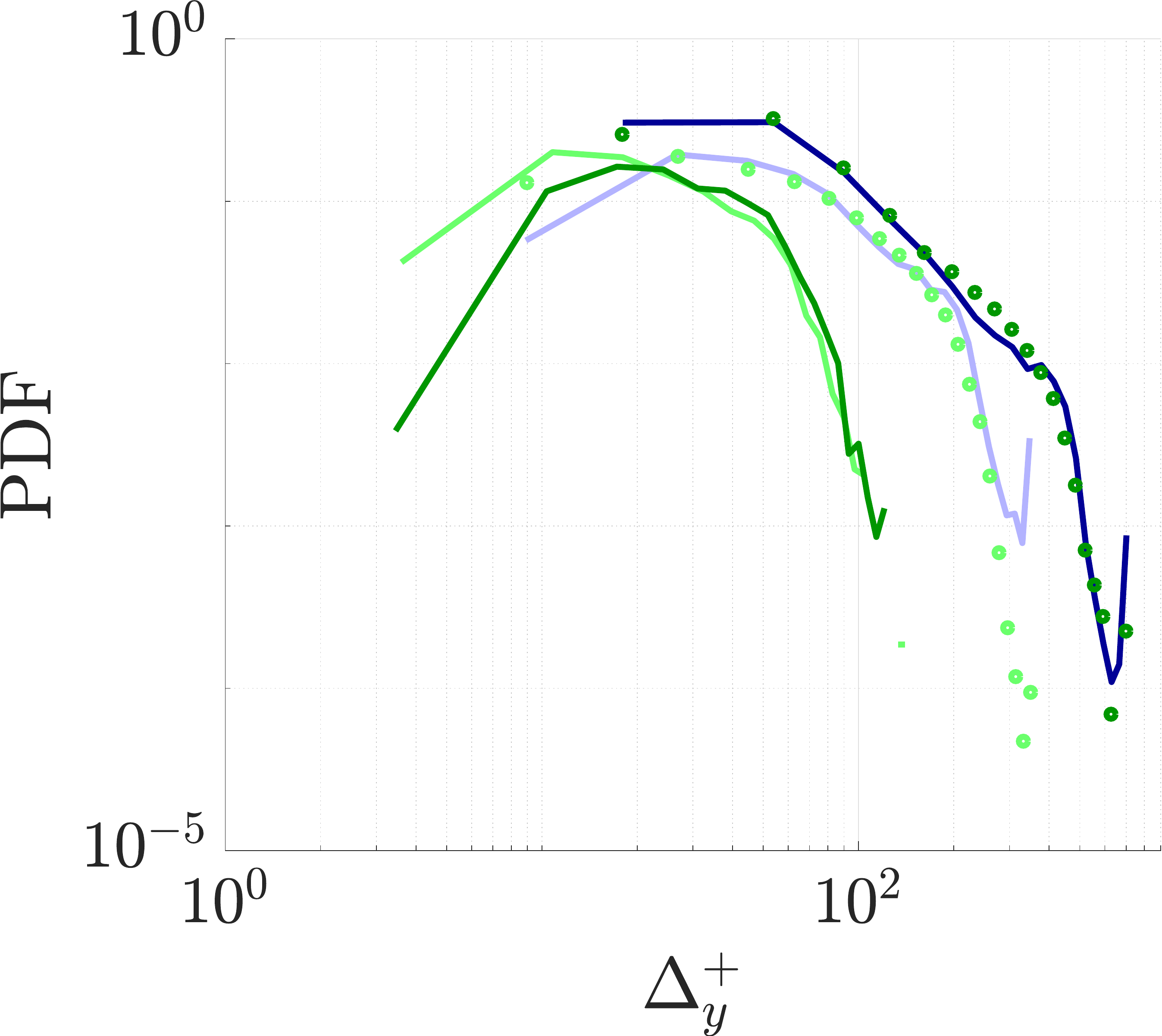}
    \includegraphics[width=0.325\textwidth]{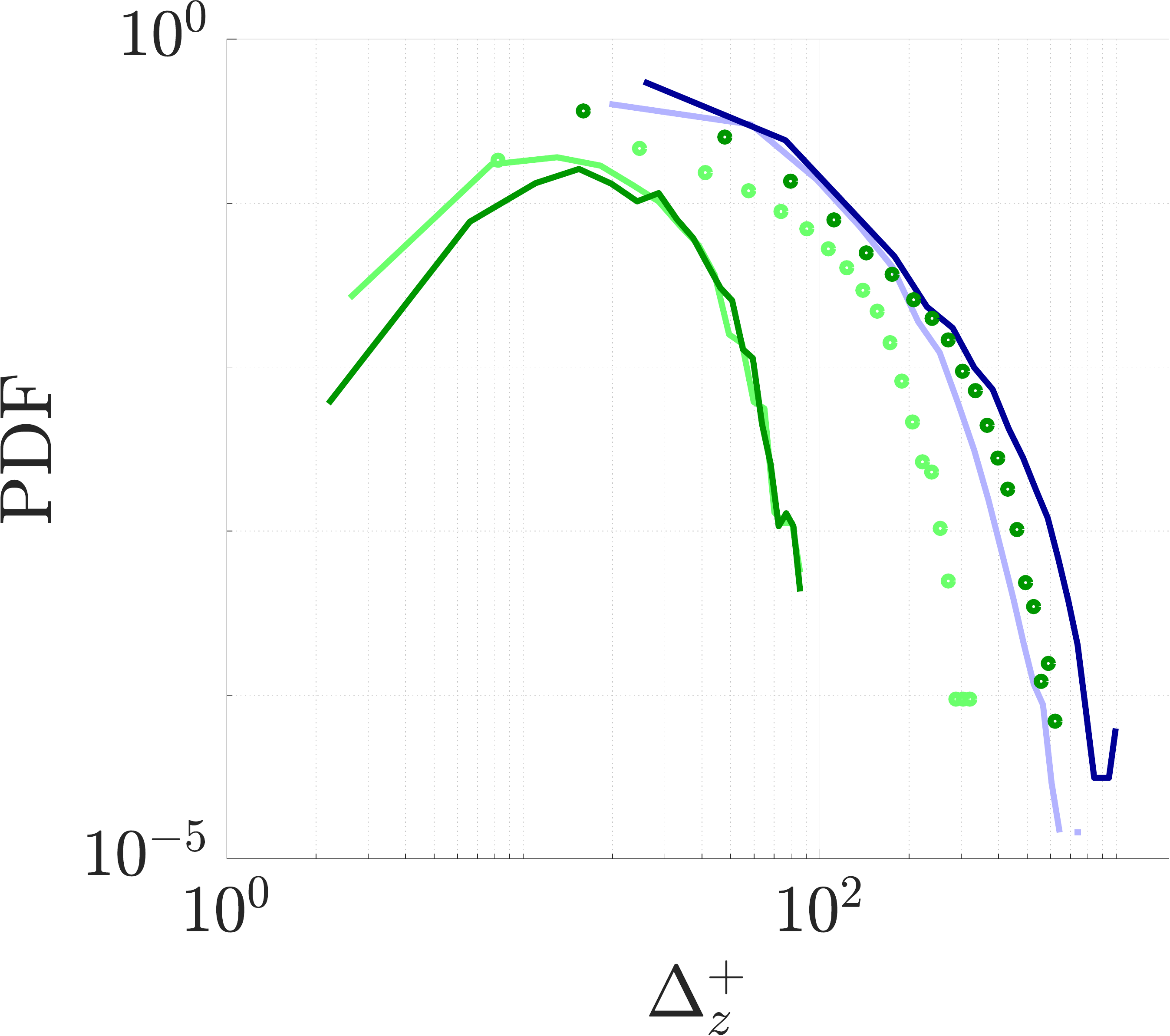}
    \includegraphics[width=0.325\textwidth]{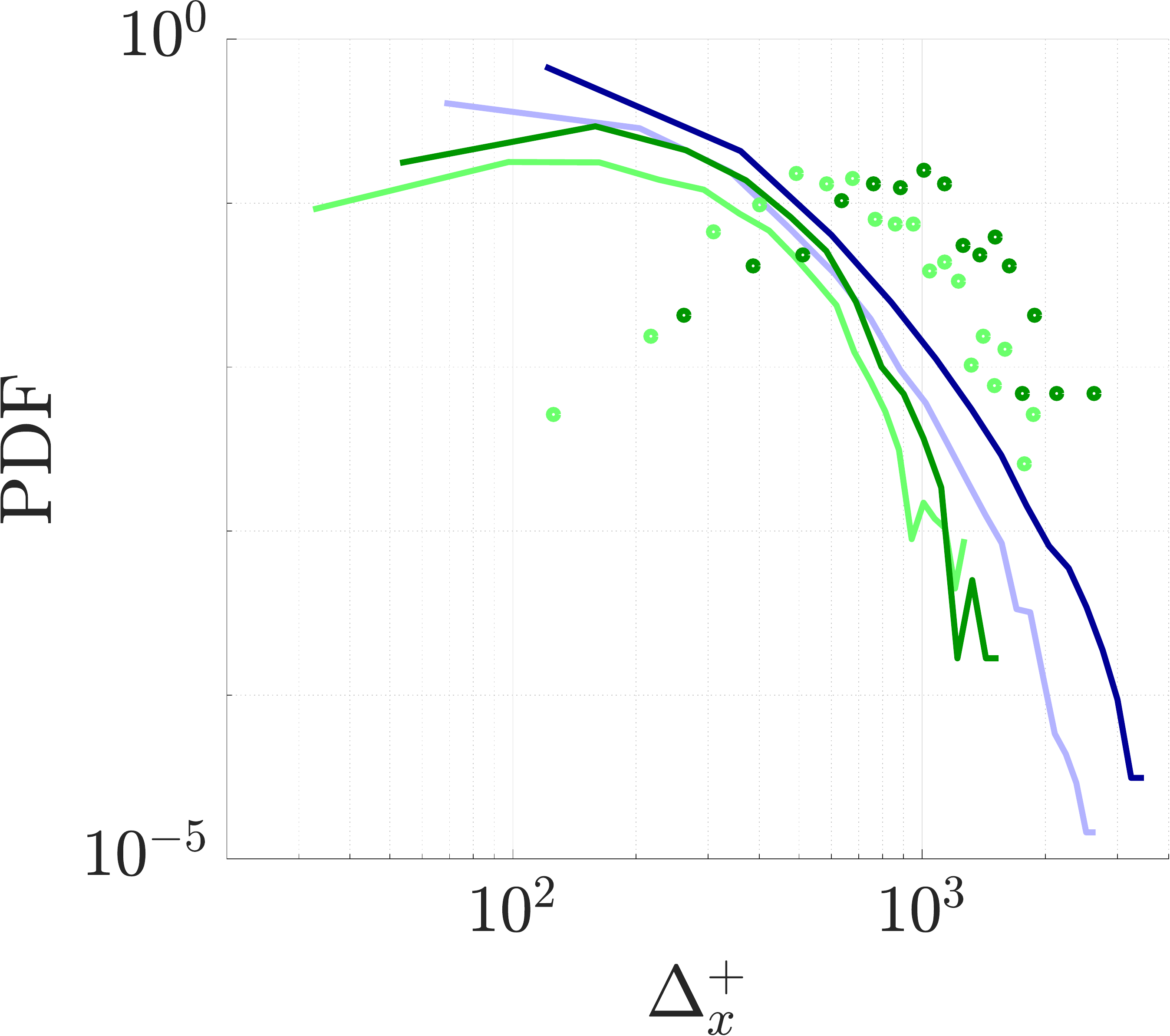}
    \includegraphics[width=0.325\textwidth]{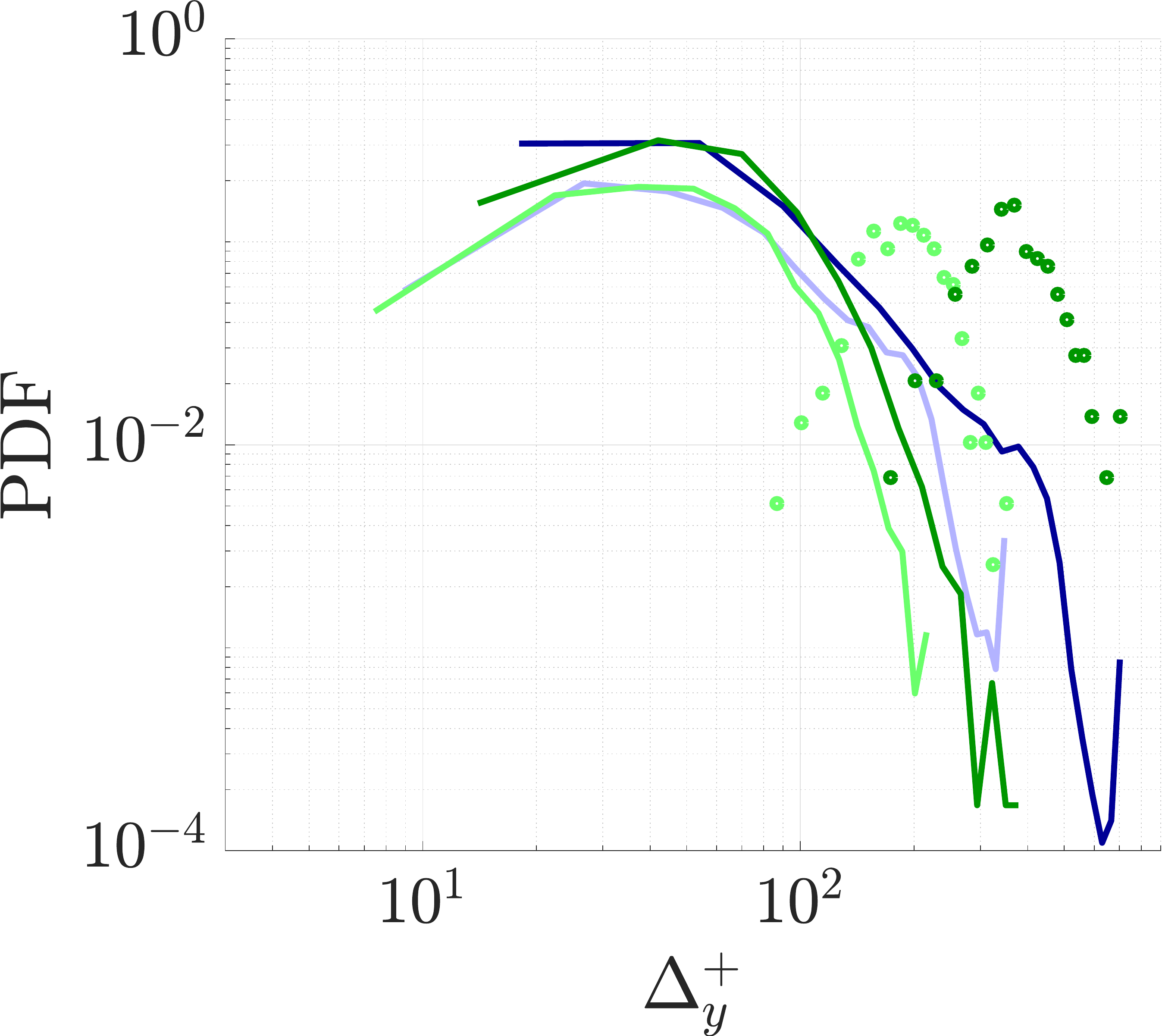}
    \includegraphics[width=0.325\textwidth]{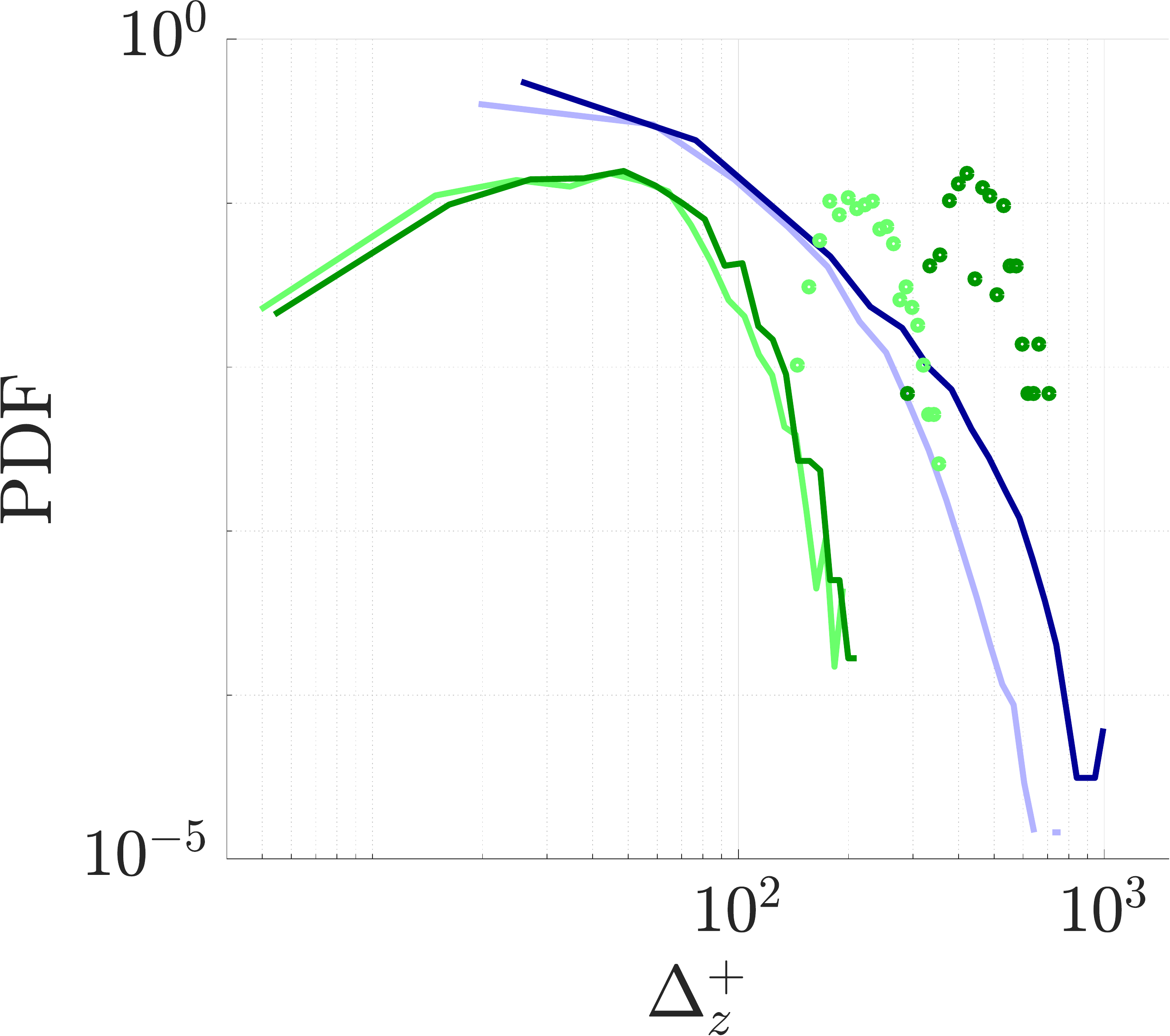}
    \caption{PDF of: (left column) $\Delta^+_x$, (middle column) $\Delta^+_y$ and (right column) $\Delta^+_z$, for $H_{uv}=2.0$. Top row: (blue solid lines) WA in channel, (green solid lines) WA in duct, region A, (green symbolds) WA in duct, region B, (dashed lines) CA in duct. Middle row: (blue solid lines) TWA in channel, (green solid lines) TWA in duct, region A, (green symbols) TWA in duct, region B. Bottom row: (blue solid lines) TWA in channel, (green solid lines) TCA in duct, region A, (green symbols) TCA in duct, region B. Bottom row, left: TCA in region B for $H_{uv}=1.0$ in symbols. Dark and light colour for $Re_\tau=360$ and $Re_\tau=180$, respectively.}
\label{fig:size_WA}
\end{figure}
Figure~\ref{fig:size_WA} (top) shows the PDF of the lengths in the three directions for the CA and the WA objects in regions A and B in the duct and the WA objects in the channel (CA are all included in region A of the duct by definition). 
There is good agreement between the data of the same family scaled in inner units for the two Reynolds numbers, as well as between the WA structures in region B of the duct and the WA objects in the channel. 
Interestingly, the WA structures in region A of the duct are longer than in region B and in the channel, despite having the same size in $y$ and $z$. 
On the other hand, the longest CA structures are as long as the longest WA objects in region B and the WA structures in channel flow, despite the fact that the most common CA structures are shorter than the WA structures for all the directions, and in particular in $z$. 
\par The presence of the vertical wall in the duct and the geometrical constraints of the domain have a strong impact on the sizes of TWA objects, which are shown in Figure~\ref{fig:size_WA} (middle) for the channel and the two spanwise regions in the duct. 
TWA structures in region A are significantly smaller than in region B or in the channel and their sizes scale in inner units in all the directions. 
On the other hand, since the largest TWA structures in region B of the duct and in the channel occupy the entire extension of the domain in the vertical direction, the distribution of $\Delta_y$ does not scale in inner units, but, interestingly, there is good agreement between channel and duct at the same Reynolds number.
\par It is important to note that the disagreement between the PDFs of $\Delta_x$ and $\Delta_y$ of the TWA structures in regions A and B in the duct is not entirely due to a fundamental difference in the geometrical properties of intense $uv$ events. 
Instead, it is in part a consequence of the fact that objects in region A that are large enough to match the size of the TWA structures in region B, also become attached to both walls and therefore are classified as TCA. 
\par The similarities between the large TCA and TWA objects in the absence of the vertical wall can be appreciated in Figure~\ref{fig:size_WA} (bottom), where the PDF of sizes are shown for the TCA structures in regions A and B of the duct and the TWA structures in the channel. 
The PDF of the sizes $\Delta^+_x$ and $\Delta^+_y$ of the TCA structures in regions A are in better agreement with those of TWA objects in the channel than the TWA sizes in the same region, although the largest TCA structures in the duct in region A remain shorter than the TWA in the channel, both in the $x$ and $y$ directions. 
The size in the spanwise direction is smaller for both TWA and TCA structures in region A, but for TCA it peaks at $\Delta_z^+=50$, with a relatively high number of structures for $\Delta^+_z>100$, namely twice the length of region A itself. 
\par The PDF distributions for TCA structures in region B is severely affected by the combined conditions of being attached to two contiguous walls and having the centre of mass at a relatively high distance from the vertical walls. 
Subsequently, this family is characterised by very large objects for which the PDFs of $\Delta^+_y$ and $\Delta^+_z$ scale well in outer units. The most common TCA structures in region B have $\Delta^+_y\approx180$ and $\Delta^+_y\approx360$ (\emph{i.e.} $\Delta_y\approx1$) for $Re_\tau=180$ and $Re_\tau=360$, respectively, and similar $\Delta^+_z$, whit both smaller and larger structures being relatively less likely. 
\par It is interesting to note that, as suggested by the dependency of the volume fraction for the different families on different $H_{uv}$, the geometrical properties of TCA objects in region B are the most affected by the choice of the threshold. 
\begin{figure} 
    \centering
    \includegraphics[width=0.325\textwidth]{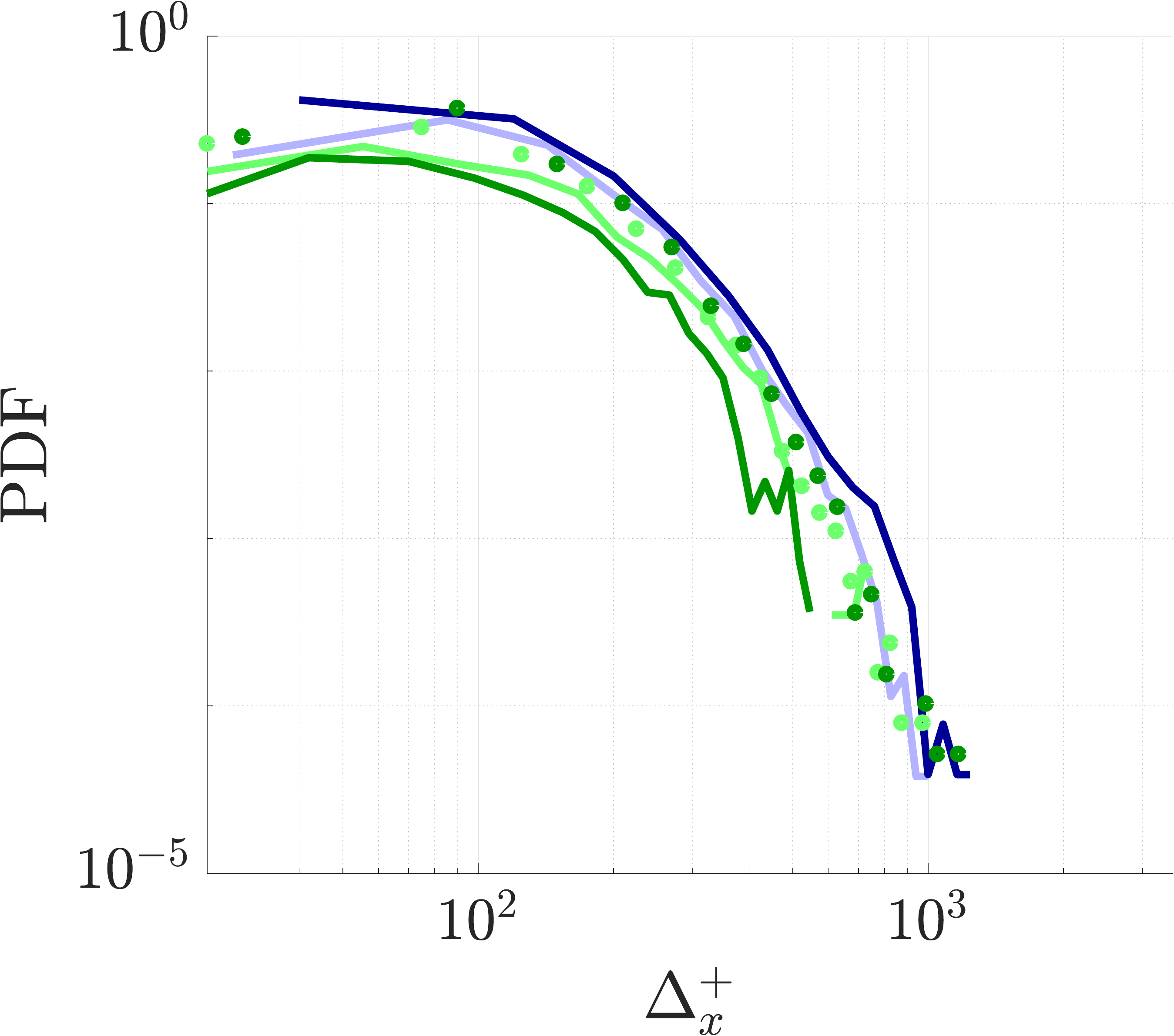}
    \includegraphics[width=0.325\textwidth]{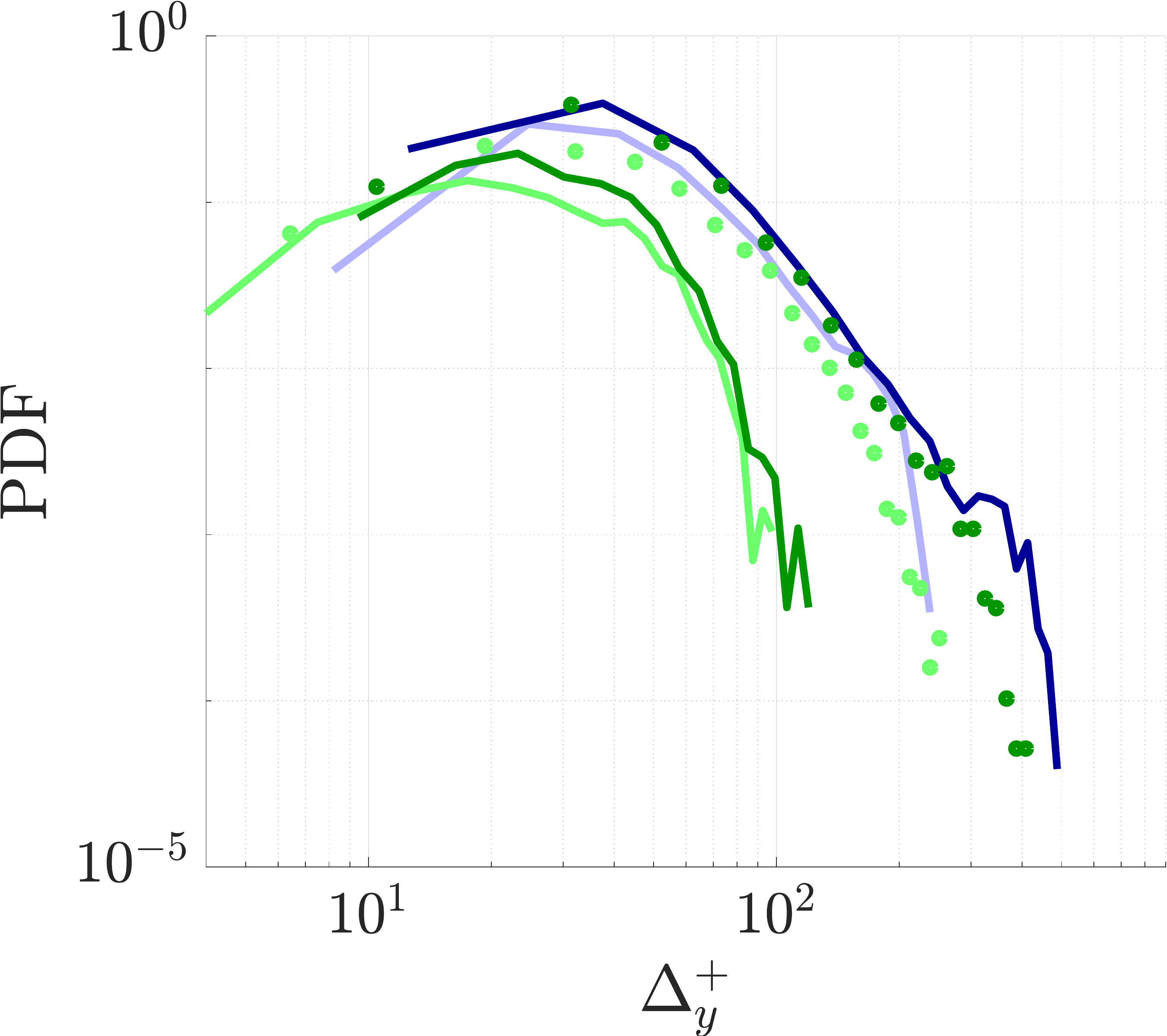}
    \includegraphics[width=0.325\textwidth]{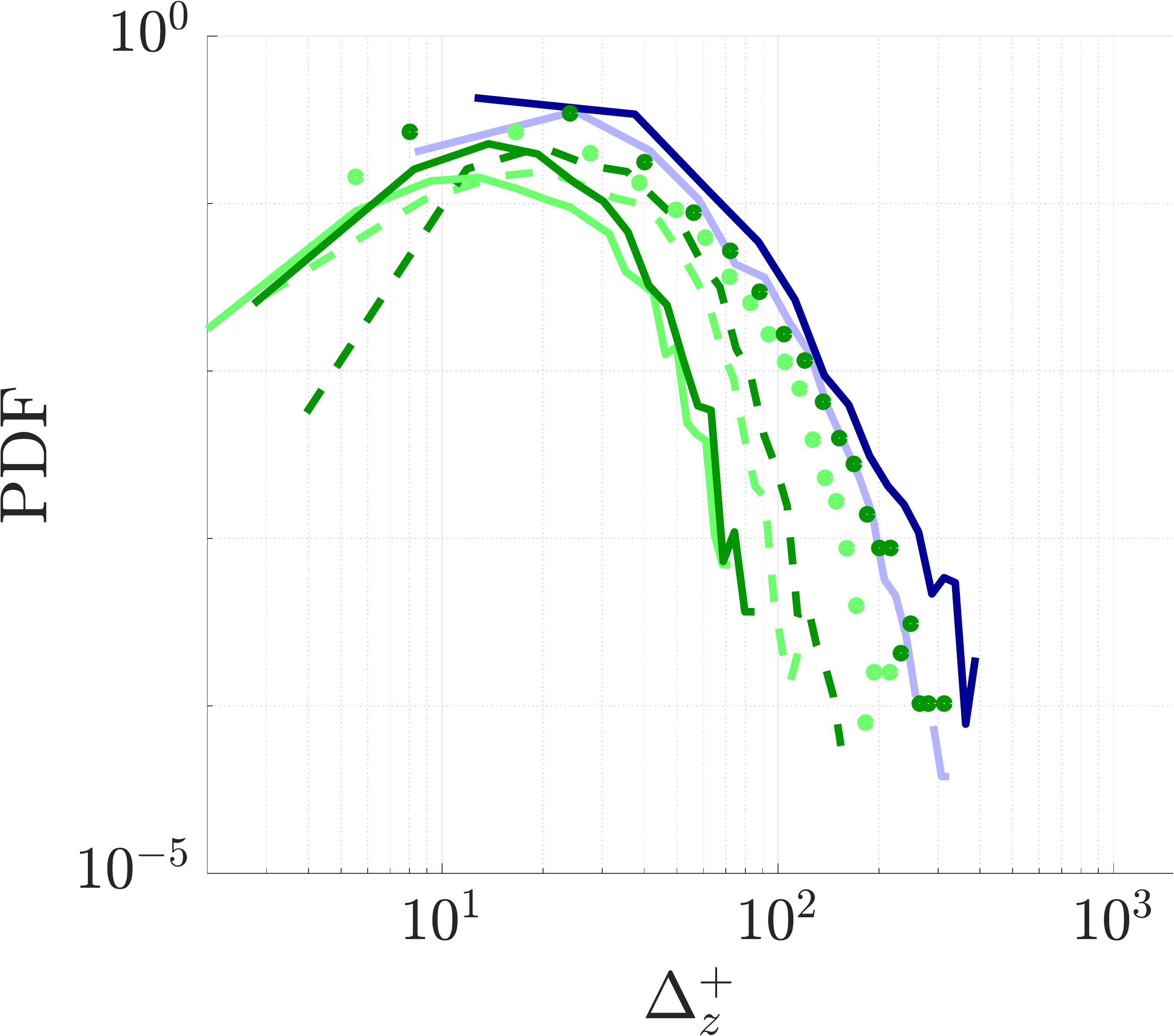}
    \includegraphics[width=0.325\textwidth]{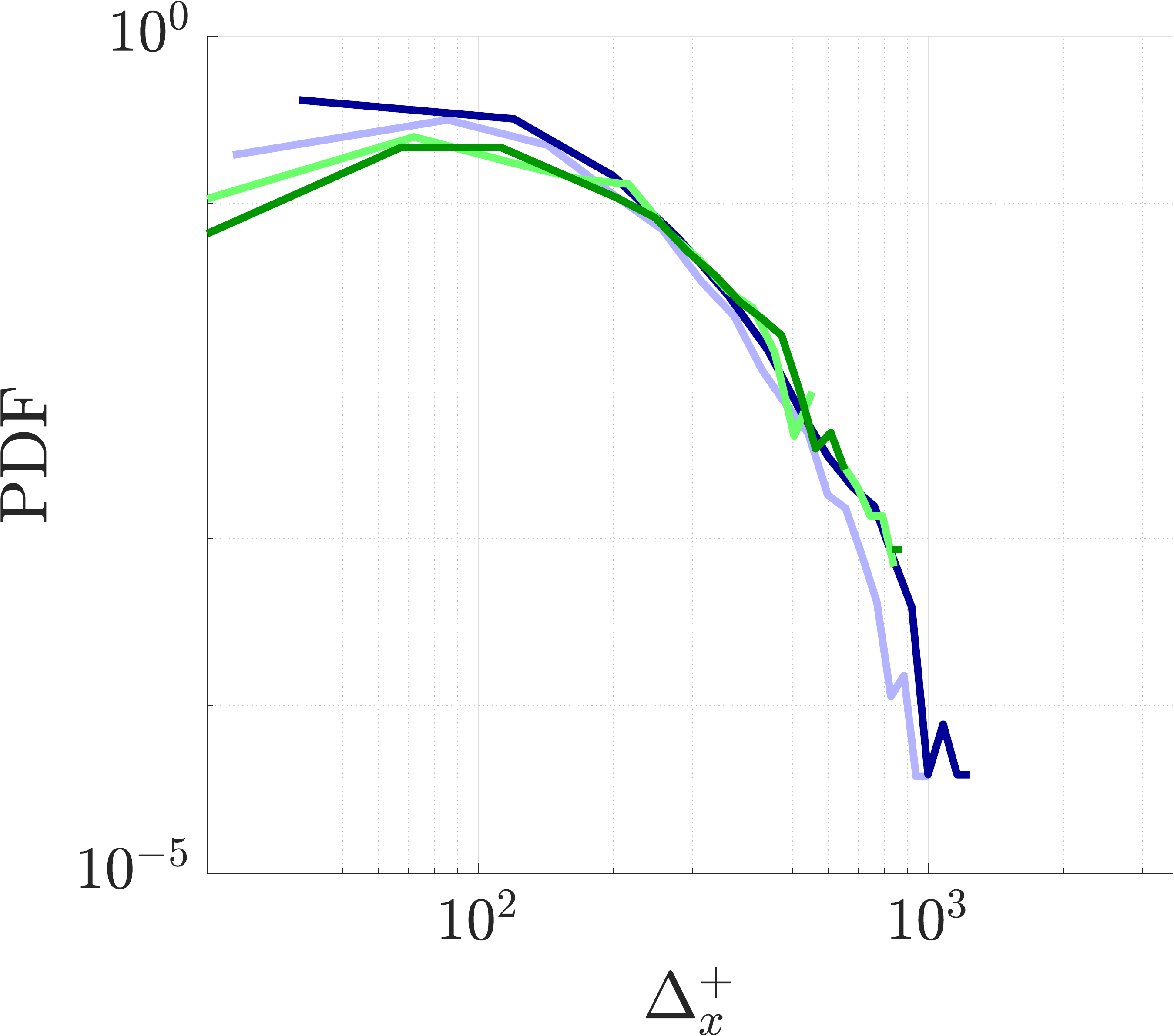}
    \includegraphics[width=0.325\textwidth]{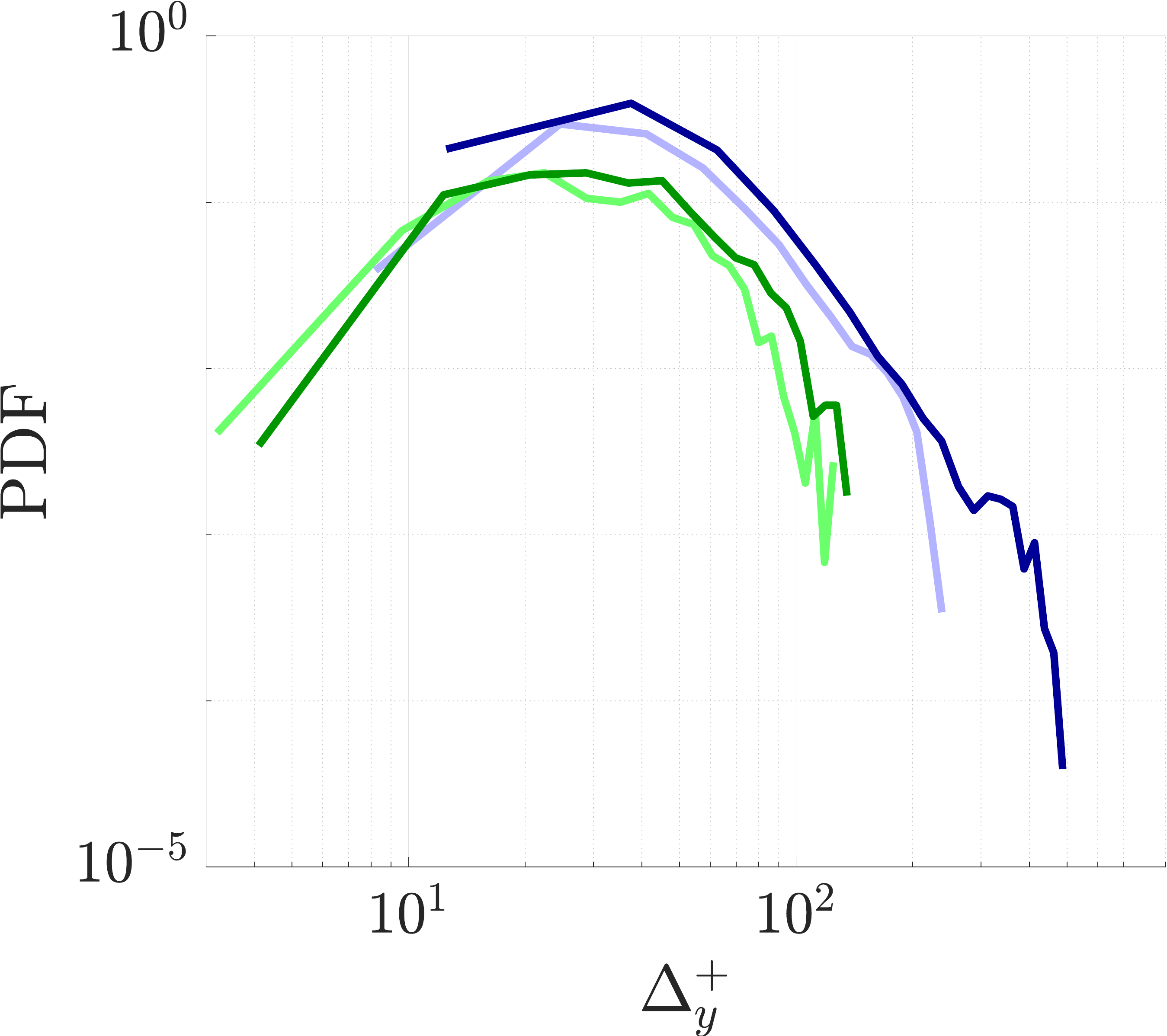}
    \includegraphics[width=0.325\textwidth]{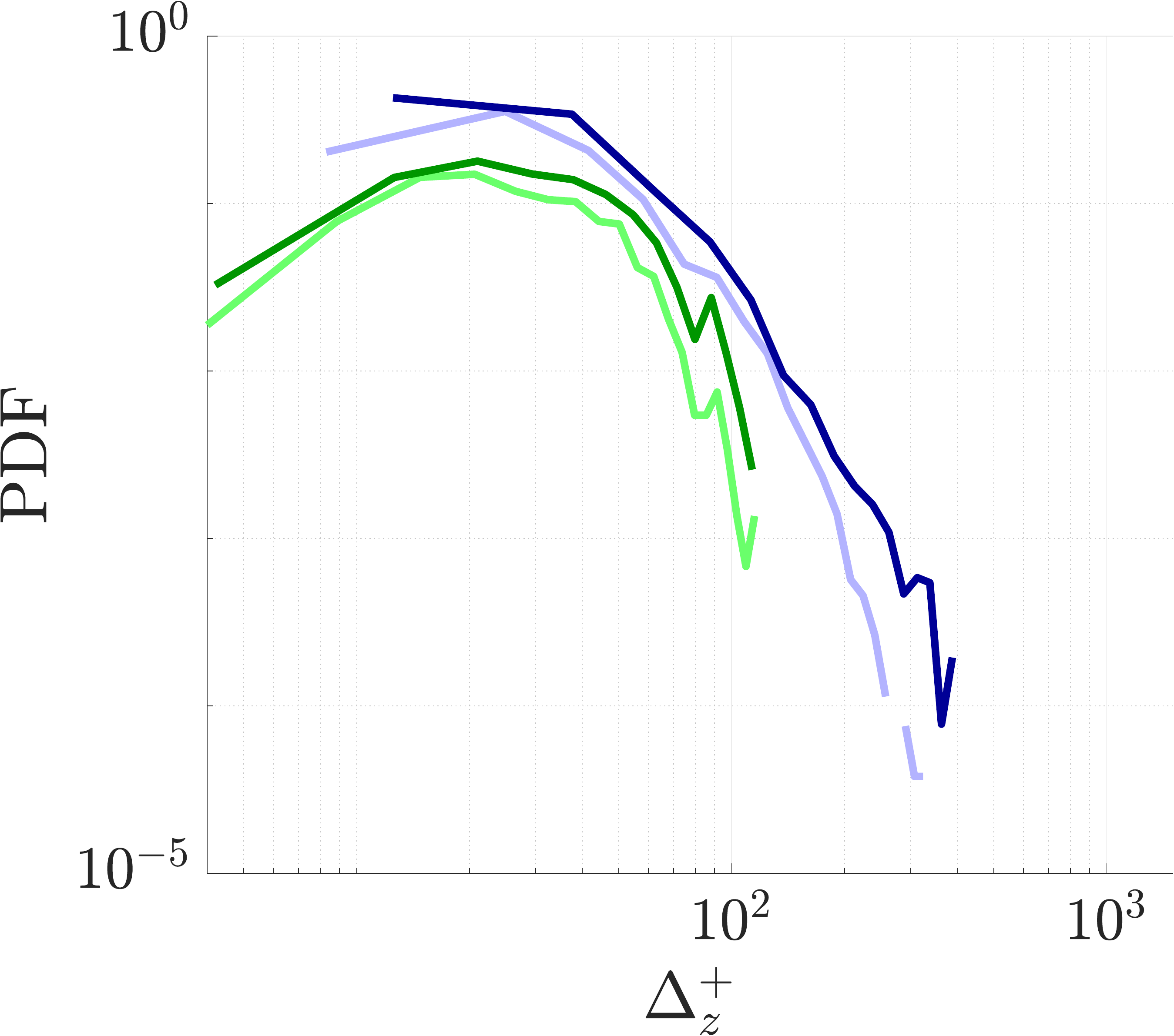}
    \caption{PDF of: (left column) $\Delta^+_x$, (middle column) $\Delta^+_y$ and (right column) $\Delta^+_z$, for $H_{uv}=4.0$. Top row: (blue solid lines) TWA in channel, (green solid lines) TWA in duct, region A, (green symbols) TWA duct, region B. Bottom row: (blue solid lines) TWA in channel, (green solid lines) TCA in duct, region A, (green symbols) TCA in duct, region B. Bottom row, left: TCA in region B for $H_{uv}=1.0$ in lines with symbols. Dark and light colour for $Re_\tau=360$ and $Re_\tau=180$, respectively.}
\label{fig:size_H40}
\end{figure}
This is shown only for $\Delta^+_x$: for $H_{uv}=1.0$, at both Reynolds numbers, several TCA structures in region B are almost as long as the entire computational domain, which is in agreement with the fact that such threshold is lower than the critical one. 
\par We describe the effects of a higher threshold considering the PDFs of the sizes for $H_{uv}=4.0$, which are shown in Figure~\ref{fig:size_H40}. 
As previously discussed, for this value of the threshold the total number of detected structures is lower than half of the total for $H_{uv}=2.0$, and $\mathcal{V}_{\rm all}$ is never above $2\%$ of the domain volume.
Nevertheless, most of the trends observed at $H_{uv}=2.0$ are still present. 
The PDFs of the sizes for small attached objects are in very good agreement with the curves at $H_{uv}=2.0$ and are not shown here, while the large attached objects are in general smaller in both channel and duct. 
The PDFs of $\Delta_x^+$ and $\Delta_y^+$ for TWA structures in region B of the duct remain in good agreement with whose of the TWA structures in the channel, and the $\Delta_z^+$ PDFs are more similar in the two flows than at $H_{uv}=2.0$. 
Furthermore, for $H_{uv}=4.0$ the TWA structures are shorter then for $H_{uv}=2.0$ both in the core of the duct and in the channel and thus the agreement of their $\Delta^+_x$ PDF with that of TCA structures in the corner region of the duct is better at $H_{uv}=4.0$ then at lower values of $H_{uv}$. 
\par The most relevant differences between the properties of the structures for $H_{uv}=2.0$ and $H_{uv}=4.0$ are related to the size the structures in the vertical direction. 
In fact, whereas in both channel and duct at $H_{uv}=2.0$ it is possible to observe the existence of objects as large in $y$ as the vertical length of the domain, at $H_{uv}=4.0$ the probability of such events to be detected is negligible. Furthermore, TCA objects in region A of the duct are more similar to the TWA structures in the same region, as opposed to what is observed for the TWA structures in region B, and are too small to be affected by geometrical constrains. Subsequently, their $\Delta_y^+$ PDFs scale in inner-units. 
\par Finally, it is interesting to note that the TCA family is virtually absent from region B of the duct at both Reynolds numbers. 
\subsubsection{Shape of structures in different families}
Further information about the geometrical properties of the different families can be obtained considering the aspect ratio of the bounding boxes. 
\begin{figure}
\centering
    \includegraphics[width=0.325\textwidth]{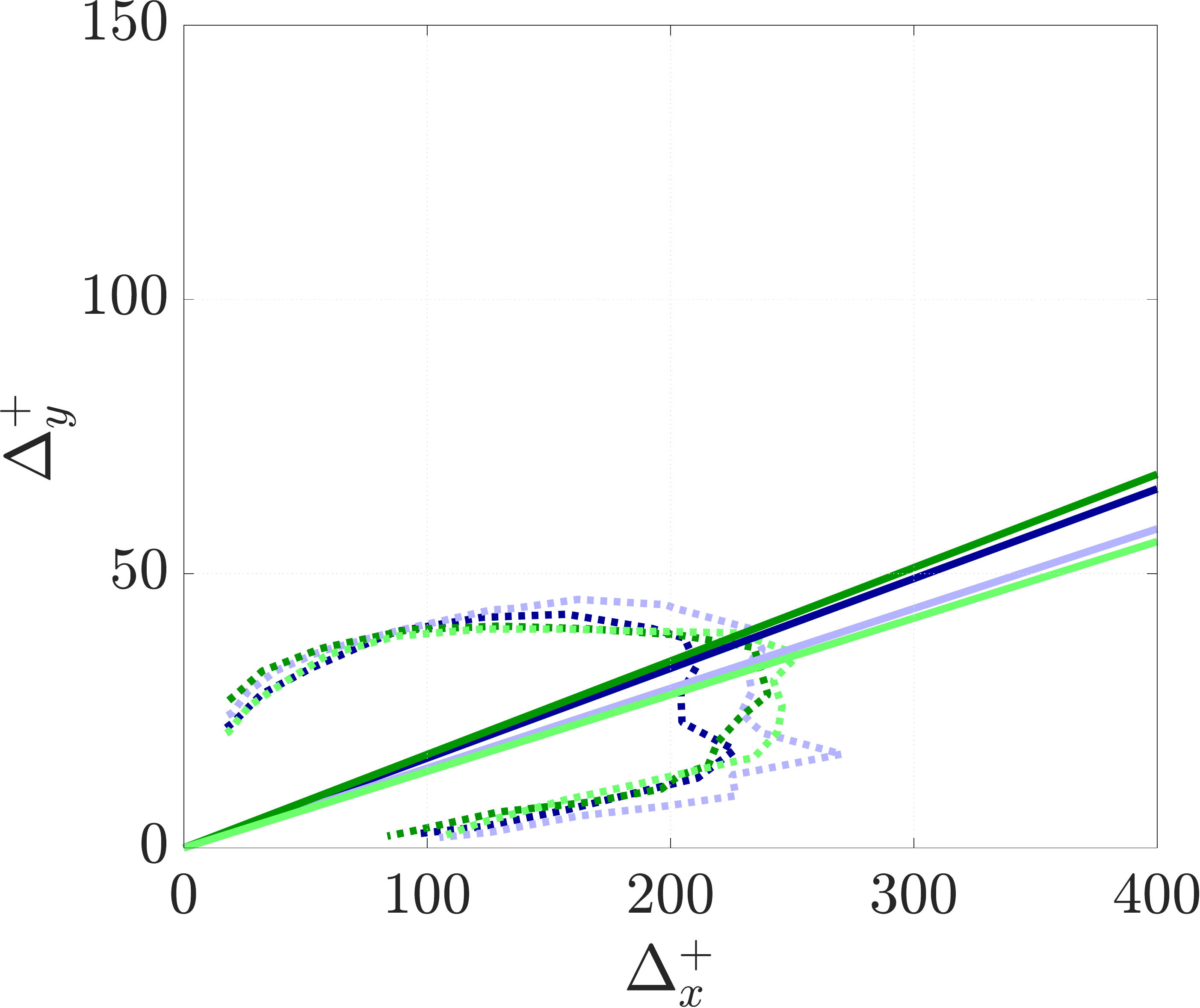}
    \includegraphics[width=0.325\textwidth]{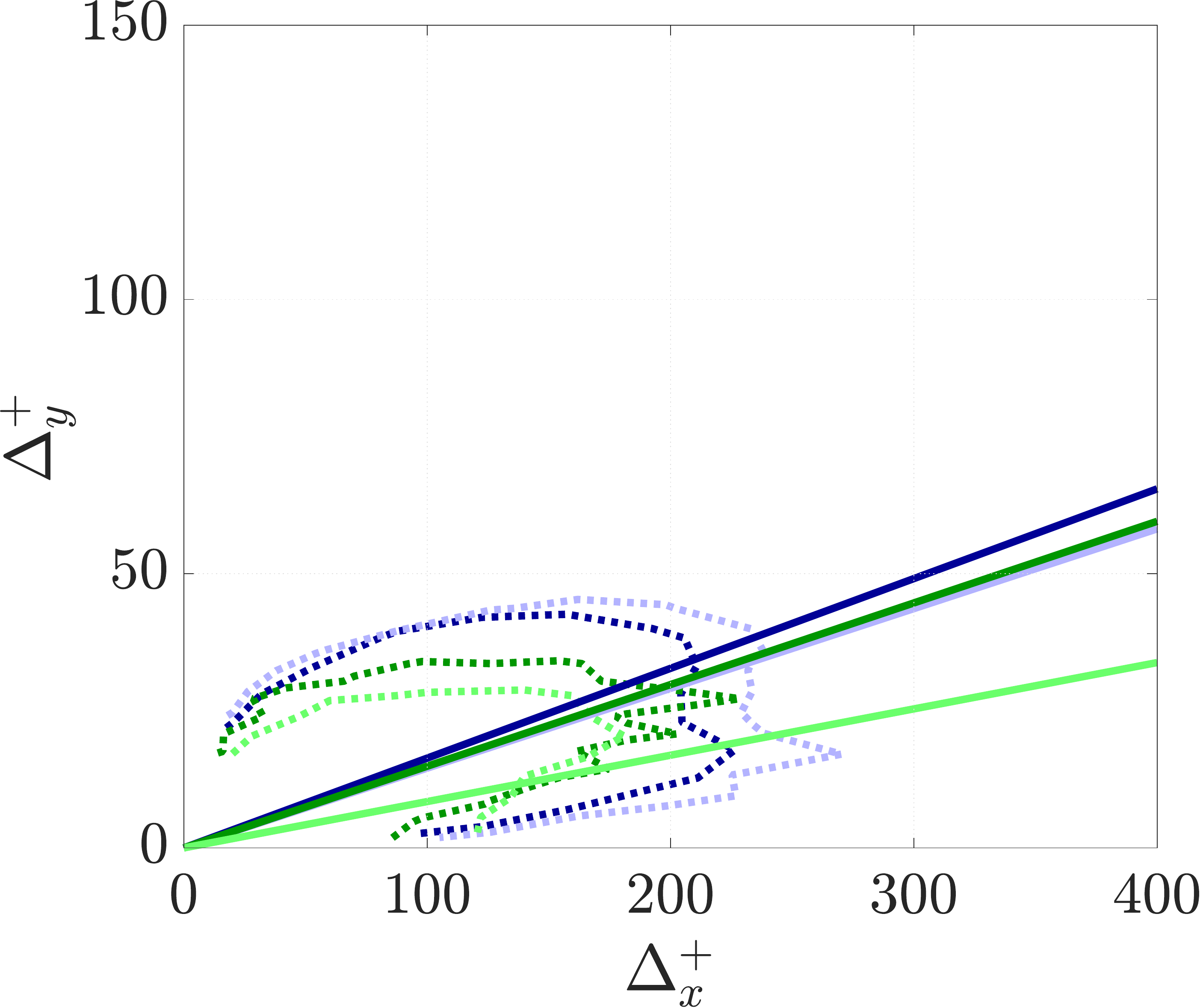} \\
    \includegraphics[width=0.325\textwidth]{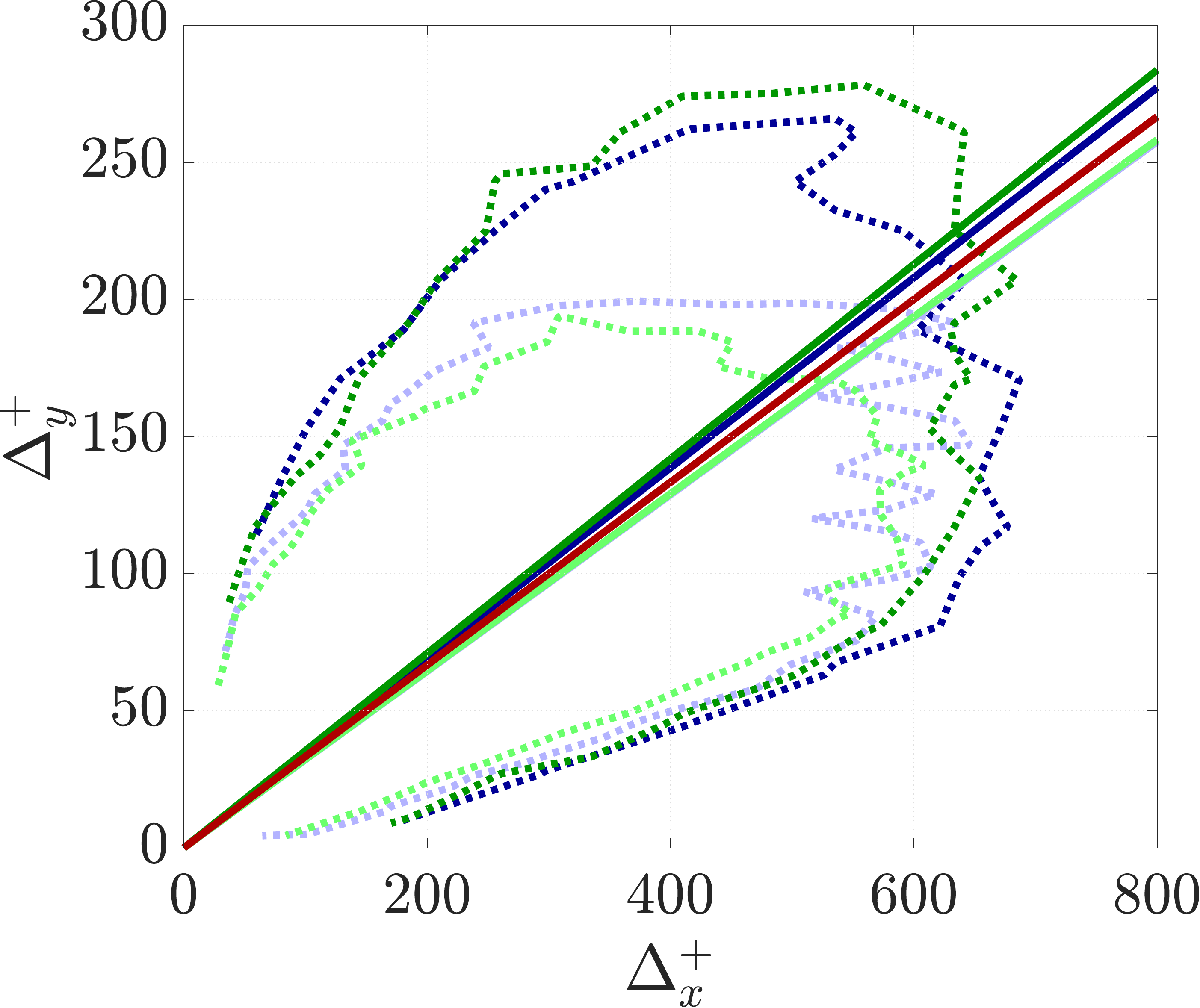}
    \includegraphics[width=0.325\textwidth]{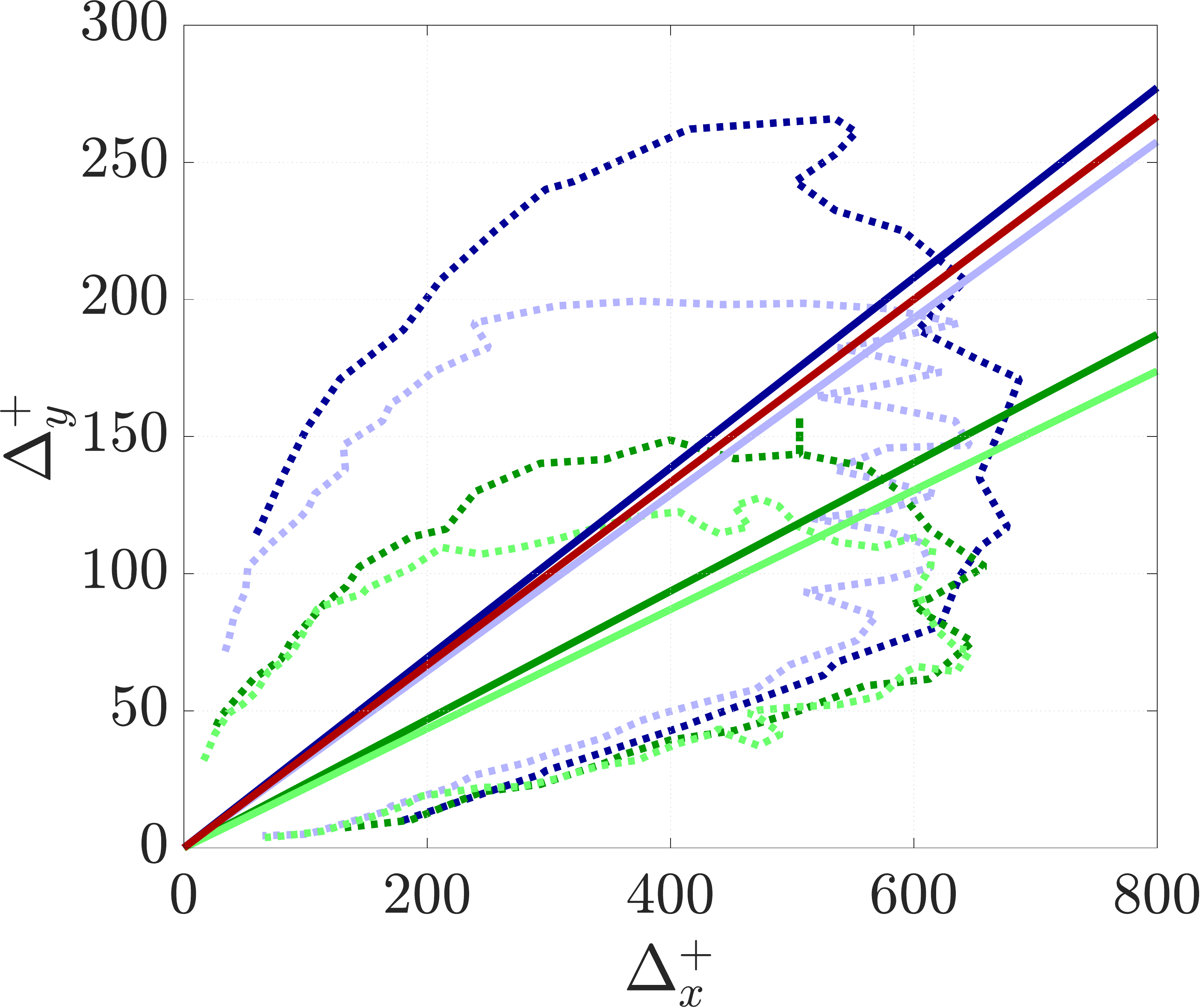}
    \includegraphics[width=0.325\textwidth]{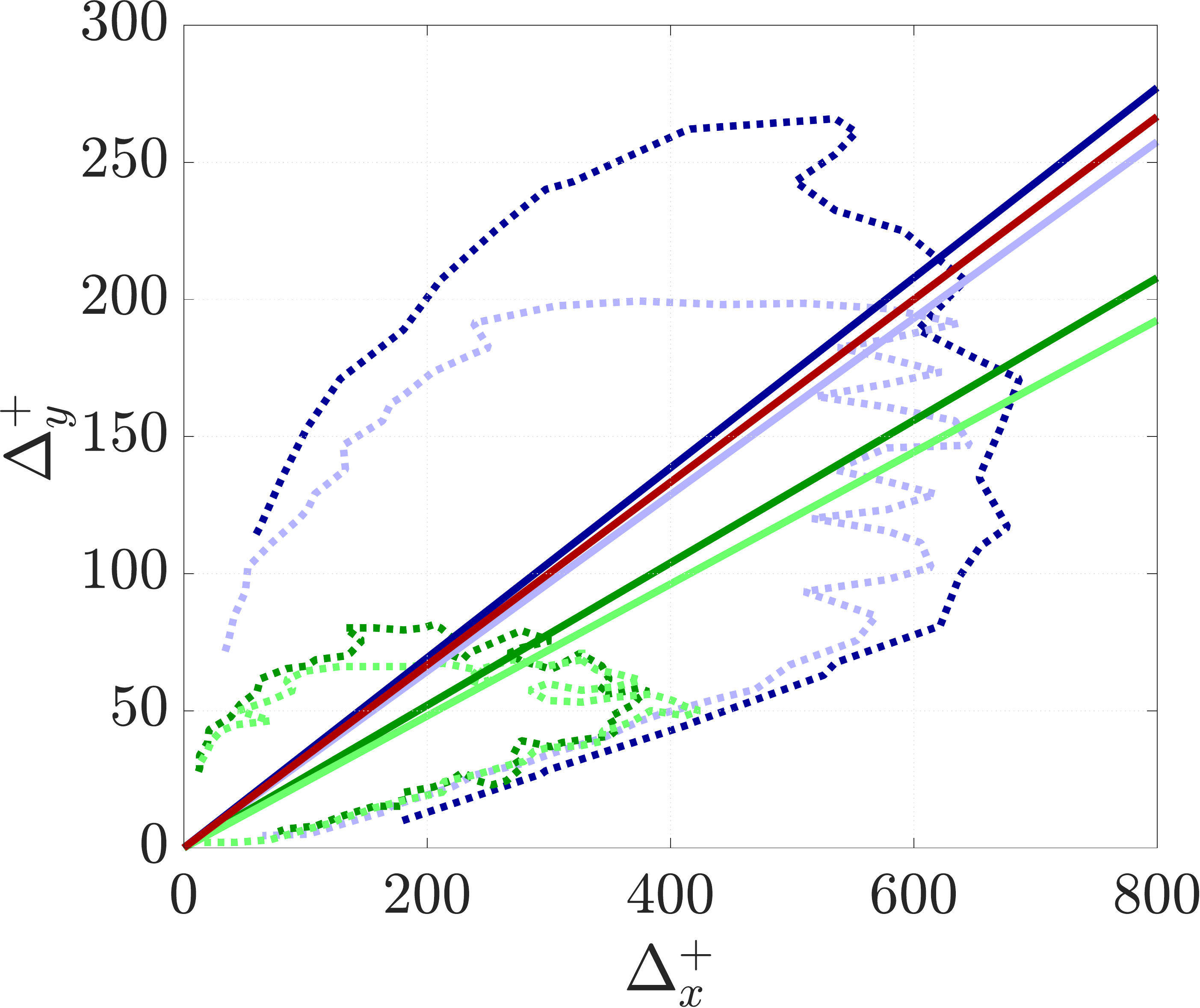}
\caption
{
Comparison between the contours of the JPDFs of $\Delta_x^+$ and $\Delta^+_y$ for different structures (dotted lines) and the corresponding lines $\Delta_x = \alpha_{xy} \Delta_y$ (solid lines), where $\alpha_{xy}$ is the average aspect ratio of the bounding box. Blue and green lines for channel and duct, respectively, and dark and light colours for $Re_\tau=360$ and $Re_\tau=180$, respectively. Top row: WA structure in the channel, compared with (left) WA structures in region A and CA structures (in region A) in the duct. 
Bottom row: TWA structured in the channel, compared with (left) TWA structures in region B, (middle) TWA structures in region A and (right) TCA structures in region A of the duct. 
The red solid line is $\alpha_{xy}=3$, reported by \cite{loza12} for TWA in channel flow at higher $Re_\tau$.
}
    \label{fig:size_aspect}
\end{figure}
Figure~\ref{fig:size_aspect} shows the JPDF of $\Delta^+_x$ and $\Delta^+_y$ for $H_{uv}=2.0$, together with the average aspect ratio $\alpha_{xy}=\Delta^+_x/\Delta^+_y$. 
\par As suggested by the PDFs examined before, small objects in the near-wall region are self similar and their shape is almost the same for both the Reynolds numbers in all the considered cases. In particular, $\alpha_{xy}$ is between $5.9$ and $7.2$ for WA structures in the channel, WA structures in the duct in region B and CA structures in the duct at $Re_\tau=360$. The CA structures in the duct at $Re_\tau=180$ are more elongated and their average aspect ratio is $\alpha_{xy}=12$. The shape of these objects resembles that of the streaks which have been widely observed in the near-wall region \citep{klin67,gupt71}. 
\par According to the previous findings on the PDFs of the sizes, the TWA structures in region B of the duct resemble more closely the ones in the channel than the TWA objects in region A (Figure~\ref{fig:size_aspect}, middle).
The aspect ratio of TWA structures in the core of the duct and in the channel is approximately $\alpha_{xy}\simeq3.1$ and $\alpha_{xy}\simeq2.8$ for $Re_\tau=180$ and $Re_\tau=360$, respectively, which are in good agreement with that reported for the same structures by \cite{loza12} in channel flow at much higher Reynolds number ($\alpha_{xy}\simeq3$). 
On the other hand, TWA structures in region A are relatively more similar to TCA structures in the same region  (Figure~\ref{fig:size_aspect}, right), although they are slightly less elongated on average. In fact, at $Re_\tau=180$, the former and the latter have $\alpha_{xy}\simeq4.2$ and $\alpha_{xy}\simeq4.6$, respectively.
\par The same trend is also observed at $Re_\tau=360$ which become $\alpha_{xy}\simeq3.8$ and $\alpha_{xy}\simeq4.3$. 
This is probably related to the fact that farther from the corner the preferential position of the streaks is not fixed, and the local flow behaviour is closer to that in turbulent channels \citep{pine10,vinu18}. 

\section{Conclusions}
In the present work we have studied the properties of coherent structures in turbulent square duct at moderate Reynolds number, with the aim of identifying the feature of the instantaneous turbulent flow responsible for the secondary motion of Prandtl's second kind. 
The coherent structures are defined as regions in the flow where the fluctuations of two components of the velocity are strongly correlated (or anti-correlated), in the spirit of the three-dimensional extension of the quadrant analysis proposed by \cite{loza12} in channel flow. 
This is, to our knowledge, the first time that this methodology is employed to investigate the secondary motion of Prandtl's second kind. 
This analysis is used to verify the hypothesis that the secondary flow can be directly connected to of the intense velocity fluctuations characteristic of wall-bounded turbulent flows.
\par We have performed a comparative analysis between duct and channel at similar Reynolds number considering two different points of view: the fractional contributions of the structures to the vertical component of velocity, and their scaling properties.
Furthermore, due to the multi-scale nature of the secondary motion, we have focused on two different spanwise regions of the domain in the duct. The corner region (region A in the text above, {\it i.e.} $z^{+}<50$) is defined to include the locations where the mean vertical component of the velocity $V$ scales if it is expressed in outer units and the wall-normal distance is expressed in inner units. 
The core region (region B, {\it i.e.} $z>0.6$) is defined to include the region where $V$ scales in outer units, although we note, according to the recent findings by \cite{gavr19}, that the topology of the secondary motion is still evolving at this range of Reynolds numbers. 
\par In the core region of the duct, the very intense events are similar to those in the channel. 
Their contribution to the mean vertical component of the velocity is in good agreement in both flows, despite the existence of the secondary motion in the duct. 
At $Re_\tau=180$ for a comparably large threshold $H_{uv}=2.0$ the structures account for only $\approx7\%$ of the volume of the computational domain but the contribution to $V$ of intense events $V_{uv}^>$ is of the same order of magnitude as the mean $V$ in the duct, in both channel and duct. 
At $Re_\tau=360$ and for the same $H_{uv}$, $V_{uv}^>$ also in good agreement with that in channel, but it is significantly smaller than $V$. Overall, in the core region, if $H_{uv}$ is such that the intense events are isolated from the rest of the flow, $V_{uv}^>$ is not related with $V$.
On the other hand, it is the contribution from the portion of the domain which does not belong to intense events, $V_{uv}^<$, that significantly differs between channel and duct. 
While in the former $V_{uv}^<=-V_{uv}^>$ since $V=0$, in the latter the dependence on the wall-normal distance of $V_{uv}^<$ is very similar to the one of $V$ and often $|V_{uv}^<|>|V_{uv}^>|$. 
Similarities between intense events also appear in their geometrical properties. 
In both the channel and the core region of the duct, wall-attached and tall-wall attached $uv$ structures are present and the largest objects tend to reach the two opposite walls. 
Furthermore, for very high values of the threshold (\textit{e.g.} $H_{uv}=4.0$), for which smaller objects are detected and the geometrical constraints are less relevant, such similarities are more pronounced. 
Not only the probability density functions of the sizes for tall-wall attached objects in channel and duct agree better, but the tall-corner attached family, which is only found in the duct, virtually disappears. 
\par The corner region of the duct, where the secondary motion reaches its highest intensity ($\approx2\%$ of the streamwise mean velocity $U$), exhibits qualitative differences compared with the core region. 
In this region the intense events are similar to those in the channel only below the corner bisector. 
This similarity leads to a peculiar consequence: at the location of the minimum of $V$, $V^>_{uv}$ has sign opposite than $V$. 
Above the corner bisector $V^>_{uv}$ in the duct is not anymore in agreement with that in the channel, and it has the same sign as $V$. 
This is the only region of the duct where the fractional contribution of intense events is clearly related to the presence of the secondary motion. 
However, $V_{uv}^>$ is a low fraction of $V$. In fact, if the same threshold $H_{uv}$ is considered, $V^>_{uv}$ is lower than in the core (except near the wall) and $V^<_{uv}$ reaches the same order of magnitude as $V$ already for $H_{uv}\simeq2.0$. 
%
%
%
%
The geometrical properties of the structures are also different than those in the core region, and in particular intense events close to the corner are on average more elongated in the streamwise direction.
\par We note that, despite the role the transverse Reynolds-stress component $\overline{vw}$ plays in the transport equation for the streamwise component of the mean vorticity, the contribution of intense $vw$ events to $V$ is lower than that of $uv$, although the (geometric) characteristics of both types of events are qualitatively similar. For this reason they have not been described in detail in this study. 
\par Our findings lead to an unexpected conclusion: intense Reynolds-stress events do not have a major contribution to the secondary flow in duct flows, neither to the core nor to the corner circulation. 
In the case of the former, because in that region intense $uv$ events are almost indistinguishable from the same type of event in channel flow, and in case of the latter because the net contribution of intense events is almost negligible if compared with the contribution from the complementary portion of the domain.
With a more general perspective, it is possible to conjecture that strong ejections, despite the fact that they play a key role in the self-sustaining process of wall-bounded turbulence observed in canonical flows, are not directly related with the secondary flow of Prandtl's second kind. 
\par Nevertheless, the differences in the geometrical properties of structures in the corner and the core region, together with the aforementioned results on vortex clusters \citep{uhlm07,pine10,vinu16}, still suggest that the secondary flow can be described in terms of instantaneous features of the turbulence flow. 
This possibility remains a motivation for future studies, which will need to answer the question of whether it is possible to isolate the events responsible for this elusive phenomenon.

\section*{Acknowledgments}
This study was funded by the Swedish Foundation for Strategic Research, project ``In-Situ Big Data Analysis for Flow and Climate Simulations'' (ref. number BD15-0082) and by the Knut and Alice Wallenberg Foundation. The simulations were performed on resources provided by the Swedish National Infrastructure for Computing (SNIC).

\bibliographystyle{jfm}
\bibliography{ms}

\end{document}